\documentclass[a4paper,11pt,nohyper]{JHEP3}
\usepackage{amsmath,latexsym,amssymb}
\usepackage{graphicx}
\usepackage{psfrag}
\usepackage[latin1]{inputenc}
\usepackage{subfigure}
\usepackage{nicefrac}
\usepackage{bbm}
\usepackage[stable]{footmisc}

\usepackage{comment}

\newcommand\nn{\nonumber}
\newcommand\bbone{\ensuremath{\mathbbm{1}}}

\newcommand{\eq}[1]{\begin{equation}#1\end{equation}}
\newcommand{\spl}[1]{\begin{split}#1\end{split}}
\newcommand{\al}[1]{\begin{align}#1\end{align}}
\newcommand{\subeq}[1]{\begin{subequations}#1\end{subequations}}

\newcommand{\arXividhepth}[1]{\href{http://arxiv.org/abs/#1}arXiv:{\tt #1} [hep-th]}
\newcommand{\arXividother}[2]{\href{http://arxiv.org/abs/#1}arXiv:{\tt #1} [#2]}

\newcommand{\bg}[1]{\hat{#1}}
\newcommand{\vols}{V_s}

\def\d{\text{d}}

\def\slashchar#1{\setbox0=\hbox{$#1$}           % set a box for #1
\dimen0=\wd0                                 % and get its size
\setbox1=\hbox{/} \dimen1=\wd1               % get siste of /
\ifdim\dimen0>\dimen1                        % #1 is bigger
\rlap{\hbox to \dimen0{\hfil/\hfil}}      % so center / in box
#1                                        % and print #1
\else                                        % / is bigger
\rlap{\hbox to \dimen1{\hfil$#1$\hfil}}   % so center #1
/                                         % and print /
\fi}

\def\Re           {{\rm Re\hskip0.1em}}
\def\Im           {{\rm Im\hskip0.1em}}

\newcommand{\E}{\text{\tiny E}}

\title{The effective theory of
type IIA AdS$_4$\\ compactifications
on nilmanifolds and cosets}

\author{Claudio Caviezel${}^{\diamondsuit}$,
Paul Koerber${}^{\diamondsuit}$, Simon K\"ors${}^{\diamondsuit}$, Dieter L\"{u}st${}^{\diamondsuit\clubsuit}$,
 Dimitrios Tsimpis${}^{\clubsuit}$ and Marco Zagermann${}^{\diamondsuit}$ \\

\begin{itemize}
  
\item  Max-Planck-Institut f\"ur Physik\\
F\"ohringer Ring 6, 80805 M\"unchen, Germany
  
\item  Arnold-Sommerfeld-Center for Theoretical Physics\\
Department f\"ur Physik, Ludwig-Maximilians-Universit\"at M\"unchen\\
Theresienstra\ss e 37, 80333 M\"unchen, Germany
  \end{itemize}

\bigskip
 E-mail:
\email{caviezel@mppmu.mpg.de}, \email{koerber@mppmu.mpg.de}, \email{koers@mppmu.mpg.de}, \email{dieter.luest@lmu.de} \& \email{luest@mppmu.mpg.de}, \email{dimitrios.tsimpis@lmu.de}, \email{zagerman@mppmu.mpg.de}}

\abstract{We consider string theory compactifications of the form
  AdS$_4\times {\cal M}_6$ with orientifold six-planes,
where ${\cal M}_6$ is a six-dimensional compact space that is either a nilmanifold or a
coset.  For all known solutions of this type we obtain the
 four-dimensional $\mathcal{N}=1$ low energy effective theory by computing the
superpotential, the K\"{a}hler potential and the mass spectrum for the
light moduli. For the nilmanifold examples we perform a cross-check on the result
for the mass spectrum by calculating it alternatively from a direct Kaluza-Klein reduction and
find perfect agreement.
We show that in all but one of the coset models all moduli are
stabilized at the classical level. As an application we show that all but one
of the coset models can potentially be used to bypass a recent no-go theorem against
inflation in type IIA theory.
}

\preprint{MPP-2008-62\\
LMU--ASC 37/08}


\begin{document}
\setcounter{footnote}{0}
\renewcommand{\thefootnote}{\arabic{footnote}}
\setcounter{section}{0}
\section{Introduction and summary}
\label{introduction}

\subsection*{Motivation}
String compactifications with non-trivial $p$-form fluxes provide
a class of backgrounds in which the stabilization of moduli fields
-- a phenomenologically problematic feature of traditional compactifications on spaces such as Calabi-Yau manifolds --
can be realized in a controlled way within the classical supergravity approximation. Motivated by this and other applications, the derivation of the four-dimensional (4D)
low energy effective actions of flux compactifications has been an active area of research during the past decade
 (for reviews and a more complete list of references, see, e.g., \cite{granareview,morereviews,blumenhagen}).

One aspect that complicates the derivation of these effective actions is that the $p$-form fluxes generally back-react on the geometry of the compact manifolds,
deforming them away from well-understood classes such as Calabi-Yau spaces.
This back-reaction can be rather mild, as, e.g., in
 type IIB orientifolds with D3/D7-branes \cite{gkp}, where the internal space is still conformal to a Calabi-Yau manifold. In these comparatively simple models, however, the fluxes turn out to stabilize only the dilaton and the complex structure moduli, while
 the K\"{a}hler moduli stabilization requires the use of quantum effects, e.g., along the lines of \cite{kklt}.

In the present paper, we will instead be interested in a different class of flux compactifications for which the
back-reaction of the fluxes on the geometry is less trivial. Concretely, we will study
$\mathcal{N}=1$ supersymmetric flux compactifications of type IIA string theory
on background spaces of the form AdS$_4\times {\cal M}_6$, where ${\cal M}_6$ is a six-dimensional compact space that is
in general not a Calabi-Yau manifold. Being compactifications to AdS space-time, these models do not appear
realistic as such, but they can serve as starting points for the construction of more realistic setups or have other applications. More specifically, some of the motivations for studying this class of string vacua are:

\begin{itemize}
\item In certain AdS$_4$ type IIA vacua, it is possible to stabilize all moduli at tree level in a controlled supergravity regime. It is then an interesting question whether these compactifications can be of phenomenological use, e.g., after the inclusion of an additional uplifting potential so as to construct meta-stable de Sitter vacua in the spirit of the IIB models discussed in \cite{kklt}. But even without an explicit uplift potential, one can investigate whether the potential already has meta-stable de Sitter vacua away from the AdS vacuum or whether there are regions suitable for slow-roll inflation.

\item Although no longer Calabi-Yau spaces, the manifolds ${\cal M}_6$ we discuss in this paper
have a surprisingly simple and explicitly known geometry, which makes them in some sense even more tractable than a Calabi-Yau manifold. In particular, it is possible to derive the low energy spectrum on AdS$_4$ directly from an explicit Kaluza-Klein reduction, as we demonstrate for some of the models. For these models, we find complete agreement with the results
from the more commonly used, but less direct, $\mathcal{N}=1$ supergravity techniques, which provides a valuable consistency check.

\item Replacing fluxes by branes, the above AdS vacua can be potentially obtained as near-horizon geometries
of intersecting branes \cite{klpt}. In view of recent developments following the Bagger-Lambert-Gustavsson theory \cite{blg},
AdS$_4$ flux vacua of the type considered here may then admit a full non-perturbative definition via a dual three-dimensional CFT. The above-mentioned brane solutions also correspond to domain walls that interpolate
between different flux vacua. The existence of these domain walls may correspond to interesting transitions in the landscape of flux vacua.
\end{itemize}

The first examples of type IIA models where all the moduli were stabilized at tree level
in a controlled classical supergravity regime were the torus orientifold models of \cite{dkpz,vill,dewolfe,fontaxions}.
Possible cosmological applications along the lines described in the first item above were subsequently explored in a number of papers, with surprisingly little success. In \cite{kalloshuplift}, for instance, a simple F-term uplift
to a meta-stable de Sitter vacuum based on an effective O'Raifeartaigh sector was found to be impossible. Using similar arguments, the authors of \cite{kachruinflation,kachrunogo} could also formulate a no-go theorem against slow-roll inflation and de Sitter vacua
for general type IIA models with only 3-form NSNS-flux, RR-fluxes, D6-branes and O6-planes.
As additional ingredients that can circumvent this no-go theorem, the authors of \cite{kachrunogo} identified geometric fluxes, NS5-branes and/or
the more exotic non-geometric fluxes.\footnote{Recent progress obtaining inflation with these ingredients appeared in \cite{silversteindS,Silverstein:2008sg}.}
We will give a short discussion of this no-go theorem in the context of the models considered in this paper, as they contain one of these additional
ingredients, namely geometric fluxes.

\subsection*{Models we will study}

As was pointed out in \cite{acha}, the model of \cite{dewolfe} belongs to a larger class of supersymmetric strict
SU(3)-structure type IIA AdS$_4$ vacua, the conditions for which were presented in \cite{lt}, generalizing the earlier work of \cite{behr}.
To obtain a ten-dimensional supergravity description of the model of \cite{dewolfe},
 a fine-tuned smeared orientifold source must be introduced. In fact, this procedure works for any Calabi-Yau space (including
K3$\times \text{T}^2$ and the torus). More generally, however, the vacua of \cite{lt}
need not be Calabi-Yau:  certain torsion classes of the SU(3)-structure can be non-zero, which in the physics literature
is sometimes called {\em geometric flux}. It is then natural to ask whether any of these more
general IIA AdS$_4$ vacua could be used in order to succeed where the model of \cite{dewolfe} has failed; in particular whether
they can be used to bypass the above-mentioned no-go theorem of \cite{kachrunogo}.

To the best of our knowledge, there are only very few explicitly known examples of $\mathcal{N}=1$ supersymmetric type IIA AdS$_4$ compactifications
{\em with geometric fluxes}. They all can be seen to belong to the class of solutions discussed in \cite{lt} (or to T-duals thereof). More explicitly,
the known examples are of the form AdS$_4 \times \mathcal{M}_6$ with the internal space $\mathcal{M}_6$
being one of the following:
\begin{itemize}
\item \textbf{The Iwasawa manifold and T-duals:}\\
 The Iwasawa is a particular nilmanifold (or `twisted torus' in the physics literature), proposed as a type IIA solution in \cite{lt}.
As for the torus examples mentioned above, smeared orientifold sources have to be introduced. In fact, for a certain regime of the parameters it is T-dual to a six-torus compactification. Essentially, the Iwasawa solution
is the twisted torus T$^6/(\mathbb{Z}_2\times\mathbb{Z}_2)$ example examined in \cite{dkpz,vill,fontaxions}\footnote{In the Iwasawa model in this paper there are four orientifolds. These can be
equivalently described as a single orientifold supplemented with its images
under a certain geometric $\mathbb{Z}_2\times\mathbb{Z}_2$ group acting on the internal manifold.}.
As shown in \cite{klpt} and reviewed in section \ref{branepicture}, for these spaces it is possible to replace
all fluxes by their corresponding brane sources (D-branes, NS5-branes, KK-monopoles) so that the original backgrounds
arise as near-horizon geometries of the intersecting branes.

We will also study a type IIB example with static SU(2)-structure on the nilmanifold 5.1 (according to the labelling of Table 4 of \cite{scan}),
which is the intermediate node in the T-duality web between the torus and the Iwasawa example.

For completeness, we mention that the Iwasawa solution is in fact a singular degeneration of a
model on K3$ \times_{t} \text{T}^{2}$, where the subscript $t$ denotes a geometric twist in the T$^2$-fiber.
In this paper we will not study this solution, which was also proposed in \cite{lt}, any further, although we expect it to be similar
to the Iwasawa solution.
\item \textbf{The coset spaces:}\\
The group manifold SU(2)$\times$SU(2) and the coset spaces
$\frac{\text{G}_2}{\text{SU(3)}}$, $\frac{\text{Sp(2)}}{\text{S}(\text{U(2)}\times \text{U(1)})}$, $\frac{\text{SU(3)}}{\text{U(1)}\times \text{U(1)}}$,
and $\frac{\text{SU(3)}\times \text{U(1)}}{\text{SU(2)}}$ provide the  remaining known examples. It will be convenient to henceforth refer to all these as `the coset models' even though
SU(2)$\times$SU(2) is a trivial coset. We remark that $\frac{\text{SU(3)}\times \text{U(1)}}{\text{SU(2)}}$ is somewhat special in that it is the only
coset model in the above list that does not have a nearly-K\"ahler limit and that does not allow for a type IIA solution without orientifold sources.
While examples in other contexts have already appeared some time ago \cite{np,Lust:1986ix},
solutions of type IIA string theory on the cosets with nearly-K\"ahler limit were proposed only in the more recent works \cite{dieter,behr,paltihouse,font,tomtwistor}.
Finally, in \cite{klt}, from which we will start the analysis in this paper, a systematic search for coset solutions was performed,
and the non-nearly-K\"ahler example of $\frac{\text{SU(3)}\times \text{U(1)}}{\text{SU(2)}}$ was added to the list of solutions. As explained in \cite{klt}, each of
these supersymmetric vacuum solutions has a number of parameters describing its scale, shape and orientifold charge.
In the full string theory these parameters are quantized.
\end{itemize}

In terms of the internal components, $\eta^{(1,2)}$,  of the 10D supersymmetry generators, a compactification on a manifold with strict SU(3)-structure corresponds to the case when $\eta^{(1)}$ and $\eta^{(2)}$ are proportional (see appendix \ref{structure} for our terminology and notation on internal spinors and the $G$-structure of manifolds).
More general compactification ans\"atze for the supersymmetry generators are also conceivable.
The corresponding general framework is commonly referred to as SU(3)$\times$SU(3)-structure. The conditions for supersymmetric solutions with this ansatz
were derived in \cite{granaN1}. In appendix \ref{structureSU2} we will provide a no-go theorem against
supersymmetric AdS$_4$ compactifications in {\em both} IIA and IIB with a {\em left-invariant} SU(3)$\times$SU(3)-structure that
is neither a strict SU(3) nor a static SU(2). A way out would be to consider $e^{2A-\Phi} \eta^{(2)\dagger} \eta^{(1)}$ non-constant in type IIA (where $A$ and $\Phi$ denote, respectively,  the warp factor and the dilaton), while in type IIB
we need a genuine type-changing dynamic SU(3)$\times$SU(3)-structure. This is beyond the scope of this paper.

\subsection*{The four-dimensional effective theory}

In this paper, we will study in detail the effective 4D field theory that is obtained from
the compactification on the above-mentioned nilmanifolds and coset spaces.
%In order to obtain an $\mathcal{N}=1$ supersymmetric, 4D theory, we will have to introduce orientifold planes.
To render the analysis tractable, we will only consider SU(3)-structures and fluxes which are constant in the basis of left-invariant one-forms,
as in \cite{klt}. Thanks to these simplifications,
we will be able to straightforwardly construct the effective action using 4D effective supergravity techniques. In more detail,
we use the expression for the K\"ahler potential and the superpotential of \cite{gl,granasup,grimmsup,effective}. For more
work see also \cite{louismicu,effectivepapers,casbil,cassanipotential}.

A general problem of this supergravity approach is that an explicit computation of the low energy theory of a given compactification
requires a suitable choice of expansion basis for the
`light' fluctuations. Unfortunately, it is still unclear how to construct such a basis {\em in general}. In generic flux compactifications, the set
of harmonic forms would be unsuitable as expansion forms,
 as the forms $J$ and $\Omega$ that define the SU(3)-structure (and which enter the supergravity expressions
for the K\"{a}hler and superpotential) are no longer closed  (see  e.g.~\cite{glmw,granasup,effectivepapers} for a few proposals). A detailed discussion of the general constraints on such a basis appeared in \cite{mkp}.  In the special case of nilmanifolds and coset manifolds, however,
the set of left-invariant forms (with the appropriate behaviour under the orientifold action) readily presents itself as the natural choice
and obeys the requirements of \cite{mkp}.\footnote{Since the left-invariant forms are constant over the moduli space, this basis satisfies
requirements *7-*9 of \cite{mkp} rather trivially. Note that left-invariant forms are not in general harmonic:
they can be combined into eigenmodes of the Laplacian to eigenvalues of the order of the geometric flux.}

Interestingly, for our models, it is also possible to derive the low energy effective action using
an alternative approach, which does not rely on supersymmetry: direct Kaluza-Klein
reduction (for a review of this approach in eleven-dimensional supergravity see \cite{duff}). In section \ref{lowennil}, we will provide an important
consistency check by calculating the mass spectrum for the six-torus and the Iwasawa examples, both by Kaluza-Klein reduction
as well as in the effective supergravity approach, obtaining exactly the same result in both cases (see also \cite{vill} for related work).
Having performed this consistency check for the six-torus and the Iwasawa examples, we will restrict ourselves to the effective supergravity approach for
the coset examples of
section \ref{lowencoset}.

\subsection*{Orientifolds}

In this paper, we introduce orientifold sources for a number of reasons. The first reason is that, in some of the models we study, the Bianchi identities cannot
be satisfied without orientifolds (to be specific, this concerns the nilmanifold examples and the $\frac{\text{SU(3)}\times \text{U(1)}}{\text{SU(2)}}$ model).
Secondly, as we discuss further in section \ref{consistency}, the orientifolds potentially allow for a hierarchy of scales
between the size of the internal manifold and the AdS$_4$ curvature, thereby providing a possibility
to decouple the tower of Kaluza-Klein modes from the light modes. The third reason is that we are interested in 4D, $\mathcal{N}=1$ supersymmetric
low energy effective theories, for which the orientifold sources are necessary.\footnote{For a discussion of the $\mathcal{N}=2$ theory arising
from IIA on nearly-K\"ahler manifolds {\em without} orientifolds see \cite{kp}.}

A somewhat delicate feature of our models is that the orientifolds have to be smeared. The reason for this is that the supersymmetry conditions of
\cite{lt} (for constant Romans mass) force the warp factor to be constant. Considering the back-reaction of a localized orientifold, on the other hand,
one would expect a non-constant warp factor, at least close to the orientifold source. A possible way around this contradiction is that taking into account $\alpha'$-corrections might allow for a non-constant warp factor (see also \cite{banks} for an alternative discussion). A helpful interpretation of the smearing of
a localized source, whose Poincar\'e dual is given roughly-speaking by a delta-function, is that it corresponds to
Fourier-expanding the delta-function and discarding all but the zero mode. In this paper, we will adopt the pragmatic point of view that
the smeared orientifolds are an unavoidable feature of our models that is consistent with a Kaluza-Klein reduction in the approximation where only the lowest modes are kept.
% so
%that upon compactification the correct four-dimensional low energy effective theory is obtained.

The question of how to associate orientifold involutions to a smeared source turns out to be somewhat subtle. We  will make
the natural assumption that the different orientifolds correspond to the decomposable (simple) terms in the orientifold current;
the rationale and details behind this are explained in appendix \ref{smearedori}.

\subsection*{Outline of the paper and summary of results}

In section \ref{IIAsusycond}, we review the general properties of the
AdS$_4$ solutions of \cite{lt} on which all the type IIA examples of the present paper are based. We also
discuss the issue of the separation of scales and the self-consistency of our analysis.
This requires that we can decouple the higher Kaluza-Klein modes in a regime with small string coupling and a sufficiently large internal manifold
(in units of the string length) so as to be able to use the classical supergravity approximation.

Section \ref{geomnil}  examines the geometry of the nilmanifold examples, namely
the six-torus and the Iwasawa manifold in the IIA theory, as well as the nilmanifold 5.1 in type IIB.
If, for the Iwasawa manifold, we restrict to the branch of moduli space where the Romans mass is zero,
these three manifolds can be shown to be T-dual to one another.

Interestingly, as shown in section \ref{branepicture}, for the same range of the
parameter space for which the T-dualities above are valid,
the solutions admit an interpretation as near-horizon geometries of intersecting brane configurations,
as in \cite{klpt}. From this point of view, the nilmanifold vacua in this range are nothing but
near-horizon geometries of intersections of KK-monopoles with other branes in flat space. This nice
feature of the `brane picture' is summarized in Table \ref{branepic}.
Each solution in this table is related to the one in the column next to it by a T-duality.
\begin{table}[h!]
\begin{center}
\begin{tabular}{|c|c|c|}
\hline
IIA & IIB & IIA\\
\hline
T$^6$ & nilmanifold 5.1 & Iwasawa \\
D4/D8/NS5& D3/D5/D7/NS5/KK & D2/D6/KK\\
\hline
\end{tabular}
\caption{Brane picture}
\label{branepic}
\end{center}
\end{table}

In section \ref{cstsoc}, we come to the analysis of the geometry of the coset examples;
this is mostly review material in which we follow closely reference \cite{klt}. The
four-dimensional low energy physics of our models is then analysed in the subsequent two sections: section \ref{lowennil} discusses the nilmanifolds, whereas section \ref{lowencoset} deals with the coset models. In each case, we compute the superpotential and the K\"{a}hler potential, as well as the
mass matrix of the scalars. As mentioned earlier, in the case of the six-torus and the Iwasawa manifold, the masses of the lightest excitations are obtained both by a direct Kaluza-Klein reduction as well as by using the general expressions for the superpotential and  K\"{a}hler potential based on the effective supergravity approach, with agreement in both cases.

In Table \ref{niltable} we list the nilmanifold geometries of section \ref{geomnil}, indicating
the number of light real scalar fields in each case. We also indicate how many real moduli remain unstabilized and whether, according to the analysis of section \ref{consistency}, it is possible to decouple the Kaluza-Klein tower.
\begin{table}[h!]
\begin{center}
\begin{tabular}{|c|c|c|c|}
\hline
& \hspace{0.6cm} T$^6$ \hspace{0.6cm} & nilmanifold 5.1 & Iwasawa\\
\hline
Light fields & 14 & 14 & 14 \\
Unstabilized & 3 & 3 & 3 \\
Decouple KK & yes & yes & yes \\
\hline
\end{tabular}
\caption{Results for the nilmanifolds}
\label{niltable}
\end{center}
\end{table}
As we show in section \ref{lowennil}, in each of the above models, three axions remain massless\footnote{The meaning of mass in AdS is somewhat subtle, however, as we review in section \ref{seckkmain}.}. In \cite{fontaxions}, it was argued that upon introducing space-time filling D-branes,
the massless axions may be `eaten' via a St\"{u}ckelberg mechanism to provide masses for pseudo-anomalous
abelian gauge fields on the world-volume of the branes; we do not pursue this here any further.

Table \ref{cosettable} lists the coset geometries of section \ref{cstsoc}, indicating in each case
the number of light real scalar fields, how many of them stay massless, whether it is possible to decouple the
tower of Kaluza-Klein modes and whether it is possible to get the internal curvature scalar $R<0$. As we will see in section \ref{inflation}, the latter property
is important for possible circumventions of the no-go theorem of \cite{kachrunogo}.
\begin{table}[h!]
\begin{center}
\begin{tabular}{|c|c|c|c|c|c|}
\hline
& \rule[1.2em]{0pt}{0pt} $\frac{\text{G}_2}{\text{SU(3)}}$& $\frac{\text{Sp(2)}}{\text{S}(\text{U(2)}\times \text{U(1)})}$ & $\frac{\text{SU(3)}}{\text{U(1)}\times \text{U(1)}}$ & SU(2)$\times$SU(2) & $\frac{\text{SU(3)}\times \text{U(1)}}{\text{SU(2)}}$\\
\hline
Light fields & 4 & 6 & 8 & 14 & 8 \\
Unstabilized & 0 & 0 & 0 & 1 & 0 \\
Decouple KK & no & yes & yes & yes & no\\
$R<0$ possible & no & yes & yes & yes & yes\\
\hline
\end{tabular}
\caption{Results for the coset spaces}
\label{cosettable}
\end{center}
\end{table}
As we show in section \ref{lowencoset}, all moduli are stabilized in each model except for SU(2)$\times$SU(2). However, it turns out to be rather hard to
decouple the tower of Kaluza-Klein modes, and in only three models there is a limit where this happens. As we will explain, for two of these three models we will have to analytically continue the shape parameters of the model to negative values, so that strictly speaking they do not describe a left-invariant SU(3)-structure on a coset anymore, but a related model based on a twistor bundle over a hyperbolic space \cite{tomtwistor,feng}. For the third model, SU(2)$\times$SU(2), if we take the limit where the Kaluza-Klein modes should decouple ($\mathcal{W}_1^- \rightarrow 0$), the other relevant torsion class $\mathcal{W}_2^-$ blows up just as the lower bound for the orientifold charge. We were not able to derive anything interesting in this singular limit.

As a general remark, we note that none of our models above contains light bulk gauge fields in the spectrum.

Section \ref{inflation} discusses an application of some of our results in the context of type IIA inflation. More concretely, we show that the
recent no-go theorem of \cite{kachrunogo} is no longer applicable when the scalar curvature $R$ can be made negative, which is the case for all the coset models  except for $\frac{\text{G}_2}{\text{SU(3)}}$.
Finally, in section \ref{conclusions} we conclude with a discussion of open questions and future directions.

In appendix \ref{sec:appb}, we summarize our conventions on supergravity, whereas appendix \ref{structure}
explains our conventions and terminology regarding SU(3), static SU(2) and SU(3)$\times$SU(3)-structures. In appendix \ref{structure}, we present
 our no-go theorem against constant intermediate SU(2)-structures on AdS$_4$-compactifications. Appendix \ref{smearedori}
 discusses in detail how one can associate orientifold involutions to a smeared orientifold current. In appendices \ref{appkk} and \ref{appeff4d},
we present the details on the Kaluza-Klein reduction and the effective supergravity approach, respectively. In appendix \ref{SU3U1qSU2special}, finally,
we discuss a special point in the moduli space of the coset model $\frac{\text{SU(3)}\times \text{U(1)}}{\text{SU(2)}}$
and show that in fact the supersymmetry there is extended to $\mathcal{N}=2$.

%%%%%%%%%%%%%%%%%%%%%%%%%%%%%%%%%%%%%%%%%%%%%%%%%%%%%%%%%%%%%%%%%%%%%%%%%%%%%%%%

\section{Supersymmetric type IIA AdS$_4$ compactifications}\label{IIAsusycond}

To date all our explicit ten-dimensional examples of ${\cal N}=1$ supersymmetric compactifications to AdS$_4$ fall within the class of type IIA SU(3)-structure compactifications and T-duals thereof. In this section we review
this class of ten-dimensional solutions. We also discuss how to obtain a controlled parameter regime in which the string coupling is small, supergravity is valid and the tower of Kaluza-Klein modes decouples. For additional background
material and a summary of our conventions the reader is referred to appendices \ref{sec:appb}, \ref{structure}
and \ref{smearedori}.

\subsection{Conditions for a supersymmetric vacuum}

The most general form of $\mathcal{N}=1$ compactifications of IIA supergravity to AdS$_4$ with the
ansatz $\eta^{(1)} \propto \eta^{(2)}$ for the internal supersymmetry generators
(the strict SU(3)-structure ansatz) was given by two of the present authors in \cite{lt}. These vacua must have
constant warp factor and constant dilaton, $\Phi$. Setting the warp factor to one, the solutions of \cite{lt} are given by\footnote{\label{convfootn}As opposed to \cite{lt} we do not use superspace conventions. Furthermore we use here the string frame and
put $m=-2 m_{\text{there}}, H=-H_{\text{there}}, J=-J_{\text{there}}, F_2=-2 m_{\text{there}} B'$ and $F_4=-G$.}:
%%%%%
\begin{subequations}
\label{ltsol}
\begin{align}
H &=\frac{2m}{5} e^{\Phi}\Re \Omega \, ,\\
F_2&=\frac{f}{9}J+F'_2 \, , \\
F_4&=f\mathrm{vol}_4+\frac{3m}{10} J\wedge J \, , \\\label{21d}
W e^{i \theta} &=-\frac{1}{5} e^{\Phi}m+\frac{i}{3} e^{\Phi }f \, .
\end{align}
\end{subequations}
%%%%%
where $H$ is the NSNS three-form, and $F_{n}$ denote the RR forms.
Furthermore, ($J$, $\Omega$) is the SU(3)-structure of the internal six-manifold,
i.e.\ $J$ is a real two-form, and $\Omega$ is a decomposable complex three form such that:
\subeq{\al{\label{suthree}
\Omega\wedge J&=0 \, , \\
\label{suthreenorm}
\Omega\wedge\Omega^*&=\frac{4i}{3}J^3\neq 0
~.}}
$f$, $m$ are constants parameterizing the solution: $f$ is the Freund-Rubin parameter, while $m$
is the mass of Romans' supergravity \cite{roma} -- which can be identified with $F_0$ in the `democratic'
formulation \cite{democratic}. $e^{i\theta}$ is a phase associated with the internal supersymmetry generators:
$\eta^{(2)}_+ = e^{i \theta} \eta^{(1)}_+$.
% the phase $\alpha$ parameterizes the `gauge-freedom'
%stemming from the fact that the SU(3)-structure is only defined up to U(1) transformations. In \cite{lt}
%$\alpha$ was gauge-fixed to zero; here we have chosen to restore it for later convenience.
The constant $W$ is defined by the following relation for the AdS$_4$ Killing spinors, $\zeta_\pm$,
\eq{
\label{defW}
\nabla_{\mu} \zeta_- = \frac{1}{2} W \gamma_{\mu} \zeta_+ \, ,
}
so that the radius of AdS$_4$ is given by $|W|^{-1}$. The two-form $F'_2$ is the primitive part of
$F_2$ (i.e.\ it is in the $\bf{8}$ of SU(3)). It is constrained by the Bianchi identity:
\eq{
\d F'_2=(\frac{2}{27}f^2-\frac{2}{5}m^2 ) e^{\Phi} \Re \Omega - j^{6} ~,\label{ltsolb}
}
where we have added a source, $j^6$, for D6-branes/O6-planes on the right-hand side.
%We immediately see that in the absence of sources the second constraint of
%table \ref{tabb} holds, i.e.\ $d \mathcal{W}^-_{2}\propto \Re\Omega$. However
%in the presence of non-zero $j^6$, this constraint may be relaxed.

The general properties of supersymmetric sources and their consequences for the integrability
of the supersymmetry equations were recently discussed by two
of the present authors in \cite{kt} within the framework of generalized geometry.
It was shown in this reference that, under certain mild assumptions,
supersymmetry guarantees that the appropriately source-modified Einstein equation and dilaton
equation of motion are automatically satisfied if the source-modified Bianchi identities are
satisfied. For this to work the source must be supersymmetric, which means it
must be generalized calibrated as in \cite{gencal}.

Finally, the only non-zero torsion classes of the internal manifold are ${\cal W}^-_1,{\cal W}^-_2$
which are defined such that (see also \eqref{torsionclassesv}):
\subeq{
\label{torsionclasses}
\al{
\label{torsionclass1}
\d J&=-\frac{3}{2}i \, \mathcal{W}_1^- \Re \Omega\, , \\
\d \Omega&= \mathcal{W}^-_1 J\wedge J+\mathcal{W}^-_2 \wedge J
~.}}
These torsion classes are given by:
\eq{
{\cal W}^-_1=-\frac{4i}{9} e^{\Phi} f  \, , \qquad
{\cal W}^-_2=-i e^{\Phi} F'_2 \,  .
\label{ltsolc}
}
For the following it will be convenient to also
introduce $c_1:= -\frac{3}{2}i \, \mathcal{W}_1^- $, which appears in \eqref{torsionclass1}.
In addition, for vanishing sources or for sources proportional to $\Re \Omega$,
we have $\d \mathcal{W}_2 \propto \Re \Omega$. We define the proportionality constant
$c_2$ by
\eq{
\label{c2def}
\d \mathcal{W}_2^- = i c_2 \, \Re \Omega \, .
}
As we review below \eqref{l1}, one can show that
\eq{
\label{c2expr}
c_2 = - \frac{1}{8} |{\cal W}^-_2|^2 \, .
}
For a given geometry to correspond
to a vacuum without orientifold sources, we find from \eqref{ltsolb}, \eqref{ltsolc}-\eqref{c2expr}
that the following bound on
(${\cal W}_1^-,{\cal W}_2^-$) has to be satisfied
\begin{align}
\frac{16}{5} e^{2\Phi} m^2 = 3|{\cal W}^-_1|^2-|{\cal W}^-_2|^2\geq 0
~,
\label{condition}
\end{align}
where we have defined $|\Theta|^2:= \Theta^*_{mn}\Theta^{mn}$, for any two-form $\Theta$.
Incidentally, let us note that
condition (\ref{condition}) turns out to be too stringent to be satisfied for
any nilmanifold whose only non-zero
torsion classes are ${\cal W}^-_{1,2}$ \cite{ktu}. This implies that without orientifolds there are no solutions on nilmanifolds.

The constraint \eqref{condition} can however be relaxed by allowing
for an orientifold source, $j^{6}\neq 0$. As a particular example, let us consider:
\begin{align}
j^{6}=-\frac{2}{5}e^{-\Phi}\mu \Re\Omega~,
\label{jo}
\end{align}
where $\mu$ is an arbitrary, {\it discrete}, real parameter of dimension (mass)$^2$, so that $-\mu$  is proportional to
the orientifold/D6-brane charge
($\mu$ is positive for net orientifold charge  and negative for net D6-brane charge).
The addition of this source term  was first considered in \cite{acha}.
Eq.~(\ref{jo}) above guarantees that the calibration conditions, which
for D6-branes/O6-planes read
\eq{
\label{calcond}
j^6 \wedge \Re \Omega = 0 \, , \qquad j^6 \wedge J = 0 \, ,
}
are satisfied and thus the source wraps supersymmetric cycles.
The bound (\ref{condition}) changes to
\begin{align}\label{boundtra}
e^{2\Phi} m^2=\mu+\frac{5}{16}\left(3|{\cal W}^-_1|^2-|{\cal W}^-_2|^2\right) \ge 0 \, .
\end{align}
Since $\mu$ is arbitrary the above equation
can always be satisfied, and therefore no longer imposes any constraint on the
torsion classes of the manifold.

Let us also note that it is possible to consider the inclusion of more general supersymmetric
orientifold six-plane sources that do {\em not} satisfy eq.~(\ref{jo}). In that case,  the constraint that
$\d \mathcal{W}^-_{2}$ should be proportional to  $\Re\Omega$
is relaxed. Requiring this source to satisfy the calibration conditions \eqref{calcond}, we find that it is
now of the following form:
\eq{\label{orinonprop}
j^{6}=-\frac{2}{5}e^{-\Phi}\mu \Re\Omega + w_3~,
}
with $w_3$ a primitive (2,1)+(1,2)-form. From the Bianchi identity \eqref{ltsolb} we find
\eq{
w_3 = -i e^{-\Phi} \d \mathcal{W}_2^-\Big|_{(2,1)+(1,2)} \, ,
}
and \eqref{boundtra} still unchanged.

In appendix \ref{smearedori} we will explain how to associate orientifold involutions to a smeared source. Under each orientifold
involution the dilaton, metric and fluxes must transform as follows:
\subeq{\label{oriall}
\eq{\label{oriflux}\spl{
& \text{\bf Even}: \qquad \sigma^* e^{\Phi} = e^{\Phi} \, , \qquad \sigma^* F_0 = F_0 \, , \qquad \sigma^* F_4 = F_4 \, , \\
& \text{\bf Odd}: \qquad \sigma^* H = -H \, , \qquad \sigma^* F_2 = -F_2  \, ,
}}
whereas the SU(3)-structure transforms as
\eq{\label{oristruc}\spl{
& \text{\bf Even}: \qquad \sigma^* \Im \Omega = \Im \Omega \, , \\
& \text{\bf Odd}: \qquad \sigma^* \Re \Omega = -\Re \Omega \, , \qquad \sigma^* J = -J \, .
}}}

%Since the geometry of the internal manifold is determined
%by the fluxes, which are quantized in the full quantum theory, it follows from
%eq.~(\ref{ipoyt}) and the fact that $\mu$ is discrete that the AdS$_4$ radius
%is also quantized {\bf [Are we sure $|{\cal W}^-_2|^2$ is quantized if $F_2$ is quantized when integrated over cycles?]}.

%%%%%%%%%%%%%%%%%%%%%%%%%%%%%%%%%%%%%%%%%%%%%%%%%%%%%%%%%%%

\subsection{Hierarchy of scales}\label{consistency}

In the full quantum theory, all fluxes have to be quantized according to
\begin{align}\label{flqu}
\frac{1}{l^{p-1}}\int_{\mathcal{C}_p}F_{p}=n_p~,
\end{align}
where $l:=2\pi\sqrt{\alpha'}$, $\mathcal{C}_p$ is a cycle in the internal manifold, and $n_p\in\mathbb{Z}$. The NSNS
three-form turns out to be exact in our models, hence its integral over any internal three-cycle vanishes; it therefore
suffices to impose (\ref{flqu}) for the RR fluxes. The issue of quantization is studied in more detail in \cite{tomtwistor}.

For the analysis of the present paper to be valid, we need to show that we can consistently take the string coupling constant
to be small ($g_s=e^{\Phi}\ll 1$), so that string loops can be safely ignored, and that the volume of the internal manifold  is large in string units ($L_{int}/l\gg 1$, where $L_{int}$ is the characteristic length of the internal manifold), so that
$\alpha'$-corrections can be neglected. This can be seen by essentially
employing the following scaling argument: Let $f_p/(g_sL_{int})$ be
the norm of the flux density $F_{p}$, for some numbers $f_p$ depending on the internal geometry (but not
on the overall scale $L_{int}$). The quantization conditions
(\ref{flqu}) imply:
\begin{align}\label{13}
g_s=(f_0^3f_4)^{\frac{1}{4}}(n_0^3n_4)^{-\frac{1}{4}};~~~
\frac{L_{int}}{l}=\left(\frac{f_0}{f_4}\right)^{\frac{1}{4}}\left(\frac{n_4}{n_0}\right)^{\frac{1}{4}};~~~
\frac{n_2}{\sqrt{n_0n_4}}=\frac{f_2}{\sqrt{f_0f_4}};~~~
\frac{n_0n_6}{n_2n_4}=\frac{f_0f_6}{f_2f_4}~.
\end{align}
It can then be easily verified that, given a solution $\{n_p\}$ to the quantization conditions $(\ref{flqu})$, there
are several different possible scalings $n_p\rightarrow N^{\lambda_p}n_p$, for $N,\lambda_p\in\mathbb{N}$, which leave the
$f_p$'s invariant and at the same time ensure that $g_s$ is parametrically small while $L_{int}/l$ is parametrically large (with large parameter $N$). This schematic argument can be made precise, by taking into account the specifics of
the geometry of each internal manifold, as in \cite{tomtwistor}. Despite the fact that we are allowing for large flux quanta,
it can be shown that higher-order flux corrections can also be neglected. Indeed it is not difficult to see that
the parameter  $|g_s F_{p}|^2$, which controls the size
of these corrections,  scales with a negative power of the large parameter $N$.

A further consistency requirement is that the Kaluza-Klein tower can be decoupled. Since the Compton wavelength of the lightest excitations
above the Breitenlohner-Freedman bound in four dimensions is of the order of the AdS$_4$ radius, we need to show that the Compton wavelength of the Kaluza-Klein excitations (which is proportional to $L_{int}$) satisfies:
\begin{align}\label{1.2}
|\Lambda_{\text{AdS}}|L_{int}^2 \ll 1~,
\end{align}
where $\Lambda_{\text{AdS}}$ is the four-dimensional cosmological constant. In models without orientifolds this is impossible to achieve,
since the characteristic length of the internal manifold turns out to be of the same order as the radius of AdS$_4$. This is the  problem of separation of scales which, for example, plagues the compactifications of eleven-dimensional supergravity on the
seven-sphere. Ultimately we would like to uplift our models to a de Sitter space with a small, positive cosmological constant, and the position could be taken that the question of the mass spectra should be re-addressed only after this uplifting. However, let us now study whether it is possible to tune the orientifold source such that there is a hierarchy between the two scales even before the uplifting and (\ref{1.2}) is obeyed.

Taking into account $|\Lambda_{\text{AdS}}|\sim |W|^2$ and using \eqref{21d}, we find that to decouple the Kaluza-Klein scale we must impose
\eq{
\label{Wsmall}
|W|^2 L_{int}^2 = \frac{1}{25} (g_s)^2 m^2 L_{int}^2 + \frac{1}{9} (g_s)^2 f^2 L_{int}^2 \ll 1 \, ,
}
which means that each of the two terms on the right-hand side of the equal sign must be separately much
smaller than one. Tuning the orientifold charge
we can accomplish $e^{2\Phi} m^2 L_{int}^2 \ll 1$. Indeed, we just need to show that we can
choose $\mu$ so that it is close to its bound \eqref{boundtra}:
\eq{
\mu L_{int}^2 + \frac{5}{16}\left(3|{\cal W}^-_1|^2-|{\cal W}^-_2|^2\right) L_{int}^2 \ll 1~.
}
In our conventions the discrete parameter $\mu$, which is proportional to the net number of orientifold planes $n_{\mathcal{O}6}$, is given by (up to numerical factors of order one): $\mu\sim g_sn_{\mathcal{O}6}l L_{int}^{-3}$. Taking into account that the
torsion classes are given by (again up to numerical factors of order one): $|\mathcal{W}^-_i|^2\sim L_{int}^{-2}$, we can rewrite
the above equation schematically as follows:
\begin{align}\label{tryighb}
n_{\mathcal{O}6} g_s\left(\frac{l}{L_{int}}\right) + a \ll 1 ~,
\end{align}
where $a$ is a number of order one. Since  $g_s\left(\frac{l}{L_{int}}\right)\ll 1$, we can then satisfy this bound by choosing some large integer $n_{\mathcal{O}6}$. Note that in the examples where we study this limit,
$a$ turns out to be negative so that we can accomplish this with positive $n_{\mathcal{O}6}$, which corresponds to net orientifold charge (as opposed to net D-brane charge).

However, we must also make sure that the second square in \eqref{Wsmall} is small, which means
that $f g_s L_{int} \propto |\mathcal{W}_1^-| L_{int}$ is small. Manifolds for which $\mathcal{W}_1^-$ vanishes (and only
$\mathcal{W}_2^-$ is possibly non-zero) are called `nearly Calabi-Yau' (NCY) see e.g.~\cite{feng}; hence for the bound \eqref{1.2}
to be satisfied, the internal manifold must admit an SU(3)-structure which is sufficiently close to the NCY limit.
For the torus model we have $\mathcal{W}^-_1=0$, while in section \ref{lowenSO5qSU2U1}-\ref{lowenSU3U1U1} we will argue that both the
$\frac{\text{Sp(2)}}{\text{S(U(2)}\times\text{U(1)})}$ and $\frac{\text{SU(3)}}{\text{U(1)}\times\text{U(1)}}$ models have
continuations that admit NCY SU(3)-structures.

Once a solution for $n_{\mathcal{O}6}$ is
obtained in this way, we are free to rescale $n_{\mathcal{O}6}\rightarrow N^{q}n_{\mathcal{O}6}$ leaving (\ref{tryighb})
invariant, provided we take: $q=(\lambda_0+\lambda_4)/2\in\mathbb{N}$. For example, the reader can verify that the rescaling
$\{ n_0\rightarrow N^{4}n_0,  n_2\rightarrow N^{6}n_2,  n_4\rightarrow N^{8}n_4,  n_6\rightarrow N^{10}n_6,  n_{\mathcal{O}6}\rightarrow N^{6}n_{\mathcal{O}6}    \}$ leave eq.~(\ref{tryighb})  and all the $f_p's$ in eq.~(\ref{13}) invariant, so that:
\begin{align}
g_s\sim N^{-5}~, \qquad \frac{L_{int}}{l}\sim N~, \qquad  |\Lambda_{\text{AdS}}|L_{int}^2=\mathrm{fixed}\ll 1
~,
\end{align}
where we can take $N$ large.

\section{Ten-dimensional geometries I: nilmanifolds}\label{geomnil}

By taking the internal six-dimensional space to be a
nilmanifold $\mathcal{M}$, it turns out that
one can construct explicit examples of the
type of compactifications reviewed in section \ref{IIAsusycond}, as we will show in the present section.
As follows from the discussion of section \ref{IIAsusycond}, it suffices to look for all possible
six-dimensional nilmanifolds
whose only non-zero torsion classes are ${\cal W}^-_{1,2}$. A systematic scan
yields exactly two possibilities in type IIA, namely the six-torus and
the nilmanifold 4.7 of Table 4 of \cite{scan} (also known as the Iwasawa manifold),
which (for some values of the parameters) turn out to be related by T-duality along two
directions. We also found a type IIB solution with static SU(2)-structure on the nilmanifold 5.1, as described in appendix
\ref{structureSU2}, which forms the intermediate step after one T-duality.
Unfortunately, in the case of type IIB we were not able to make a systematic scan so there might be
more solutions of this type.

Before we  turn to the description of each of those possibilities in the next section, let us also mention that closely related
solutions can be obtained by replacing the six-torus
by a direct product K3 $\times$ T$^2$, the Iwasawa manifold by the T$^2$ fibration over K3 constructed in \cite{lt}, and the 5.1 nilmanifold by an
S$^1$ fibration over K3$\times$S$^1$.
This relation arises from the fact that on the boundary of its
moduli space, the K3 surface degenerates to a discrete quotient of T$^4$.
 Just as the previous cases, all these three solutions are connected by T-dualities.

\subsection{The type IIA T$^6$ solution}
\label{sixtorusex}

Our first IIA solution is obtained by taking the internal  manifold to be a six-dimensional torus.
Let us define a left-invariant basis $\{e^i\}$ such that:
\eq{
\d e^i = 0, \qquad i=1,\dots, 6~.
}
On the torus we can just choose $e^i = \d y^i$, where $y^i$ are the internal coordinates.
The SU(3)-structure is given by
\begin{align}\label{jot}
J&=e^{12}+e^{34}+e^{56} \, ,\nn\\
\Omega&=(ie^1+e^2)\wedge(ie^3+e^4)\wedge(ie^5+e^6)~,
\end{align}
which can indeed be seen to satisfy eqs.~(\ref{OmegaJ}, \ref{voln}) and \eqref{torsionclasses} for $f=0$,
putting $\mathrm{vol}_6=e^{1\dots 6}$. It readily follows that all torsion
classes vanish in this case. Note, however, that there are non-vanishing $H$ and $F_{4}$
fields given by \eqref{ltsol}
\eq{
\begin{split}
H & = \frac{2}{5} e^{\Phi} m \left(e^{246}-e^{136}-e^{145}-e^{235}\right) \, , \\
F_4 & = \frac{3}{5} m \left( e^{1234} + e^{1256} + e^{3456} \right) \, .
\end{split}
}
{} From \eqref{boundtra} we find that there is an orientifold source of the type \eqref{jo}
with $\mu=e^{2\Phi} m^2$, which corresponds to smeared orientifolds along $(1,3,5)$, $(2,4,5)$, $(2,3,6)$
and $(1,4,6)$. The corresponding orientifold involutions are
\eq{\spl{
\label{torusinv}
O6: & \qquad e^2 \rightarrow -e^2 \, , \quad e^4 \rightarrow -e^4 \, , \quad e^6 \rightarrow -e^6 \, , \\
O6: & \qquad e^1 \rightarrow -e^1 \, , \quad e^3 \rightarrow -e^3 \, , \quad e^6 \rightarrow -e^6 \, , \\
O6: & \qquad e^1 \rightarrow -e^1 \, , \quad e^4 \rightarrow -e^4 \, , \quad e^5 \rightarrow -e^5 \, , \\
O6: & \qquad e^2 \rightarrow -e^2 \, , \quad e^3 \rightarrow -e^3 \, , \quad e^5 \rightarrow -e^5 \, .
}}

\subsection{The type IIA Iwasawa-manifold solution}\label{iwasawaex}

The second IIA solution is obtained by taking the internal manifold to be
the Iwasawa manifold. The left-invariant basis
is defined by:
\eq{\spl{
\d e^a&=0,~~a=1,\dots, 4 \, ,  \\
\d e^5&=e^{13}-e^{24} \, , \\
\d e^6&=e^{14}+e^{23}~,
}}
and is usually denoted by $(0,0,0,0,13-24,14+23)$. Up to basis transformations there
is a unique SU(3)-structure satisfying the supersymmetry conditions of section \ref{IIAsusycond}:
\eq{\spl{
J & = e^{12}+e^{34}+ \beta^2 e^{65} \, , \\
\Omega&=\beta \, (ie^5-e^6)\wedge(ie^1+e^2)\wedge(ie^3+e^4)~,
\label{lolut}
}}
In the left-invariant basis, the metric is  given by $g=\text{diag} (1,1,1,1,\beta^2,\beta^2)$, and the torsion classes
can be read off from $\d J$, $\d \Omega$, taking eqs.~(\ref{torsionclasses}) into account:
\eq{\spl{
{\cal W}^-_1 & = -\frac{2i}{3} \beta \, , \\
{\cal W}^-_2 & = -\frac{4i}{3} \beta \, \left(e^{12}+e^{34}+2\,\beta^2 \, e^{56}\right) \, ,
\label{expltord}
}}
while all other torsion classes vanish. The fluxes can be read off from \eqref{ltsol} by plugging in $f=\frac{3}{2} e^{-\Phi} \beta$,
while we can find $m$ from \eqref{boundtra}.
We can verify that $\d{\cal W}^-_2$ is
proportional to $\Re\Omega$:
\begin{align}
\d{\cal W}^-_2=-\frac{8i}{3}\beta^2\Re\Omega~.
\label{dw2d}
\end{align}
{}From the second line of (\ref{expltord}) we can read off: $|{\cal W}^-_2|^2=64\beta^2/3$.
Comparing with eq.~(\ref{boundtra}), taking $|{\cal W}^-_1|^2=4\beta^2/9$ into account -- as follows from the first line of (\ref{expltord}) -- we therefore
find a non-zero net orientifold six-plane charge:
\begin{align}
\mu\geq\frac{25}{4}\beta^2 ~.
\label{toriwa}
\end{align}
The solution (\ref{lolut}) has one continuous parameter, $\beta$, corresponding essentially to the first torsion class ${\cal W}^-_{1}$. An additional second parameter can be introduced by noting that the defining SU(3)-structure equations (\ref{a1}) are invariant under the rescaling
\begin{align}
J\rightarrow \gamma^2 J;~~~~~ \Omega\rightarrow \gamma^3 \Omega~.
\end{align}
The additional scalar $\gamma$ is related to the volume modulus via $\mathrm{vol}_6=-\gamma^6 \beta^2 e^{1\dots 6}$, as can be seen
from eq.~(\ref{voln}).

For the case $m=0$, for which the bound \eqref{toriwa} is saturated, the above example can also be obtained by performing two T-dualities on the torus solution of section \ref{sixtorusex}, as can be checked explicitly. We find then that $\beta=\frac{2}{5} m_{T} e^{\Phi}$ where $m_T$ is the mass parameter of the dual torus solution.

\subsection{The type IIB nilmanifold 5.1 solution}\label{nilIIBsol}

This solution is related, via a single T-duality, to both T$^6$ and
the Iwasawa manifold of \ref{iwasawaex}. Indeed, let us perform a T-duality on the six-torus example of section \ref{sixtorusex}
using the T-duality rules of e.g.~\cite{hassan} (see also \cite{florian} for a discussion of the
action of T-duality on the pure spinors of a SU(3)$\times$SU(3)-structure).\footnote{Note that
it does not matter along which direction one performs the T-duality since
all six perpendicular directions are equivalent. For the second T-duality (from which we obtain the Iwasawa
solution of the previous section), only one direction leading to a geometric background is possible. We will not pursue the interesting case of non-geometric backgrounds in the present paper.} After rescaling and relabelling the left-invariant forms
we find the nilmanifold 5.1 described by (0,0,0,0,0,12+34). For the SU(2)-structure quantities described in
appendix \ref{structureSU2} we obtain
\eq{\spl{
e^{i\theta} V & = \frac{1}{2} \left(\beta e^6 +i e^5 \right) \, , \\
\omega_2 & = e^{13} - e^{24} \, , \\
\Omega_2 & = -i e^{i \theta} (i e^1 +e^3)\wedge (i e^4 + e^2 ) \, .
}}
The metric is given by $g=\text{diag} (1,1,1,1,\beta^2,\beta^2)$, and
for the fluxes we have
\eq{\spl{
H & = - \beta \left( e^{235} + e^{145}\right) \, , \\
e^{\Phi} F_1 & = \frac{5}{2} \, \beta^2 e^6 \, , \\
e^{\Phi} F_3 & = \frac{3}{2} \, \beta \left( e^{135}-e^{245}\right) \, , \\
e^{\Phi} F_5 & = \frac{3}{2} \, \beta^2 e^{12346} \, .
}}
Again we find that $\beta$ is related to the mass parameter of the torus example via $\beta=\frac{2}{5} m_T e^{\Phi}$.

%%%%%%%%%%%%%%%%%%%%%%%%%%%%%%%%%%%%%%%%%%%%%%%%%%%%%%%%%%%

\subsection[The brane picture]{The brane picture\footnote{This section is somewhat outside the main line of the paper and may be omitted in a first reading.}}
\label{branepicture}

Following \cite{klpt}, it is possible to interpret the solutions presented in sections \ref{sixtorusex}-\ref{nilIIBsol},
from the perspective of intersecting branes. Namely, we would like to recover these
solutions as near-horizon limits of domain walls in four noncompact dimensions,
corresponding to systems of (orthogonally)
intersecting branes (we will henceforth use the term `brane' to refer to either a
D$p$-brane, an NS5-brane, or a KK-monopole).

%%%%%%%%%%%%%%%%%%%%%%%%%%%%%%%%%%%%%%%%%%%%%%%%%%%%%%%%%%%

 More specifically, we will
impose the following requirements on our brane configurations:
\begin{enumerate}
\item All configurations should consist of branes in ten-dimensional flat space, of which four directions are noncompact
and six directions form a six-torus.
\item All branes should have exactly the same two spatial directions along the noncompact space.
\item All branes should intersect orthogonally, and we do not consider world-volume gauge fields.
\item The resulting configuration should preserve $\mathcal{N}=1$ supersymmetry in D=3, \\
and should admit a regular near-horizon geometry with an AdS$_4$ factor.
\item Each configuration should include the maximum number of branes compatible
with requirements 1-4.
\end{enumerate}

Before we come to the description of explicit configurations satisfying the above requirements, let us note that, as we will see in the following,  only brane configurations that lead to strict SU(3)-structure (as well as their T-dual configurations leading to static SU(2)-structures) arise in this way; this is the same class of backgrounds considered in section \ref{IIAsusycond}. The easiest way to arrive at this conclusion is to first determine which types of SU(3)$\times$SU(3)-structure
are compatible with each brane separately . Indeed, using their corresponding $\kappa$-symmetry projectors, it is straightforward to
analyse what relations between the internal supersymmetry generators $\eta^{(1)}$ and $\eta^{(2)}$ of \eqref{spinansatz} are possible, which leads to the following table of branes and their corresponding compatible types of structure:\footnote{We also refer to Table 1 of \cite{kt}
which represents the allowed types of structure too, but now for space-filling orientifolds. Orientifolds have the same supersymmetry
properties as D-branes with vanishing world-volume gauge field, however the difference of space-filling versus domain wall basically shifts the table.}
%\begin{table}[b]
\begin{center}
  \begin{tabular}{|c|c|}
    \hline
     Brane & Structure type\\
    \hline
    \hline
     D2  & strict SU(3) \\
     D3  & static SU(2) \\
     D4  & SU(3)$\times$SU(3) \\
     D5  & SU(3)$\times$SU(3) \\
     D6  & SU(3)$\times$SU(3) \\
     D7  & static SU(2) \\
     D8  & strict SU(3) \\
     NS5  & SU(3)$\times$SU(3) \\
     KK  & SU(3)$\times$SU(3) \\
\hline
   \end{tabular}
\end{center}
%\caption{\label{tableGstructure} D$p$-branes vs $G$-structure.}
%\end{table}
See appendix \ref{structure} for the terminology. It turns out, that the configuration always needs to have D-branes to get a regular near-horizon AdS limit. From the above table it follows, that if one of these D-branes is a D2, D3, D7 or D8 we already find
strict SU(3)- or static SU(2)-structure. If not, let us consider the SU(3)-structure associated to $\eta^{(1)}$
as in \eqref{defJOm}. Let us also define the complex coordinates $z^i$ associated with this SU(3)-structure
as well as their real and imaginary parts: $z^i = x^i + i y^i$.
Because all the branes defining this SU(3)-structure intersect orthogonally (requirement 3), for
each brane the $x^i$ and $y^i$ directions will be either along or perpendicular to the brane,
i.e., there are no angles other than right angles. Now the relation between $\eta^{(1)}$
and $\eta^{(2)}$, which we can get from the $\kappa$-symmetry conditions of one of the D-branes, will contain
gamma-matrices for directions that are also parallel or orthogonal
to the $x^i$ and $y^i$ directions. Exhausting then all possibilities for the resulting structure shows that it can only be strict SU(3) or static SU(2).
It follows that if one is interested in constructing a configuration with general SU(3)$\times$SU(3)-structure, one
should restrict to D4, D6, D5, NS5 and KK-branes {\em and} put these branes at non-orthogonal angles.

Let us make a few comments concerning the requirements 1-5 above. The first one
anticipates the fact that, as it will turn out,
the internal nilmanifolds in the solutions of section \ref{geomnil}
can be thought of as intersections of KK-monopoles in flat space. It therefore
suffices to consider branes in flat space.
The second requirement is of course
just the requirement that the configuration should correspond to a domain wall
in four space-time dimensions.
The requirement of orthogonality was imposed for simplicity. It would be
interesting to consider branes/monopoles intersecting at angles, but it would be quite difficult
to construct the corresponding geometry because one could no longer use the harmonic superposition rules
for branes \cite{harmonicsup}.
The first part of the
fourth requirement is equivalent to demanding that the domain wall, viewed from the
point of view of four-dimensional  space-time, should be supersymmetric. Indeed, the minimal supersymmetry
a domain wall in four dimensions can preserve, is one-half of
$\mathcal{N}=1$ in $D=4$. This is equal to two real supercharges, i.e.\ $\mathcal{N}=1$ in $D=3$. Note that
this implies that exactly one-sixteenth of the original supersymmetry
of type II supergravity in $D=10$ should be preserved. As each brane breaks supersymmetry
by (at most) one-half, there will be (at least) four branes in the configuration.
The final requirement is imposed because
a configuration that does not include the maximum number of branes
compatible with requirements 1-4, turns out not to have a regular AdS$_4$ near-horizon limit.

The rules for supersymmetric, orthogonally-intersecting branes were formulated some time ago \cite{harmonicsup,intersectionrefs}.
For the type of configurations  we are considering in the present paper, they can be summarized as follows:
\begin{center}
  \begin{tabular}{|c|c|}
    \hline
     intersecting branes& $\#$ of relative transverse directions  \\
    \hline
    \hline
    Dp/Dq & $0\ \mbox{mod}\ 4$  \\
    NS5/NS5 &  $0\ \mbox{mod}\ 4$ \\
    Dp/NS5 & $7-p$ or $11-p$ \\
    Dp/KK & $5-p$ or $9-p$   \\
    KK/KK &  $0\ \mbox{mod}\ 4$ \\
    NS5/KK & 4 or 8 \\
\hline
   \end{tabular}
\end{center}
The requirements 1-5 listed above severely restrict the set of admissible
intersecting-brane configurations. It is in fact straightforward to show that
all possible such configurations are related to each other by T-dualities. The brane configurations
comprising the `nodes' of this T-duality web, listed in Table \ref{branepic}, are analysed
in the following\footnote{Without the second
part of the fourth requirement there are three more configurations connected to each other by T-duality:
D5/NS5, D6/D4/NS5/KK and D5/KK. Because they do not admit a regular near-horizon limit with AdS$_4$ factor
they are not of interest to us here and we do not consider them.}.

%%%%%%%%%%%%%%%%%%%%%%%%%%%%%%%%%%%%%%%%%%%%%%%%%%%%%%%%%%%

\subsubsection*{D4/D8/NS5}\label{sec1}

This is the IIA solution given in \cite{klpt} and corresponds to
the following system of intersecting D4/NS5/D8-branes:
\bigskip
\begin{center}
  \begin{tabular}{|c||c|c|c|c|c|c|c|c|c|c|}
    % after \\: \hline or \cline{col1-col2} \cline{col3-col4} ...
    \hline
    & $x^0$ & $x^1$  & $x^2$  & $x^3$ & $y^1$  & $y^2$  & $y^3$  & $y^4$  & $y^5$  & $y^6$ \\
    \hline
    \hline
    $\mathrm{D}4$ & $\bigotimes$ & $\bigotimes$  & $\bigotimes$ &  &$\bigotimes$  &  $\bigotimes$ &   &
                &   &  \\

    \hline
    $\mathrm{D}4^{\prime}$ & $\bigotimes$ & $\bigotimes$  & $\bigotimes$ &  &  &  & $\bigotimes$  &
               $\bigotimes$  &   &  \\

    \hline
    $\mathrm{D}4^{\prime\prime}$ & $\bigotimes$ & $\bigotimes$  & $\bigotimes$ &   &   &  &   &
                 & $\bigotimes$  & $\bigotimes$ \\
\hline
    $\mathrm{NS}5$ & $\bigotimes$ & $\bigotimes$  & $\bigotimes$ &   &  $\bigotimes$  &   &  $\bigotimes$ &
                 & $\bigotimes$  & \\
\hline
    $\mathrm{NS}5^{\prime}$ & $\bigotimes$ & $\bigotimes$  & $\bigotimes$ &   &  $\bigotimes$  &  &   &  $\bigotimes$
                 &  & $\bigotimes$ \\
 \hline
    $\mathrm{NS}5^{\prime\prime}$ & $\bigotimes$ & $\bigotimes$  & $\bigotimes$ &   &  & $\bigotimes$ &   &
               $\bigotimes$    &  $\bigotimes$  & \\
\hline
    $\mathrm{NS}5^{\prime\prime\prime}$ & $\bigotimes$ & $\bigotimes$  & $\bigotimes$ &   &  & $\bigotimes$ &  $\bigotimes$     &
               &  &  $\bigotimes$ \\
 \hline
    $\mathrm{D}8$ & $\bigotimes$ & $\bigotimes$  & $\bigotimes$ &   &  $\bigotimes$ & $\bigotimes$ & $\bigotimes$  &  $\bigotimes$
                 & $\bigotimes$  & $\bigotimes$ \\
\hline
   \end{tabular}
\end{center}
\bigskip
The full solution of \cite{klpt} patches two asymptotic regions: a
near-horizon AdS$_4\times$T$^6$ region and a flat  region
at infinity. Here we will concentrate on the near-horizon limit of the
solution where the brane system above is replaced by fluxes.
After rescaling of the coordinates, it can be written as:
\eq{\spl{
\d s_{10}^2 & =
\d s^2_{\mathrm{AdS}_4}+\sum_{i=1}^6(\d y^i)^2 ;~~~\Phi=\mathrm{const.} \, ; \\
H_{y^2y^4y^6}&=H_{y^2y^5y^3}=H_{y^1y^6y^3}=H_{y^1y^5y^4} = a \, , \\
F_{y^3y^4y^5y^6}&= F_{y^1y^2y^5y^6}=F_{y^1y^2y^3y^4}= \frac{3}{2} e^{-\Phi} a \, , \qquad F_0 = \frac{5}{2} e^{-\Phi} a~,
\label{rtyu}
}}
where $a$ and $e^{\Phi}$ are given in terms of the brane quanta in \cite{klpt}, and
the SU(3)-structure is given by:
\eq{\spl{
J&=\d y^{1}\wedge\d y^{2}
+\d y^{3}\wedge\d y^{4}
+\d y^{5}\wedge\d y^{6} \, , \\
\Omega&=
\left(i\d y^{1}+\d y^{2}\right)\wedge
\left(i\d y^{3}+\d y^{4}\right)\wedge
\left(i\d y^{5}+\d y^{6}\right)~.
%\d y^{136}+\d y^{145}+\d y^{235}+\d y^{246}
%+i\left(\d y^{135}+\d y^{146}+\d y^{236}+\d y^{245}\right)~.
}}
We can readily see that, in the language of section \ref{IIAsusycond},
the present solution corresponds to setting $F'_2=0$, $f=0$
and $m=a$ with a source term:
\begin{align}
j^{O6}=- \frac{2a^2}{5} e^{-\Phi} \Re\Omega~.
\end{align}
So while the original brane configuration has disappeared in the near-horizon limit we have to introduce
a set of smeared orientifold sources in order to satisfy the tadpole conditions:
\bigskip
\begin{center}
  \begin{tabular}{|c||c|c|c|c|c|c|c|c|c|c|}
    % after \\: \hline or \cline{col1-col2} \cline{col3-col4} ...
    \hline
    & $x^0$ & $x^1$  & $x^2$  & $x^3$ & $y^1$  & $y^2$  & $y^3$  & $y^4$  & $y^5$  & $y^6$ \\
    \hline
    \hline
    $\mathrm{O}6$ & $\bigotimes$ & $\bigotimes$  & $\bigotimes$ & $\bigotimes$  &$\bigotimes$  &    &$\bigotimes$ &
               &    $\bigotimes$ &  \\

    \hline
    $\mathrm{O}6^{\prime}$ & $\bigotimes$ & $\bigotimes$  & $\bigotimes$ &   $\bigotimes$ & $\bigotimes$ &  & &
               $\bigotimes$  &   &    $\bigotimes$\\

    \hline
    $\mathrm{O}6^{\prime\prime}$ & $\bigotimes$ & $\bigotimes$  & $\bigotimes$ &    $\bigotimes$ &     &  $\bigotimes$ &   &
                 $\bigotimes$  & $\bigotimes$  &  \\
\hline
    $\mathrm{O}6^{\prime\prime\prime}$ & $\bigotimes$ & $\bigotimes$  & $\bigotimes$ &   $\bigotimes$  &   &
 $\bigotimes$ &  $\bigotimes$ &
                 &   &   $\bigotimes$\\
\hline
   \end{tabular}\label{tabela}
\end{center}
\bigskip

Indeed, as follows from (\ref{ltsolc}), in this limit, all torsion classes of the internal manifold vanish,
as they should for T$^6$. Moreover, this is exactly the solution of section \ref{sixtorusex}.

%%%%%%%%%%%%%%%%%%%%%%%%%%%%%%%%%%%%%%%%%%%%%%%%%%%%%%%%%%%

\subsubsection*{D3/D5/D7/NS5/KK}\label{subsecbrane2}

By applying a T-duality on the solution of the previous subsection, we obtain the following
configuration (we do not display the noncompact directions anymore, but let us keep in mind that they
form domain walls):
\bigskip
\begin{center}
%\begin{tabular}{cc}

  \begin{tabular}{|c|c|c|c|c|c|c|}
    % after \\: \hline or \cline{col1-col2} \cline{col3-col4} ...
    \hline
    & $y^1$  & $y^2$  & $y^3$  & $y^4$  & $y^5$  & $y^6$ \\
    \hline
     \hline
    $\mathrm{D}7$    &   & $\bigotimes$ & $\bigotimes$  &  $\bigotimes$
                 & $\bigotimes$  & $\bigotimes$ \\
    \hline
    $\mathrm{D}3$  &&  $\bigotimes$ &   &
                &   &  \\

    \hline
    $\mathrm{D}5^{\prime}$  & $\bigotimes$   &  & $\bigotimes$  &
               $\bigotimes$  &   &  \\

    \hline
    $\mathrm{D}5^{\prime\prime}$  &  $\bigotimes$   &  &   &
                 & $\bigotimes$  & $\bigotimes$ \\
\hline
    $\mathrm{NS}5$ &   $\bigotimes$  &   &  $\bigotimes$ &
                 & $\bigotimes$ &  \\
\hline
    $\mathrm{NS}5^{\prime}$    &  $\bigotimes$  &  &   &  $\bigotimes$
                 &  & $\bigotimes$ \\
 \hline
    $\mathrm{KK}^{\prime\prime}$    & $\bullet$  & $\bigotimes$ &   &
               $\bigotimes$    &  $\bigotimes$  & \\
\hline
    $\mathrm{KK}^{\prime\prime\prime}$    & $\bullet$ & $\bigotimes$ &  $\bigotimes$     &
               &  &  $\bigotimes$ \\
\hline
   \end{tabular}
%\ \ \ \ \ \ \ \ \

 %  \end{tabular}
   \end{center}
Without loss of generality, we have taken the T-duality to be along $y^{1}$.
Let us only describe the salient features of this model.

First of all, an analysis
of the $\kappa$-symmetry conditions of the D-branes reveals that for this configuration
the internal spinors satisfy:
\eq{
\eta^{(2)}_+ = - e^{-i \theta} \gamma_{\underline{1}} \eta^{(1)}_- ~,
}
where $e^{-i\theta}$ is a phase describing the supersymmetry preserved by the domain
wall in four dimensions, or, after taking the near-horizon limit, the phase of the
superpotential $W$ of AdS. So we see that we have static SU(2)-structure, which is also
the only possibility for type IIB as explained in appendix \eqref{structureSU2}.

Secondly, when one goes to the near-horizon limit, the effect of the KK-monopoles is
to twist the $S^1$ of direction $1$ over the T$^4$ corresponding to the directions $(3,4,5,6)$, which is indicated with a bullet in the tables.
This means that we find for the metric, after rescaling,
\eq{
\d s_{10}^2 =
\d s^2_{\mathrm{AdS}_4}+\sum_{i=1}^6(e^i)^2 \, ,
}
with
\eq{\spl{
e^1 &:= \d y^1 + a(y^6\d y^3 + y^5\d y^4) \, , \\
e^i &:= \d y^i~;~~ i=2,\ldots,6 \, ,
}}
where $a$ is the same parameter as in the T-dual.
This means we have
\eq{\spl{
\d e^1 & = a (e^{63} + e^{54}) \, , \\
\d e^i & = 0 \, ,
}}
which, in fact,  is equivalent to nilmanifold 5.1. So we see that just like the other branes the KK-monopoles
disappear in the near-horizon limit and are replaced by flux, in this case the geometric flux $a$.

It turns out that in addition to the fluxes we have O$5$/O$7$ orientifold planes along the following directions:
\begin{center}
  \begin{tabular}{|c||c|c|c|c|c|c|c|c|c|c|}
    % after \\: \hline or \cline{col1-col2} \cline{col3-col4} ...
    \hline
    & $x^0$ & $x^1$  & $x^2$  & $x^3$ & $y^1$  & $y^2$  & $y^3$  & $y^4$  & $y^5$  & $y^6$ \\
    \hline
    \hline
    $\mathrm{O}5$ & $\bigotimes$ & $\bigotimes$  & $\bigotimes$ & $\bigotimes$  &  &    &$\bigotimes$ &
               &    $\bigotimes$ &  \\

    \hline
    $\mathrm{O}5^{\prime}$ & $\bigotimes$ & $\bigotimes$  & $\bigotimes$ &   $\bigotimes$ &  &  & &
               $\bigotimes$  &   &    $\bigotimes$\\

    \hline
    $\mathrm{O}7^{}$ & $\bigotimes$ & $\bigotimes$  & $\bigotimes$ &    $\bigotimes$ &  $\bigotimes$   &  $\bigotimes$ &   &
                 $\bigotimes$  & $\bigotimes$  &  \\
\hline
    $\mathrm{O}7^{\prime}$ & $\bigotimes$ & $\bigotimes$  & $\bigotimes$ &   $\bigotimes$  & $\bigotimes$  &
 $\bigotimes$ &  $\bigotimes$ &
                 &   &   $\bigotimes$\\
\hline
   \end{tabular}
\end{center}
After appropriate rescaling and relabelling, this solution corresponds to the solution on the nilmanifold 5.1 of section \ref{nilIIBsol}.

\subsubsection*{D2/D6/KK}\label{subsecbrane3}

Starting from the configuration of section \ref{subsecbrane2},
there is exactly one possibility left for a T-duality, i.e.\ along $y^{2}$. This is because
 T-dualizing along a direction perpendicular to a KK-monopole would result in a nongeometric background.

\begin{center}

    \begin{tabular}{|c|c|c|c|c|c|c|}
    % after \\: \hline or \cline{col1-col2} \cline{col3-col4} ...
    \hline
    & $y^1$  & $y^2$  & $y^3$  & $y^4$  & $y^5$  & $y^6$ \\
    \hline
     \hline
    $\mathrm{D}6$    &  & & $\bigotimes$  &  $\bigotimes$
                 & $\bigotimes$  & $\bigotimes$ \\
    \hline
    $\mathrm{D}2$  & &  &   &
                &   &  \\

    \hline
    $\mathrm{D}6^{\prime}$  & $\bigotimes$  & $\bigotimes$  & $\bigotimes$  &
               $\bigotimes$  &   &  \\

    \hline
    $\mathrm{D}6^{\prime\prime}$  & $\bigotimes$   & $\bigotimes$  &   &
                 & $\bigotimes$  & $\bigotimes$ \\
\hline
    $\mathrm{KK}$ &   $\bigotimes$  & $\bullet$ &  $\bigotimes$ &
                 & $\bigotimes$ &  \\
\hline
    $\mathrm{KK}^{\prime}$    &  $\bigotimes$  & $\bullet$ &   &  $\bigotimes$
                 &  & $\bigotimes$ \\
 \hline
    $\mathrm{KK}^{\prime\prime}$    & $\bullet$ & $\bigotimes$ &   &
               $\bigotimes$    &  $\bigotimes$  & \\
\hline
    $\mathrm{KK}^{\prime\prime\prime}$    & $\bullet$ & $\bigotimes$ &  $\bigotimes$     &
               &  &  $\bigotimes$ \\
\hline
   \end{tabular}

\end{center}

An analysis of the $\kappa$-symmetry conditions of the branes
reveals that this model has again strict SU(3)-structure. The four KK-monopoles
result in a near-horizon geometry for which the T$^2$ along the directions $(1,2)$
is twisted over the base T$^4$ along $(3,4,5,6)$. The metric reads
\eq{
\d s_{10}^2 = \d s^2_{\mathrm{AdS}_4}+\sum_{i=1}^6(e^i)^2~,
\label{rtyu2}
}
where we have defined
\eq{\spl{
e^1 &:=  \d y^1 + a(y^6\d y^3 + y^5\d y^4) \, ,\\
e^2 &:= \d y^2 + a(y^5\d y^3 - y^6\d y^4) \, , \\
e^i & := \d y^i~;~~ i=3,\ldots,6 ~,
}}
such that
\eq{\spl{
\d e^1 & = a (e^{63}+e^{54}) \, , \\
\d e^2 & = a (e^{53}+e^{46}) \, , \\
\d e^i & := \d y^i~;~~ i=3,\ldots,6 ~.
}}
After rescaling and relabelling we find the solution of section \ref{iwasawaex}
for $m=0$. For $m \neq 0$ the latter solution does not have a dual brane picture.

Finally note that in order to satisfy the tadpole conditions we have again O$6$-planes along the following directions:
\begin{center}
  \begin{tabular}{|c||c|c|c|c|c|c|c|c|c|c|}
    % after \\: \hline or \cline{col1-col2} \cline{col3-col4} ...
    \hline
    & $x^0$ & $x^1$  & $x^2$  & $x^3$ & $y^1$  & $y^2$  & $y^3$  & $y^4$  & $y^5$  & $y^6$ \\
    \hline
    \hline
    $\mathrm{O}6$ & $\bigotimes$ & $\bigotimes$  & $\bigotimes$ & $\bigotimes$  &  &  $\bigotimes$  &$\bigotimes$ &
               &    $\bigotimes$ &  \\

    \hline
    $\mathrm{O}6^{\prime}$ & $\bigotimes$ & $\bigotimes$  & $\bigotimes$ &   $\bigotimes$ &  & $\bigotimes$ & &
               $\bigotimes$  &   &    $\bigotimes$\\

    \hline
    $\mathrm{O}6^{\prime\prime}$ & $\bigotimes$ & $\bigotimes$  & $\bigotimes$ &  $\bigotimes$   &  $\bigotimes$ &  &   &
                 $\bigotimes$  & $\bigotimes$  &  \\
\hline
    $\mathrm{O}6^{\prime\prime\prime}$ & $\bigotimes$ & $\bigotimes$  & $\bigotimes$ & $\bigotimes$  & $\bigotimes$ &
 &  $\bigotimes$ &
                 &   &   $\bigotimes$\\
\hline
   \end{tabular}
\end{center}
This completes the overview of brane configurations of Table \ref{branepic}.

%%%%%%%%%%%%%%%%%%%%%%%%%%%%%%%%%%%%%%%%%%%%%%%%%%%%%%%%%%%

%\input{noads.tex}

%%%%%%%%%%%%%%%%%%%%%%%%%%%%%%%%%%%%%%%%%%%%%%%%%%%%%%%%%%%

%%%%%%%%%%%%%%%%%%%%%%%%%%%%%%%%%%%%%%%%%%%%%%%%%%%%%%%%%%%

\section{Ten-dimensional geometries II: coset spaces}\label{cstsoc}

A large class of IIA solutions of the type described in section \ref{IIAsusycond} was
given recently in \cite{klt}, also incorporating solutions that were already known 
\cite{np,behr,tomtwistor} into a single unifying framework of 
left-invariant SU(3)-structures on coset spaces. In other words, 
the solutions described in \cite{klt} are all of the form AdS$_4\times\mathcal{M}_6$ where the internal manifold
is a coset, $\mathcal{M}_6=G/H$, equipped with a left-invariant SU(3)-structure. 
In Tomasiello's recent work \cite{tomtwistor} an alternative description in terms
of twistor bundles is used for the cosets of sections \ref{SO5qSU2U1} and \ref{SU3qU1U1}.
Although this description does not allow to describe the complete parameter space
on the coset $\frac{\text{SU(3)}}{\text{U(1)}\times \text{U(1)}}$, it is more accurate
for the nearly Calabi-Yau limit in which, as we will see, the shape parameters take
negative values and the coset description is not valid anymore. 

Before we come to the description of the individual
coset solutions listed in \cite{klt}, let us review some well-known facts
about coset spaces. For more details see, e.g., \cite{cosetrev0,cosetrev1}.

In dealing
with coset spaces of the form $G/H$ it suffices for our purposes to examine the corresponding Lie algebras
$\mathfrak{g}$, $\mathfrak{h}$. Let $\{ \mathcal{H}_a\}$ be a basis of
generators of the algebra $\mathfrak{h}$, and let $\{ \mathcal{K}_i\}$ be a basis of the
complement $\mathfrak{k}$ of $\mathfrak{h}$ inside $\mathfrak{g}$, i.e.\ $a=1,\dots,$ dim($H$) and
$i=1,\dots,$ dim($G$)$-$dim($H$). We define the structure constants as follows:
\eq{\spl{\label{commut}
[\mathcal{H}_a,\mathcal{H}_b]&=f^c{}_{ab}\mathcal{H}_c \, , \\
[\mathcal{H}_a,\mathcal{K}_i]&=f^j{}_{ai}\mathcal{K}_j+f^b{}_{ai}\mathcal{H}_b \, , \\
[\mathcal{K}_i,\mathcal{K}_j]&=f^k{}_{ij}\mathcal{K}_k+f^a{}_{ij}\mathcal{H}_a~.
}}
If $H$ is connected and semisimple, or compact, one can always find a basis of generators $\{ \mathcal{K}_i\}$
such that the structure constants $f^b{}_{ai}$ vanish \cite{cosetrev0}. In other
words: $[\mathcal{H}, \mathcal{K}]\subset \mathcal{K}$, and in this case
the coset $G/H$ is called {\it reductive}.

Let $y^m$, $m=1,\dots,$ dim($G$)$-$dim($H$), be local coordinates on $G/H$ and let
$L(y)$ be a coset representative. The decomposition of the Lie-algebra valued one-form
\al{
L^{-1}dL= e^i\mathcal{K}_i+\omega^a\mathcal{H}_a
~,}
defines a coframe $e^i(y)$ on $G/H$. Moreover, using the commutation relations
(\ref{commut}), we find
\al{\label{dcommut}
de^i=-\frac{1}{2}f^i{}_{jk}e^j\wedge e^k-f^i{}_{aj}\omega^a\wedge e^j
~.}
We are interested in expanding in forms that are {\em left-invariant} under the action
of $G$ on $G/H$. One can show that this is the case if and only if for a $p$-form
\eq{
\phi=\frac{1}{p!}\phi_{i_1\dots i_p}e^{i_1}\wedge\dots \wedge e^{i_p} \, ,
}
its components $\phi_{i_1\dots i_p}$ are constants {\em and}
\eq{\label{leftinv}
f^j{}_{a[i_1}\phi_{i_2\dots i_p]j}=0~.
}
If we then take the exterior derivative $d\phi$, condition \eqref{leftinv} ensures
that the part coming from the second term in \eqref{dcommut} drops out and we get again
a left-invariant form. One can show that harmonic forms must be left-invariant and
thus the cohomology of the coset manifold is isomorphic to the cohomology of left-invariant
forms.

Similarly, a metric $g = g_{ij} e^i \otimes e^j$ is left-invariant if and only if
its components $g_{ij}$ are constants and
\eq{
f^k{}_{a(i} g_{j)k}=0 \, .
}
The Riemann tensor for such a metric is calculated in e.g.~\cite{cosetrev1}. We display
here the Ricci scalar, which we find by contracting indices:
\eq{
\label{riccicoset}
R = - g^{ij} f^k{}_{ai} f^a{}_{kj} - \frac{1}{2} g^{ij} f^k{}_{li} f^l{}_{kj}
- \frac{1}{4} g_{ij} g^{kl} g^{mn} f^i{}_{km} f^j{}_{ln} \, .
}
If we introduce orientifolds the structure constant tensor
\eq{\label{structensor}
f = \frac{1}{2}f^i{}_{jk} E_i \otimes e^j\wedge e^k + f^i{}_{aj} E_i \otimes \omega^a\wedge e^j +
\frac{1}{2} f^a{}_{ij} U_a \otimes e^i \wedge e^j + \frac{1}{2} f^{a}{}_{bc} U_a \otimes \omega^b \wedge \omega^c \, ,
}
where the $E_i,U_a$ are dual to the $e^i,\omega^a$, has to be even under the orientifold involution (for some suitable extension of the involution to the $\omega^a$) in order to ensure that the exterior derivative is even.

We are now ready to  proceed to the description of the individual
coset solutions listed in \cite{klt}.

\subsection{The $\frac{\text{G}_2}{\text{SU(3)}}$ solution}\label{G2qSU3}

The $G_2$ structure constants are given by:
\eq{\spl{
& f^1{}_{63}=f^1{}_{45}=f^2{}_{53}=f^2{}_{64} = \frac{1}{\sqrt{3}} \, , \\
& f^7{}_{36}=f^7{}_{45}=f^8{}_{53}=f^8{}_{46}=f^9{}_{56}=f^9{}_{34}=f^{10}{}_{16}=f^{10}{}_{52}\\
& =f^{11}{}_{51} =f^{11}{}_{62}=f^{12}{}_{41}=f^{12}{}_{32}
=f^{13}{}_{31}=f^{13}{}_{24}=\frac{1}{2} \, , \\
& f^{14}{}_{43}=f^{14}{}_{56}=\frac{1}{2\sqrt{3}} \, , \qquad f^{14}{}_{21}=\frac{1}{\sqrt{3}} \, , \\
& f^{i+6}{}_{j+6,k+6} = f_{\text{GM}ijk} \, ,
}}
where $f_{\text{GM}ijk}$ are
the Gell-Mann structure constants.

The $G$-invariant two-forms and three-forms are spanned by
\eq{
\{e^{12}-e^{34}+e^{56}\}\, , \qquad \{\rho=e^{245}+e^{135}+e^{146}-e^{236},\hat{\rho}=-e^{235}-e^{246}+e^{145}-e^{136}\} \, ,
}
respectively, and there are no invariant one-forms. With only these two invariant
three-forms\footnote{$\hat{\rho}$ can be found by lowering one index of the purely $\mathcal{K}_i$-part of the structure constant tensor with
the Cartan-Killing metric and $\rho$ is its Hodge dual, so they are both left-invariant. Moreover, since the structure constant tensor should be even
under all orientifold involutions and the Hodge dual is odd, we find that $\hat{\rho}$ is even and $\rho$ odd. We can immediately conclude that
they should be proportional to $\Im \Omega$ and $\Re \Omega$ respectively. Of course a priori there could have been more left-invariant
three-forms.} there is no room for a source not proportional to $\Re \Omega$.

The most general solution is then given by
\eq{\spl{
J & = a (e^{12} - e^{34} + e^{56}) \, , \\
\Omega & = d \left[ (e^{245}+e^{146}+e^{135}-e^{236}) + i (e^{145}-e^{246}-e^{235}-e^{136}) \right] \, ,
}}
with $a$, the overall scale, the only free parameter, and
\eq{\spl{\label{G2sol}
 a & > 0 \, ,  \qquad \text{metric positivity} \, , \\
 d^2 & = a^{3}, \qquad \text{normalization of } \Omega \, , \\
 c_1 & := -\frac{3i}{2} {\cal W}_1^-= -\frac{2}{3} e^{\Phi} f = -\frac{\sqrt{3}a}{d} \, , \\
 {\cal W}_2^- & = 0 \, , \\
  e^{2\Phi} m^2 - \mu & = \frac{5}{12} c_1^2  \,  .
}}
We conclude that the only possibility for this coset is the nearly-K\"ahler geometry.
It will be convenient to isolate the scale $a$ and introduce the reduced flux parameters
\eq{\label{reduced}
\tilde{m}=a^{1/2} e^{\Phi} m \, , \qquad  \tilde{f}=a^{1/2} e^{\Phi} f \, , \qquad \tilde{\mu}=a \mu \, , \qquad \tilde{c}_1 = a^{1/2} c_1 \, ,
}
in terms of which the background fluxes take the form:
\eq{\spl{
H &= \frac{2 \tilde{m}}{5} a  (e^{245}+e^{135}+e^{146}-e^{236}) \, , \\
e^{\Phi} F_2 &= \frac{a^{1/2}}{2 \sqrt{3} } \left(e^{12}-e^{34}+e^{56} \right) \, , \\
e^{\Phi}  F_4 & = a^{-1/2} \tilde{f} \text{vol}_4 - \frac{3}{5} \tilde{m} a^{3/2} \left(e^{1234}- e^{1256}+ e^{3456} \right) \, , \\
e^{\Phi} j^6& = - \frac{2}{5} a^{1/2} \tilde{\mu} (e^{245}+e^{135}+e^{146}-e^{236}) \, .
}}

As mentioned before, $\mu>0$ corresponds to net orientifold charge. Solutions with $\mu \le 0$ --- i.e. with net D-brane charge --- are possible,
but in that case we still assume that smeared orientifolds are present, which then should be compensated by introducing enough smeared D-branes.
It can be easily read off from $j^6$ that the orientifolds are along the directions $(1,3,6), (2,4,6), (2,3,5)$ and $(1,4,5)$, leading to
four orientifold involutions. One can check that all fields and the SU(3)-structure transform as in \eqref{oriall} under {\em each} of the
orientifold involutions. Also, the structure constant tensor \eqref{structensor} is even.

\subsection{The $\frac{\text{Sp(2)}}{\text{S}(\text{U(2)}\times \text{U(1)})}$ solution}\label{SO5qSU2U1}

The structure constants are totally antisymmetric. The non-zero ones are given by:
\begin{eqnarray}
f^5{}_{41}=f^5{}_{32}=f^6{}_{13}=f^6{}_{42}=\frac{1}{2} \, , ~~~~~~~~~f^7{}_{56}=f^{10}{}_{89}=-1 \, ,\nn\\
f^7{}_{21}=f^7{}_{43}=f^8{}_{14}=f^8{}_{32}=f^9{}_{13}=f^9{}_{24}=f^{10}{}_{34}=f^{10}{}_{21}=\frac{1}{2} \, ,
\end{eqnarray}
corresponding to the nonmaximal embedding.
The $G$-invariant two-forms and three-forms are spanned by
\eq{
\{e^{12}+e^{34},e^{56}\}\, , \qquad \{\rho=e^{245}-e^{135}-e^{146}-e^{236},\hat{\rho}=e^{235}+e^{246}+e^{145}-e^{136}\} \, ,
}
respectively, and there are no invariant one-forms. The source (if present) must be proportional to $\Re \Omega$.

The most general solution is then given by
\eq{\spl{
J & = a (e^{12} + e^{34}) - c e^{56} \, , \\
\Omega & = d \left[ (e^{245}-e^{236}-e^{146}-e^{135}) + i (e^{246}+e^{235}+e^{145}-e^{136}) \right] \, ,
}}
with $a$ and $c$ two free parameters and
\eq{\spl{\label{SO5qSU2U1sol}
 a & > 0 \, , \quad c>0, \qquad \text{metric positivity} \, , \\
 d^2 & = a^2 c, \qquad \text{normalization of } \Omega \, , \\
 c_1 & := -\frac{3i}{2} {\cal W}_1^-= -\frac{2}{3} e^{\Phi} f = \frac{2a+c}{2d} \, , \\
 {\cal W}_2^- & = -\frac{2i}{3d} \left[ a(a-c) (e^{12} + e^{34}) + 2c(a-c) e^{56} \right] \, , \\
 c_2 & := - \frac{1}{8} |{\cal W}_2^-|^2 = - \frac{2}{3a^2c}  (a-c)^2 \, , \\
 \frac{2}{5}  (e^{2\Phi} m^2 - \mu) & = c_2 + \frac{1}{6} c_1^2 = \frac{1}{8a^2c} \left(- 4a^2-5c^2+12ac \right) \,  .
}}

The nearly-K\"ahler limit corresponds to setting $a=c$. The two parameters correspond to the overall scale $a$
and a parameter $\sigma \equiv c/a$ that measures the deviation from the nearly-K\"ahler limit, and we can make contact with the results of \cite{tomtwistor} as in \cite{klt}. 

For the background fluxes and source we find in terms of the reduced flux parameters \eqref{reduced}:
\eq{\spl{
H &= \frac{2 \tilde{m}}{5} a  \sigma^{1/2} (e^{245}-e^{135}-e^{146}-e^{236}) \, , \\
e^{\Phi} F_2 &= \frac{a^{1/2}}{4} \sigma^{-1/2} \left[ (2-3\sigma) (e^{12}+e^{34})+(6\sigma-5\sigma^2)e^{56} \right] \, , \\
e^{\Phi} F_4 &= a^{-1/2} \tilde{f} \text{vol}_4 + \frac{3}{5} a^{3/2} \tilde{m}  \left( e^{1234}- \sigma e^{1256}- \sigma e^{3456} \right) \, , \\
e^{\Phi} j^6 &= - \frac{2}{5} a^{1/2} \tilde{\mu} \sigma^{1/2} (e^{245}-e^{135}-e^{146}-e^{236}) \, .
}}
We introduce the same orientifold involutions as in section \ref{G2qSU3} and check that all fields and the structure constants transform appropriately.

\subsection{The $\frac{\text{SU(3)}}{\text{U(1)}\times \text{U(1)}}$ solution}\label{SU3qU1U1}

We choose a basis such that the structure constants of SU(3) are given by
\eq{
f^1{}_{54}=f^1{}_{36}=f^2{}_{46}=f^2{}_{35}=f^3{}_{47}=f^5{}_{76}=\frac{1}{2} \, , \,
\quad f^1{}_{27}=1 \, , \quad f^3{}_{48}=f^5{}_{68}=\frac{\sqrt{3}}{2} \, ,
~\mathrm{cyclic}~.}
The $G$-invariant two-forms and three-forms are spanned by
\eq{
\{e^{12},e^{34},e^{56}\}\, , \qquad \{\rho=e^{245}+e^{135}+e^{146}-e^{236},\hat{\rho}=e^{235}+e^{136}+e^{246}-e^{145}\} \, ,
}
respectively, and there are no invariant one-forms.  The source (if present) must again be proportional to $\Re \Omega$.

The most general solution is then given by
\eq{\spl{
J & = - a e^{12} + b e^{34} - c e^{56} \, , \\
\Omega & = d \left[ (e^{245}+e^{135}+e^{146}-e^{236}) + i (e^{235}+e^{136}+e^{246}-e^{145}) \right] \, ,
}}
with $a,b$ and $c$ three free parameters and
\eq{\spl{\label{SU3qU1U1sol}
 a & > 0 , \, b>0, \, c>0 \, , \qquad \text{metric positivity} \, , \\
 d^2 & = abc, \qquad \text{normalization of } \Omega \, , \\
 c_1 & := -\frac{3i}{2} {\cal W}_1^-= -\frac{2}{3} e^{\Phi} f = -\frac{a+b+c}{2d} \, , \\
 {\cal W}_2^- & = -\frac{2i}{3d} \left[ a(2a-b-c) e^{12} + b (a-2b+c)e^{34} + c(-a-b+2c) e^{56} \right] \, , \\
 c_2 & := - \frac{1}{8} |{\cal W}_2^-|^2 = - \frac{2}{3abc} \left( a^2 + b^2 + c^2 - (ab+ac+bc)\right) \, , \\
\frac{2}{5}  (e^{2\Phi} m^2 - \mu) & = c_2 + \frac{1}{6} c_1^2 = \frac{1}{8abc} \left[- 5(a^2+b^2+c^2) + 6 (ab+ac+bc) \right]  \,  .
}}
Putting $a=b$ we end up with a model that is very similar to the one of section \eqref{SO5qSU2U1}, while
further putting $a=b=c$ corresponds to the nearly-K\"ahler limit. Next to the overall scale $a$ we have this time
two shape parameters $\rho \equiv b/a$ and $\sigma \equiv c/a$. For a comparison 
with the results of \cite{tomtwistor} see \cite{klt}.

Introducing again the reduced flux parameters \eqref{reduced}
we find for the fluxes and source
\eq{\spl{
H &= \frac{2 \tilde{m}}{5} a (\rho \sigma)^{1/2} (e^{245}+e^{135}+e^{146}-e^{236}) \, , \\
e^{\Phi} F_2 &= \frac{a^{1/2}}{4} (\rho \sigma)^{-1/2} \left[ (5-3\rho-3\sigma) e^{12}+(3\rho-5\rho^2+3\rho\sigma)e^{34}+(-3\sigma-3\rho\sigma+5\sigma^2)e^{56} \right] \, , \\
e^{\Phi} F_4 &= a^{-1/2} \tilde{f} \text{vol}_4 - \frac{3}{5} a^{3/2} \tilde{m}  \left(\rho e^{1234}-\sigma e^{1256}+ \rho\sigma  e^{3456} \right) \, , \\
e^{\Phi} j^6 &= - \frac{2}{5} a^{1/2} \tilde{\mu} (\rho \sigma)^{1/2} (e^{135}+e^{146}+e^{245}-e^{236}) \, ,
}}
while the orientifold involutions are still as in section \ref{G2qSU3}. 

\subsection{The SU(2)$\times$SU(2) solution}
\label{SU2SU2}

The structure constants in this case are
\eq{
f^1{}_{23} = f^4{}_{56}=1 \, , \qquad \mathrm{cyclic}
~.}
The most general solution to eqs.~\eqref{ltsolb}, \eqref{torsionclasses}, \eqref{ltsolc}, \eqref{boundtra} and \eqref{orinonprop}
is
\eq{\spl{
J & = a e^{14}+ b e^{25}+c e^{36} \, , \\
\Omega & = -\frac{1}{c_1} \Bigg\{a (e^{234} - e^{156}) + b (e^{246} - e^{135}) + c (e^{126} - e^{345}) \\
& -\frac{i}{h} \Big[ -2 \, abc (e^{123}+e^{456}) + a(b^2+c^2-a^2) (e^{234}+e^{156}) + b(a^2+c^2-b^2) (e^{153}+e^{426}) \\
& + c(a^2+b^2-c^2) (e^{345}+e^{126}) \Big] \Bigg\} \, ,
}}
with $a,b$ and $c$ three free parameters and
\eq{\spl{
abc & > 0 \, ,  \qquad \text{metric positivity} \, , \\
h & = \sqrt{2 \, a^2 b^2 + 2 \, b^2 c^2 + 2 \, a^2 c^2 - a^4 - b^4 - c^4} \, , \\
& \text{and thus} \quad 2 \, a^2 b^2 + 2 \, b^2 c^2 + 2 \, a^2 c^2 - a^4 - b^4 - c^4 > 0 \, , \\
c_1^2 & = \frac{4}{9} e^{2\Phi} f^2 = \frac{h}{2abc} \, , \\
 {\cal W}_2^- & = -\frac{2i}{3 h c_1} \Bigg[\frac{(b^2-c^2)^2 + a^2(-2a^2 + b^2 +c^2)}{bc} e^{14}
 +\frac{(c^2-a^2)^2 + b^2(-2b^2 + c^2 +a^2)}{ac} e^{25}  \\
&  +\frac{(a^2-b^2)^2 + c^2(-2c^2 + a^2 +b^2)}{ab} e^{36}
 \Bigg] \, .
}}
By a suitable change of basis we can always arrange for $a>0,b>0$ and $c>0$, which we will assume from now on.
In terms of the reduced flux parameters \eqref{reduced}, to which
we add
\eq{
\tilde{h} = a^{-2} h \, ,
}
we find for the fluxes
\eq{\label{fluxessu2expl}\spl{
H & =  \frac{2\tilde{m}}{5\tilde{c}_1} a  \left[(e^{156} - e^{234}) + \rho(e^{135} - e^{246}) + \sigma(e^{345} - e^{126}) \right] \, , \\
F_2 & =  \frac{\tilde{c}_1 a^{1/2}}{2\tilde{h}^2} \Big\{
\left[ 3 (\rho^4 + \sigma^4 )  - 5 + 2(\rho^2 + \sigma^2) - 6\rho^2 \sigma^2\right]e^{14} \\
& \phantom{a^{1/2}\frac{\tilde{c}_1}{2\tilde{h}^2} (}
 +\rho\left[3(1+\sigma^4) -5 \rho^4 + 2\rho^2(1 + \sigma^2 ) -6 \sigma^2 \right]e^{25} \\
& \phantom{a^{1/2}\frac{\tilde{c}_1}{2\tilde{h}^2} (}
+ \sigma\left[3(1+\rho^4) -5 \sigma^4 + 2\sigma^2(1 + \rho^2 ) -6 \rho^2 \right]e^{36} \Big\} \, , \\
F_4 & =  a^{-1/2} \tilde{f} \text{vol}_4 -a^{3/2} \frac{3\tilde{m}}{5}  (\rho e^{1245} + \sigma e^{1346} + \rho\sigma e^{2356}) \, .
}}
Computing $j^6$ gives
\eq{\spl{
\label{SU2SU2source}
e^\Phi j & = - i\d \mathcal{W}^-_2 + \left(\frac{2}{27} f^2- \frac{2}{5}m^2\right)e^{2\Phi}\Re\Omega  \, , \\
& = j_1 (e^{234} - e^{156}) + j_2 (e^{246} - e^{135}) + j_3 (e^{126} - e^{345}) \, ,
}}
with $j_1, j_2$ and $j_3$ some complicated factors depending on $a,b$ and $c$ whose exact form does not
matter for the moment. It contains the same terms as $\Re \Omega$ but with different coefficients. In fact, one can check
that $j^6$ is {\em not} proportional to $\Re \Omega$ unless $|a|=|b|=|c|$, which reduces the solution to a nearly-K\"ahler geometry.
This time it is not immediately obvious how to choose the orientifold projection. Choosing them naively along the six terms
leads to the fields and structure constants having the wrong transformation properties. In appendix \ref{smearedori}
we outline how to find the orientifold involutions associated to a smeared source in general and then apply the procedure to the case at hand. In order to present the resulting involutions, it is convenient
to define complex one-forms as follows
\eq{\label{complbasis}\spl{
e^{z^1} & = \pm \frac{e^{\frac{i 3\pi}{4}}}{2 c_1 \sqrt{bc(2bc-h)}} \left\{[2bc -h + i (a^2-b^2-c^2)]e^1 + [a^2-b^2-c^2 +i(2bc-h)]e^4\right\}\, , \\
e^{z^2} & = \pm \frac{e^{\frac{i 3\pi}{4}}}{2 c_1 \sqrt{ac(2ac-h)}} \left\{[2ac -h + i (b^2-a^2-c^2)]e^2 + [b^2-a^2-c^2 +i(2ac-h)]e^5\right\}\, , \\
e^{z^3} & = \pm \frac{e^{\frac{i \pi}{4}}}{2 c_1 \sqrt{ab(2ab-h)}} \left\{[2ab -h + i (c^2-a^2-b^2)]e^3 + [c^2-a^2-b^2 +i(2ab-h)]e^6\right\}\, ,
}}
where the signs must be chosen such that $\Omega=e^{z^1z^2z^2}$. Defining further the associated $x$ and $y$ one-forms $e^{z^i}=e^{x^i}-i e^{y^i}$, the orientifold involutions are
given as in \eqref{standinv}.

\subsection{The $\frac{\text{SU(3)}\times \text{U(1)}}{\text{SU(2)}}$ solution}

We construct the algebra by taking
\eq{\spl{
 E_{i}&=G_{i+3}, \quad i=1,\dots, 5; \quad E_6=M; \\
E_7&=G_1; \quad E_8=G_2;\quad E_9=G_3
~,}}
where the $G_i$'s are the Gell-Mann matrices generating su(3), $M$ generates a u(1), and
the su(2) subalgebra is generated by $E_7,E_8$ and $E_9$. It follows that
the SU(2) subgroup is embedded entirely inside the SU(3),
so that the total space is given by $\frac{\text{SU(3)}}{\text{SU(2)}} \times \text{U(1)}\simeq S^5\times S^1$.
The structure constants are
\eq{
f^7{}_{89}=1, \quad f^7{}_{14}=f^7{}_{32}=f^8{}_{13}=f^8{}_{24}=f^9{}_{12}=f^9{}_{43}=1/2,
\quad f^{5}{}_{12}=f^{5}{}_{34}=\frac{\sqrt{3}}{2}, \quad\text{cyclic}
%,\nn\\
%f^6{}_{14}&=f^6{}_{32}=\frac{a_1}{2},\quad f^6{}_{13}=f^6{}_{24}=\frac{a_2}{2},\quad
%f^6{}_{78}=a_3,\quad  f^6{}_{12}=f^6{}_{43}=\frac{a_3}{2}
~.}
Invariant one-forms are generated by $\{e^5,e^6\}$, invariant two-forms by
$$
\{e^{12}+e^{34}, e^{13}-e^{24},e^{14}+e^{23}, e^{56}\}~,$$
and invariant
three-forms are given by $$\{e^{145}+e^{235},e^{135}-e^{245},e^{126} + e^{346},e^{146}+e^{236},e^{136}-e^{246},e^{125}+e^{345} \}~.$$
There is a solution for non-zero source:
\eq{\spl{
J & =  -a (e^{13}-e^{24})+b(e^{14}+e^{23})+c e^{56} \, , \\
\Omega & = -\frac{\sqrt{3}}{2c_1} \Big\{\left[ 2a (e^{145}+e^{235}) + 2b (e^{135}-e^{245}) + c(e^{126} + e^{346})\right]  \\
& - \frac{i}{\sqrt{a^2+b^2}} \left[ac(e^{146}+e^{236}) +bc (e^{136}-e^{246}) -2 (a^2+b^2)(e^{125}+e^{345}) \right]
\Big\} \, ,
}}
with $a,b$ and c three free parameters and
\eq{\label{su3qu1su2sol}
\spl{
c & > 0 \, , \quad a^2+b^2 \neq 0 \, , \qquad \text{metric positivity} \, , \\
\frac{1}{(c_1)^2} & = \frac{2}{3}{\sqrt{a^2+b^2}}, \qquad \text{normalization of } \Omega \, , \\
c_1 & := -\frac{3i}{2} {\cal W}_1^- = -\frac{2}{3} e^{\Phi} f \, , \\
{\cal W}_2^- & = \frac{i}{2\, c_1\sqrt{a^2+b^2}} \left[ -a(e^{13}-e^{24})+b(e^{14}+e^{23}) - 2 c e^{56}\right] \, , \\
d {\cal W}_2^- & = -\frac{i\sqrt{3}}{2\,c_1\sqrt{a^2+b^2}} \left[ a(e^{145}+e^{235}) + b(e^{135}-e^{245}) - c (e^{126}+e^{346})\right] \, , \\
 3|{\cal W}_1^-|^2 - |{\cal W}_2^-|^2 & = 0 \, .
}}
By a suitable change of basis we can always arrange for $a>0$ and $b>0$, which we will assume from now on.
Note that $d {\cal W}_2^-$ is not proportional to $\Re\Omega$, hence
the source is not of the form (\ref{jo}). Interestingly, if we take the part of the source along $\Re \Omega$
to be zero, i.e.\ $j^6 \wedge \Im \Omega=0$,  we find from the last equation in \eqref{su3qu1su2sol}
that $m=0$. This would amount to a combination of smeared D6-branes and O6-planes such that the total tension
is zero. Allowing for negative total tension (more orientifolds), we could have $m > 0$.
For an arbitrary $m$ we find the background
\eq{\spl{
H &= -\frac{\sqrt{3} \tilde{m}}{5 \tilde{c}_1} a \left[ 2(e^{145}+e^{235}) + 2 \rho (e^{135}-e^{245}) + \sigma (e^{126} + e^{346}) \right] \, , \\
e^{\Phi} F_2 &= \frac{1}{2} a^{1/2} \tilde{c}_1 \left[ (e^{13}-e^{24}) - \rho (e^{14} + e^{23} ) + \sigma e^{56} \right] \, , \\
e^{\Phi} F_4 &= a^{-1/2} \tilde{f} \text{vol}_4 + \frac{3}{5} a^{3/2} \tilde{m}  \left[ (1+\rho^2) e^{1234}- \sigma (e^{1356}-e^{2456}) + \rho \sigma (e^{1456}+e^{2356}) \right] \, ,
}}
where we defined $\rho=b/a$ and $\sigma=c/a$ and used again (\ref{reduced}).
{}From (\ref{ltsolb}) we compute for the source
\eq{\spl{\label{aaa}
e^{\Phi} j^{O6} & = -\frac{\sqrt{3}}{10 \tilde{c}_1} a^{1/2} \left(5 \tilde{c}_1^2 - 4 \tilde{m}^2 \right)\left[e^{145} + e^{235} + \rho(e^{135} - e^{245})\right] \\
& + \frac{\sqrt{3}}{20 c_1} a^{1/2} \sigma \left(5 \tilde{c}_1^2 + 4 \tilde{m}^2 \right)  \left(e^{126} + e^{346}\right) \, .
}}
One can check that for the background the source satisfies the calibration conditions \eqref{calcond}.
If we make the following coordinate transformation
\eq{\spl{
e^{1'}=e^1 \, , \quad
e^{2'}=e^2 \, , \quad
e^{3'}=e^3 + \rho^{-1} e^4 \, , \quad
e^{4'}=e^3 - \rho e^4 \, , \quad
e^{5'}=e^5 \, , \quad
e^{6'}=e^6 \, ,
}}
we see clearly that $j$ is a sum of four decomposable terms
\eq{\spl{
\label{specialsource}
e^{\Phi} j^{6} & = -\frac{\sqrt{3}}{10 \tilde{c}_1} a^{1/2} \left(5 \tilde{c}_1^2 - 4 \tilde{m}^2 \right) (e^{2'4'5'} + \rho e^{1'3'5'}) \\
&+ \frac{\sqrt{3}}{20 \tilde{c}_1} a^{1/2}  \sigma \left(5 \tilde{c}_1^2 + 4 \tilde{m}^2 \right) \left(e^{1'2'6'} - \frac{\rho}{1+\rho^2} e^{3'4'6'}\right) \, ,
}}
to which we can associate four orientifold involutions.

%%%%%%%%%%%%%%%%%%%%%%%%%%%%%%%%%%%%%%%%%%%%%%%%%%%%%%%%%%%

\section{Low energy physics I: nilmanifolds}\label{lowennil}

In this section, we will first explicitly perform the Kaluza-Klein reduction on the torus solution of section \ref{sixtorusex}
and the Iwasawa solution with $m=0$ of section \ref{iwasawaex} and calculate the mass spectrum.  Next, we will use the effective supergravity approach and construct the K\"ahler potential and the superpotential. From there we can get the potential and compare
the mass spectrum in both approaches. We find exact agreement. From then on, we will only use the effective supergravity approach and study the Iwasawa solution with $m \neq 0$ and the type IIB solution of section \ref{nilIIBsol} in this section as well as the coset models in the next section.\footnote{As a general remark, we will not
consider blow-up modes associated to the fixed points of the orientifold involutions. Ideally, we would like to argue that the blow-up modes will be stabilized by flux through the blown-up cycle at a size much smaller than the size of the internal manifold. Unfortunately, such an analysis is beyond the scope of the present paper. It may be possible, however,
 to argue for the stabilization of the blow-up modes using a local analysis of the singularities as in \cite{dewolfe}.}

\subsection{Kaluza-Klein reduction}\label{seckkmain}

We are interested in performing a Kaluza-Klein reduction on each of the AdS$_4\times\mathcal{M}_6$ solutions described
in  sections \ref{sixtorusex} and \ref{iwasawaex}. Let $x$ and $y$ be space-time and internal-manifold coordinates, respectively.
Moreover, let $\bg{\Phi}(x,y)$ be a `vacuum', i.e.\ a particular solution of
the equations of motion of ten-dimensional supergravity.
The Kaluza-Klein reduction (see \cite{duff} for a review) consists in expanding all ten-dimensional fields $\Phi(x,y)$ in `small' fluctuations
around the vacuum:
\al{
\Phi(x,y)=\bg{\Phi}(x,y)+\delta\Phi(x,y)~,
\label{kkansatza}
}
keeping only  terms up to linear order in $\delta\Phi(x,y)$ in the equations of motion (corresponding to
at most quadratic terms in the Lagrangian). From now on the hats indicate background quantities and the $\delta$s
fluctuations. The fluctuations are Fourier-expanded in the
internal space:
\al{
\delta\Phi(x,y)=\sum_n\phi_n(x)\omega_n(y)
~,
\label{kkansatzb}
}
where $\phi_n(x)$ are four-dimensional space-time fields,
and the
$\omega_n(y)$'s
form a basis of eigenforms of the Laplacian operator $\Delta=\d \d^\dagger+\d^\dagger \d$ in the six-dimensional space $\mathcal{M}$ (the internal
part of the vacuum solution).

In the following we will truncate
all the higher Kaluza-Klein modes in the harmonic expansion (\ref{kkansatzb}) and keep
only those $\omega_n(y)$'s in (\ref{kkansatzb}) that are left-invariant
on $\mathcal{M}_6$. The resulting modes are not in general harmonic, but can be combined into
eigenvectors of the Laplacian whose eigenvalues are of order of the geometric fluxes.

Plugging the ansatz (\ref{kkansatza})-(\ref{kkansatzb}) into the ten-dimensional equations of motion and keeping at most linear-order
terms in the fluctuations, one can read off the masses of the space-time fields, i.e.\ the `spectrum'.
In the present case, this is accomplished by comparing with the equations of motion
for non-interacting
fields propagating in AdS$_4$. Let $M$ and $\Lambda$ be the mass of the field and
 the cosmological constant of the AdS space, respectively, such that
\subeq{
\al{
\label{scalar}\mathrm{Scalar:}& \qquad \Delta\phi+\left(M^2+\frac{2}{3}\Lambda\right)\phi=0~,\\
\label{vector}\mathrm{Vector:}& \qquad \Delta\phi_{\mu}+\nabla_{\mu}\nabla^{\nu}\phi_{\nu}+M^2\phi_{\mu}=0~,\\
\label{metric}\mathrm{Metric:}& \qquad \Delta_Lh_{\mu\nu}+2\nabla_{(\mu}\nabla^{\rho}h_{\nu)\rho}
-\nabla_{(\mu}\nabla_{\nu)}h^{\rho}{}_{\rho}+(M^2-2\Lambda)h_{\mu\nu}=0~,
}}
where $\Delta_L$ is the Lichnerowicz operator defined by:
\al{\label{Lichnerowicz}
\Delta_Lh_{\mu\nu}=-\nabla^2h_{\mu\nu}-2R_{\mu\rho\nu\sigma}h^{\rho\sigma}+2R_{(\mu}{}^{\rho}h_{\nu)\rho}
~.
}
With the above definitions, the Breitenlohner-Freedman bound \cite{bf} is simply
\eq{
M^2\geq0
~,}
for the metric and the vectors. For the scalars, however, a negative mass-squared is allowed:
\eq{
M^2\geq\frac{\Lambda}{12}=-\frac{|W|^2}{4}
~,}
where $W$ was defined in eq.~(\ref{defW}). Actually, we will present the results for the mass spectrum of the scalars
in terms of
\eq{
\label{modmass}
\tilde{M}^2= M^2 + \frac{2}{3}\Lambda \, ,
}
for which the Breitenlohner-Freedman bound reads
\eq{
\label{BFbound}
\tilde{M}^2 \geq -\frac{9|W|^2}{4} \, .
}
We will take $\tilde{M}=0$ as the definition of an unstabilized modulus since
from \eqref{scalar} we see that then, if it were not for the boundary conditions of
AdS$_4$, a constant shift of $\phi$ would be a solution to the equations of motion.
Therefore a constant shift of $\phi$ leads to a new vacuum solution.

We would also like to express the fluctuations of the RR field strengths $\delta F$
in terms of the fluctuations of the potentials $\delta C$ in such a way that the
 Bianchi identity $d_HF=-j$ is automatically satisfied.
Indeed, as explained in appendix \ref{appkk}, this is achieved for
\eq{
\label{Fsolvedpot}
e^{\delta B} \delta F = (\d + \bg{H}) \delta C - (e^{\delta B}-1) \bg{F} \, ,
}
where we have set $\delta F_0=0$. For the NSNS flux we can just write
\eq{
\label{Hsolvedpot}
H = \bg{H} + \delta H = \bg{H} + \d \delta B \, .
}

\subsubsection{IIA on AdS$_4\times$T$^6$}
\label{torusKKresult}

By direct computation of the Kaluza-Klein reduction on the six-torus solution of section \ref{sixtorusex},
we obtain the following mass eigenvalues $\tilde{M}^2/|W|^2$ for the scalar fields:\footnote{The calculations
in appendix \ref{torusKK} were made in the {\em ten-dimensional} Einstein frame, while the effective supergravity
approach followed in later sections will lead to a result in the {\em four-dimensional} Einstein frame. By dividing
out with $|W|^2$ we avoid conversion problems, since $\tilde{M}^2$ and $|W|^2$ transform in the same way under change of
frame.}
\begin{center}
  \begin{tabular}{|c|c|}
    \hline
    Complex structure & $-2$, $-2$, $-2$  \\
    \hline
    K\"ahler \& dilaton & $70$, $18$, $18$, $18$ \\
    \hline
    Three axions of $\delta C_3$ & $0$, $0$, $0$ \\
    \hline
    $\delta B$ \& one more axion & $88$, $10$, $10$, $10$\\
    \hline
    \end{tabular}
\end{center}
Although the Kaluza-Klein procedure, as outlined
 in \ref{seckkmain} is straightforward, many
of the intermediate steps are rather subtle. The interested reader
may consult appendix \ref{appkk} for more details on the derivation and
on the exact mass eigenvectors.

Even without these details we can make a number of interesting observations. First of all three axions
correspond to massless moduli. This is a feature that is also discussed in \cite{fontaxions}. It is argued there that, when one introduces D6-branes, these axions can provide St\"uckelberg masses to some of the U(1) gauge fields on
the D-brane. In any case, we will see later that most of the coset examples do have all moduli
stabilized. Secondly, we notice that some masses are tachyonic, which is allowed because they are
still above the Breitenlohner-Freedman bound \eqref{BFbound}. And finally, scalars that are in the same supermultiplet, like
the complex structure moduli and the three corresponding axions, the dilaton and the remaining axion,
the K\"ahler moduli and the $B$-field moduli have different masses. This is in fact a subtlety of the supersymmetry algebra of AdS$_4$ that no longer  allows a definition for the mass as an invariant Casimir operator.

For this model, we can decouple the tower of Kaluza-Klein masses (see the discussion below \eqref{1.2} in section \ref{consistency}) when
we take $m^2 (e^{2\Phi} L_{int}^2)\ll1$.

\subsubsection{IIA on the Iwasawa manifold}

As explained in detail in appendix \ref{appkk}, performing the Kaluza-Klein reduction on the Iwasawa manifold we obtain the exact
same mass spectrum as in the case of the Kaluza-Klein reduction on the six-torus solution of the previous section.
This is of course the expected result, since the two solutions are related by T-duality. The limit for decoupling the Kaluza-Klein tower
corresponds to taking $\beta \ll 1$.

\subsection{Effective supergravity}\label{WKAnalysis}

In this section we derive the masses of the scalar fields by means of the superpotential and K\"ahler potential for the three explicit examples of compactification manifolds we found. Comparing these results with the results of the explicit Kaluza-Klein reduction in the previous section may be seen as a cross-check for the expressions for the superpotential and K\"ahler potential.

\subsubsection{Superpotential and K\"ahler potential}

The superpotential and K\"{a}hler potential of the effective $\mathcal{N}=1$ supergravity
have been derived in various ways in \cite{granasup,grimmsup,effective} (based on earlier work of \cite{GVW,gl}).
Here we summarize the main formul\ae{} which will be used in the following; more details
on the derivation can be found in appendix \ref{appeff4d}. We present first the superpotential and K\"ahler potential appropriate
for general SU(3)$\times$SU(3)-structure and then specialize to strict SU(3) and static SU(2)-structure.

The part of the effective four-dimensional action containing the graviton and
the scalars reads:
\eq{%\label{4Daction}
S =  \int \d^4 x \sqrt{-g_4}  \left( \frac{M_P^2}{2} R  - M_P^2\mathcal{K}_{i \bar{\jmath}} \partial_{\mu} \phi^i \partial^{\mu}  \bar{\phi}^{\bar{\jmath}} - V (\phi,\bar{\phi}) \right)\, ,
}
where $M_P$ is the four-dimensional Planck mass.
The scalar potential is given in terms of the superpotential via:\footnote{In \cite{cassanipotential} the scalar potential
was for general type II SU(3)$\times$SU(3) compactifications directly derived from dimensional reduction of the action.}
\eq{
V(\phi,\bar{\phi}) = M_P^{-2} e^{\mathcal{K}} \left( \mathcal{K}^{i\bar{\jmath}} D_i \mathcal{W}_{\E} D_{\bar{\jmath}} \mathcal{W}^*_{\E} - 3 |\mathcal{W}_{\E}|^2 \right) \, ,
}
where the superpotential in the Einstein frame $\mathcal{W}_{\E}$ reads
\eq{
\label{suppoteinstein}
\mathcal{W}_{\E}  = \frac{-i}{4 \kappa_{10}^2} \int_M\langle \Psi_2,F+i\,\d_H(\Re \mathcal{T})\rangle\ ,
}
and $\langle \cdot, \cdot \rangle$ indicates the Mukai pairing \eqref{mukai}, $\Re \mathcal{T}= e^{-\Phi} \Im \Psi_1$, and $\Psi_1$ and $\Psi_2$ are the pure spinors describing the geometry. Using the expansion in background and fluctuations of \eqref{Fsolvedpot} and \eqref{Hsolvedpot} we can rewrite this
as
\eq{
\mathcal{W}_{\E}  = \frac{-i}{4 \kappa_{10}^2} \int_M\langle \Psi_2 e^{\delta B},\bg{F}+i\,\d_{\bg{H}}(e^{\delta B} \Re \mathcal{T}-i\delta C)\rangle\ ,
}
where we used property \eqref{mukaiprop}.
This shows how the fields organize in complex multiplets $\Psi_2 e^{\delta B}$ and $\Re \mathcal{T}-i\delta C$, which will
be clearer in concrete examples.

The K\"ahler potential reads
\eq{
\label{kahlereinstein}
\mathcal{K}  =  - \ln i \int_M \langle \Psi_2, \bar{\Psi}_2 \rangle - 2 \ln i \int_M \langle t , \bar{t} \rangle +3 \ln(8 \kappa_{10}^2 M_P^2)\, ,
}
where we defined $t=e^{-\Phi} \Psi_1$.
Note that  $\Re t$ should be thought of as a function of $\Im t$ so that $t$ can be seen as (non-holomorphically)
dependent on $\mathcal{T}$. This is explained in more detail in appendix \ref{appeff4d}.

\subsubsection*{IIA SU(3)}

Specializing to the IIA SU(3) case with pure spinors \eqref{SU3pure}, the superpotential takes the form
\eq{\label{WSU3}
\mathcal{W}_{\E}  = \frac{-i e^{-i \theta}}{4 \kappa_{10}^2} \int_M \langle e^{i(J-i\delta B)}, \bg{F} - i \d_{\bg{H}} \left( e^{\delta B} e^{-\Phi} \Im \Omega +i \delta C_3 \right) \rangle \, ,
}
and the K\"ahler potential is given by
\eq{\label{KahlerJOmega}
\mathcal{K}  = - \ln \int_M \, \frac{4}{3} J^3 - 2 \ln \int_M \, 2 \, e^{-\Phi}\Im \Omega \wedge e^{-\Phi} \Re \Omega + 3 \ln(8 \kappa_{10}^2 M_P^2) \, ,
}
where $e^{-\Phi }\Re \Omega$ should be seen as a function of $e^{-\Phi}\Im \Omega$.
On the fluctuations we must impose the orientifold projections \eqref{oriall}. It turns out that for all our examples (except
for a special case of the $\frac{\text{SU(3)}\times \text{U(1)}}{\text{SU(2)}}$-model):
\eq{
\delta B \wedge \Im \Omega = 0 \, ,
}
since there are no odd five-forms.
By expanding in a suitable basis of even and odd expansion forms (which have to be identified separately for each case), we find that the fluctuations organize naturally in complex scalars
\subeq{\label{expansionSU3}\al{
& J_c = J - i \delta B = (k^i - i b^i)Y^{(2-)}_i = t^i Y^{(2-)}_i \, , \\
& e^{-\Phi} \Im \Omega + i \delta C_3 = (u^i +i c^i) e^{-\hat{\Phi}} Y^{(3+)}_i = z^i e^{-\hat{\Phi}} Y^{(3+)}_i \, ,
}}
where we took out the background $e^{-\hat{\Phi}}$ from the definition of $z^i$ for further
convenience.

\subsubsection*{IIB SU(2)}

Specializing to the case of type IIB SU(2) with pure spinors \eqref{SU2pure}, the superpotential becomes
\eq{\label{WSU2}
\mathcal{W}_{\E} = \frac{i}{4 \kappa_{10}^2} \int_M \langle 2 V\wedge e^{i(\omega_2-i\delta B)}, \bg{F} - i \d_{\bg{H}} \left( e^{\delta B} e^{-\Phi} \Im(e^{2V\wedge \bar{V}}\wedge \Omega_2 )+ i\delta C\right) \rangle \, ,
}
and the K\"ahler potential
\eq{\label{KahlerPotIIB}
 \mathcal{K} = - \ln \left( -2i \int_M  \,  2V\wedge 2\bar{V} \wedge \omega_2^2\right) - 2 \ln \int_M 2 \, \langle \Re t,\Im t \rangle + 3 \ln(8 \kappa_{10}^2 M_P^2) \, ,
}
where again $\Re t$ should be considered as a function of $\Im t = - \Im \left(e^{-\Phi} e^{2 V \wedge \bar{V}} \Omega_2 \right)$.

Under the orientifold projections we find from eq.~(3.5a) and (3.5b) of \cite{kt} for the NSNS-sector
\subeq{\al{
& O5: \qquad \sigma^* V =-V \, , \qquad \sigma^* \omega_2=-\omega_2 \, , \qquad \sigma^* \Omega_2 = -\Omega^*_2 \, , \qquad \sigma^* \delta B = -\delta B \, , \\
& O7: \qquad \sigma^* V =V \, , \qquad \sigma^* \omega_2=-\omega_2 \, , \qquad \sigma^* \Omega_2 = \Omega^*_2 \, , \qquad \sigma^* \delta B = -\delta B \, ,
}}
and for the RR-sector
\subeq{\al{
& O5: \qquad \sigma^* \delta C_2 = \delta C_2 \, , \qquad \sigma^* \delta C_4 = -\delta C_4 \, , \\
& O7: \qquad \sigma^* \delta C_2 = -\delta C_2 \, , \qquad \sigma^* \delta C_4 = \delta C_4 \, .
}}
Again we find that the fluctuations organize naturally in complex scalars
\subeq{\label{expansionSU2}
\al{
\omega_c = \omega_2 - i \delta B & = (k^i - i b^i)Y^{(2--)}_i = t^i Y^{(2--)}_i \, , \\
 e^{-\Phi} \Im \Omega_2 + i \delta C_2 & = (u^i + i c ^i) e^{-\hat{\Phi}} Y^{(2+-)}_i = z^i e^{-\hat{\Phi}} Y^{(2+-)}_i \, , \\
 -i e^{-\Phi} 2 V \wedge \bar{V} \wedge \Re \Omega_2 + i \delta C_4 & = (v^i + i h ^i) e^{-\hat{\Phi}} Y^{(4-+)}_i = w^i e^{-\hat{\Phi}} Y^{(4-+)}_i \, , \\
 2 V & = C(iY_1^{(1-+)} - \tau Y_2^{(1-+)}) \, ,
}}
where we define $\tau = x + i y$, and each time the first/second sign of the $Y_{i}$ indicates the behaviour under the O5/O7-involution. Note that $C$ is a complex overall factor that can be
scaled away together with the warp factor and the arbitrary $U(1)$ phase.

\subsubsection{IIA on AdS$_4\times$T$^6$}

For convenience we choose a slightly different expansion basis as in appendix \ref{torusKK}:
\eq{\spl{
Y^{(2-)} & : \qquad e^{12}, e^{34}, e^{56} \, ; \\
Y^{(3+)} & : \qquad -e^{135}, e^{146}, e^{236}, e^{245} \, .
}}
We then find the superpotential
\eq{
\mathcal{W}_{\E, \text{Torus}} = \frac{e^{-i \theta}}{4 \kappa_{10}^2} \vols m \left[ - t^1 t^2 t^3 + \frac{3}{5} (t^1+t^2+t^3) -\frac{2}{5} (z^1+z^2+z^3+z^4) \right] \, ,
\
}
where $\vols$ is a standard volume $\vols= \int e^{1\ldots 6}$, which does not depend
on the moduli. Moreover,
the K\"{a}hler potential reads:
\subeq{
\label{kp}
\eq{%\label{kpa}
\mathcal{K} = \mathcal{K}_k + \mathcal{K}_c +3 \ln(8 \kappa_{10}^2 M_P^2 \vols^{-1} e^{4\bg{\Phi}/3}) \, ,
}
where
\eq{\label{kpa}
\mathcal{K}_k = -\ln \left(\prod_{i=1}^3 (t^i+\bar{t}^i) \right) \,
}
is the K\"ahler potential in the K\"ahler-moduli sector
and
\eq{\label{kpb}
\mathcal{K}_c  = - \ln \left(4 \prod_{i=1}^4 \left( z^i+\bar{z}^i\right) \right) \,
}}
is the K\"ahler potential in the complex structure moduli sector.

Using the expressions for the superpotential and the K\"ahler potential it is straightforward to calculate the masses for the scalar fields
from the quadratic terms in the potential. To perform this calculation we made use of \cite{supercosmo}. Upon noting that in the Kaluza-Klein analysis we set the background values for the warp factor and the dilaton equal to zero
and $\text{Vol}=\vols$, we find exactly the same result as in section \ref{torusKKresult}.

\subsubsection{IIA on the Iwasawa manifold}

We choose the following expansion basis:
\eq{\spl{
Y^{(2-)} & : \qquad \beta^2 e^{65}, e^{12}, e^{34} \, ; \\
Y^{(3+)} & : \qquad -\beta e^{135}, -\beta e^{146}, -\beta e^{236}, \beta e^{245} \, .
}}
This implies that $\d Y^{(3+)}_i = - \beta e^{1234}$  for all $i=1,\ldots,4$.
 We find the superpotential
\eq{
\label{supiw}
\mathcal{W}_{\E, \text{Iwasawa}} = \frac{-i e^{-i \theta}}{4 \kappa_{10}^2} m_T \vols \left[ \frac{3}{5} - \frac{2}{5} t^1 (z^1+z^2+z^3+z^4) +\frac{3}{5} (t^1 t^2 + t^1 t^3)- t^2 t^3 \right] \, ,
}
where $\vols=\int -\beta^2 e^{1\ldots 6}$ is again a standard volume and $m_T=\frac{5}{2} e^{-\bg{\Phi}}\beta$ the Romans
mass of the T-dual torus solution.
We note here the following relation
\eq{
\mathcal{W}_{\E, \text{Iwasawa}} = -i t^1 \mathcal{W}_{\E, \text{Torus}}(t^1 \rightarrow \frac{1}{t^1}) \, ,
}
which follows from T-duality.
The K\"ahler potential for the Iwasawa manifold is
the same as in \eqref{kp}.

In the end, we find exactly the same masses as on the torus, as expected from T-duality, and thus
also the same masses as in the Kaluza-Klein approach for the Iwasawa. This provides a consistency check
on the ability of the superpotential/K\"ahler potential approach to handle geometric fluxes.

If we now turn on $m\neq$ 0 in the Iwasawa solution, we get extra terms in the superpotential
that look exactly like the torus superpotential, so we find:
\eq{
\mathcal{W}_{\E, \text{Iwasawa}, m\neq 0} = \mathcal{W}_{\E, \text{Iwasawa}}(m_T) + \mathcal{W}_{\E, \text{Torus}}(m) \, .
}
The mass spectrum is the same upon replacing $m_T^2 \rightarrow m^2+m_T^2$. Also, this time it is possible to decouple
the Kaluza-Klein tower: in the limit $(m^2+m_T^2) (e^{2\Phi} L_{int}^2) \ll 1$.

\subsubsection{IIB on the nilmanifold 5.1}

For our analysis we will need expansion forms with the following behaviour under $O5$ and $O7$-planes
\bigskip
\begin{center}
  \begin{tabular}{|c|c|c|}

    \hline
    type under O5/O7& basis & name \\
    \hline
    \hline
  odd/even 1-form  &  $ e^{5}, \beta e^{6} $ & $Y^{(1-+)}_i$ \\
 \hline
  even/odd 2-form  &  $-e^{23}, e^{14}$ & $Y^{(2+-)}_i$\\
 \hline
   odd/odd 2-form  &  $-e^{13}, e^{24}$ & $Y^{(2--)}_i$\\
    \hline
 odd/even 4-form &  $ e^{1256}, e^{3456}$ & $Y^{(4-+)}_i$ \\
    \hline
   \end{tabular}
\end{center}
and choose the standard volume $\vols= \int \beta e^{123456}$.

So together with \eqref{expansionSU2} we see that there are two complex ``four-dimensional'' K\"ahler moduli in $\omega_c$, two complex
moduli in $\delta C_2 - i e^{-\Phi} \Im \Omega_2$ and two in $e^{-\Phi} 2 V \wedge \bar{V} \wedge \Re \Omega_2 + i \delta C_4$.
These four moduli include the axions, the dilaton, the ``four-dimensional'' complex structure moduli and the ``two-dimensional'' K\"ahler modulus in $2 V \wedge \bar{V}$.

The superpotential is given by:
\eq{
\mathcal{W_{\E, \text{nil}}} = -\frac{m_T \vols C}{4 \kappa_{10}^2} \left( \frac{3}{5} -\frac{2}{5} \tau (z^1 - z^2 + w^1 + w^2) + \frac{3}{5} \tau (t^1 + t^2) -  t^1 t^2 \right) \,  ,
}
where $\vols= \int \beta e^{123456}$ is the standard volume.
The K\"ahler potential reads:
\eq{\spl{
\mathcal{K}  = &  -\ln \left((\tau^i+\bar{\tau}^i)\prod_{i=1}^2 (t^i+\bar{t}^i) \right)
- \ln \left(-4 \prod_{i=1}^2 ( z^i+\bar{z}^i)\prod_{i=1}^2 ( w^i+\bar{w}^i)\right) \\
&+ 3 \ln(8 \kappa_{10} M_P^2 \vols^{-1} e^{4\bg{\Phi}/3})
  - \ln |C|^{2}   \, .
}}
We can eliminate the complex scalar $C$ by performing a K\"ahler transformation \eqref{ktrans}.
Using the above, we derive the expected (due to T-duality) result that
the masses for the scalar fields are the same as for the T$^6$ and the Iwasawa manifold.

\section{Low energy physics II: coset spaces}\label{lowencoset}

In this section we study the low energy effective theory of the coset spaces described in section \ref{cstsoc}.

\subsection{IIA on $\frac{\text{G}_2}{\text{SU(3)}}$}
\label{lowenG2qSU3}

We choose the expansion forms in \eqref{expansionSU3} as follows:
\eq{\spl{\label{basisG2qSU3}
Y^{(2-)} & : \qquad a (e^{12}-e^{34}+e^{56}) \, ; \\
Y^{(3+)} & : \qquad a^{3/2} (-e^{235}-e^{246}+e^{145}-e^{136}) \, ,
}}
and the standard volume $\vols=-\int a^3 \, e^{123456}$.

The superpotential reads:
\eq{
\mathcal{W}_{\E} = \frac{i e^{-i\theta}e^{-\bg{\Phi}}}{4 \kappa_{10}^2} \vols a^{-1/2} \left( -\frac{3\sqrt{3}}{2} +\frac{8\tilde{m}i}{5}z^0-\frac{9\tilde{m}i}{5}t^1+ 4 \sqrt{3}z^0 t^1 -\frac{\sqrt{3}}{2}(t^1)^2 +i\tilde{m} (t^1)^3  \right) \, ,
}
whereas the K\"ahler potential is
\eq{
\mathcal{K} = -\ln \left((t^1+\bar{t}^1)^3 \right) - \ln \left(4 (z^0+\bar{z}^0)^4 \right)+3 \ln(8\kappa_{10}^2 M_P^2 \vols^{-1}e^{4\bg{\Phi}/3}) \, .
}

\begin{figure}[tp]
\centering
\subfigure[Behaviour for small $\tilde{\mu}$]{
\psfrag{mu}{\footnotesize$\tilde{\mu}$}
\psfrag{mass}{\footnotesize$\tilde{M}^2/|W|^2$}
\includegraphics[width=7cm]{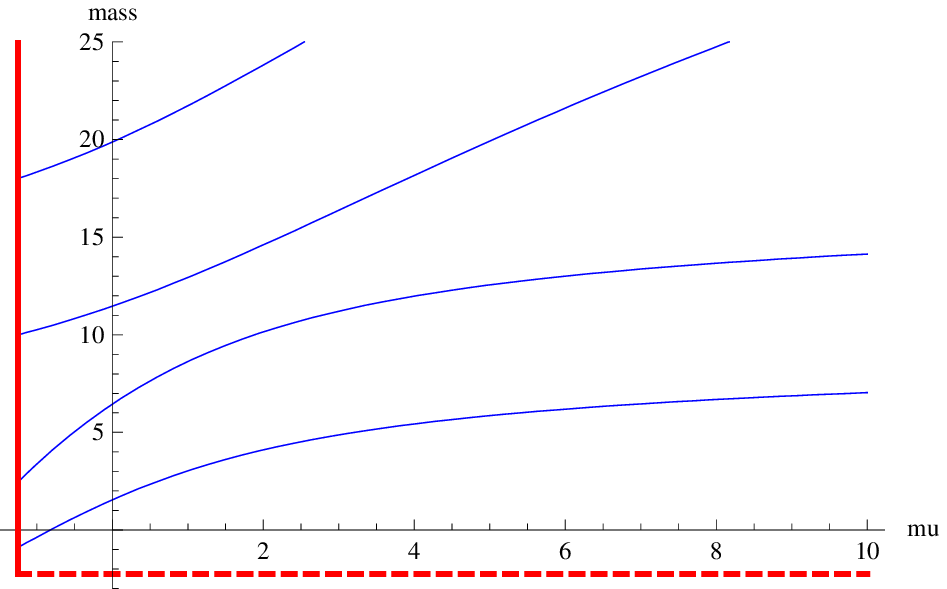}
}\hspace{0.2cm}
\subfigure[Behaviour for large $\tilde{\mu}$]{
\psfrag{mu}{\footnotesize$\tilde{\mu}$}
\psfrag{mass}{\footnotesize$\tilde{M}^2/|W|^2$}
\includegraphics[width=7cm]{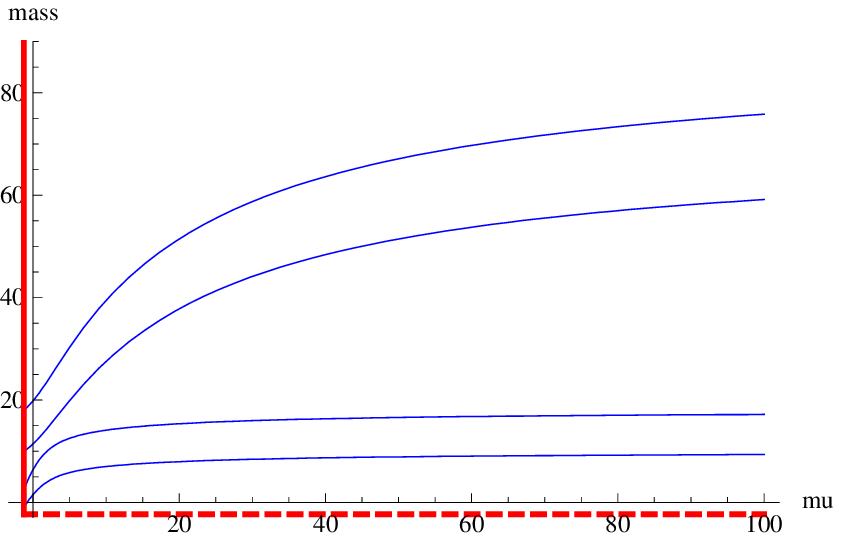}
}
\caption{Mass spectrum of $\frac{\text{G}_2}{\text{SU(3)}}$.}
\label{plotG2qSU3}
\end{figure}

If we plot $\tilde{M}^2/|W|^2$ the overall scale $a$ drops out and the only parameter
is the reduced orientifold tension $\tilde{\mu}$: see Figure \ref{plotG2qSU3}, where the dashed and solid red line represent the Breitenlohner-Freedman bound (\ref{BFbound}) and the bound (\ref{condition}) for $\tilde{\mu}$, respectively. We see that all four moduli masses are above the Breitenlohner-Freedman bound. Moreover, all masses are positive for $\tilde{\mu}>-0.82$. For $\tilde{\mu} \rightarrow \infty$ the masses asymptote to $\tilde{M}^2/|W|^2=(10,18,70,88)$, which
are the same as for the torus in section \ref{torusKKresult} (except there are no complex structure moduli and corresponding axions).
In fact, this is universal behaviour for all models we studied. Indeed, for $\tilde{\mu} \rightarrow \infty$ we find from \eqref{boundtra} that
$m \rightarrow \infty$ regardless of the details $\mathcal{W}^-_1,\mathcal{W}^-_2$ of the model, and exactly only the same terms as in
the torus example are relevant in the superpotential.

In section \ref{consistency} we have seen that $|\mathcal{W}_1^-| L_{int} \ll 1$ is one way to obtain a separation of scales between the light masses and the Kaluza-Klein masses even before the uplifting. However, as can be seen from eq.~\eqref{G2sol}, this is impossible to achieve for this coset.

\subsection{IIA on $\frac{\text{Sp(2)}}{\text{S}(\text{U(2)}\times \text{U(1)})}$}
\label{lowenSO5qSU2U1}

\begin{figure}[tp]
\centering
\subfigure[$\sigma=1$: nearly-K\"ahler ($\Rightarrow$ Einstein)]{
\psfrag{mu}{\footnotesize$\tilde{\mu}$}
\psfrag{mass}{\footnotesize$\tilde{M}^2/|W|^2$}
\includegraphics[width=7cm]{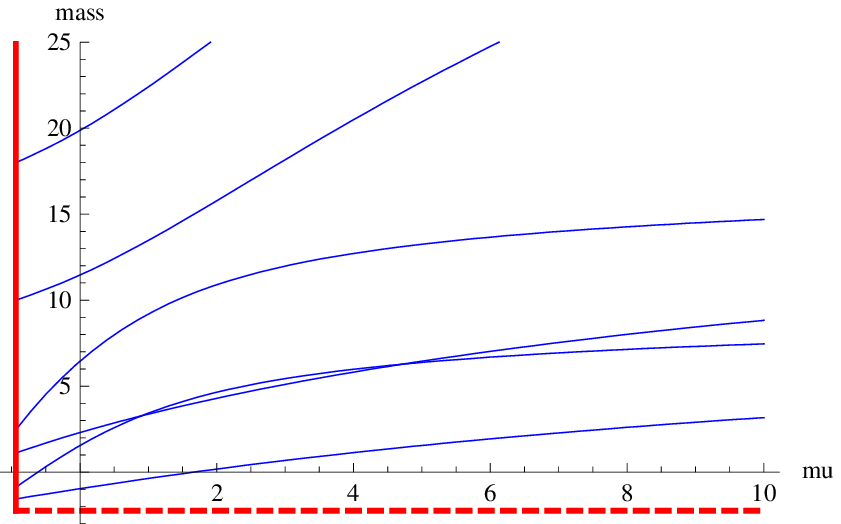}
}\hspace{0.1cm}
\subfigure[$\sigma=\frac{2}{5}$]{
\psfrag{mu}{\footnotesize$\tilde{\mu}$}
\psfrag{mass}{\footnotesize$\tilde{M}^2/|W|^2$}
\includegraphics[width=7cm]{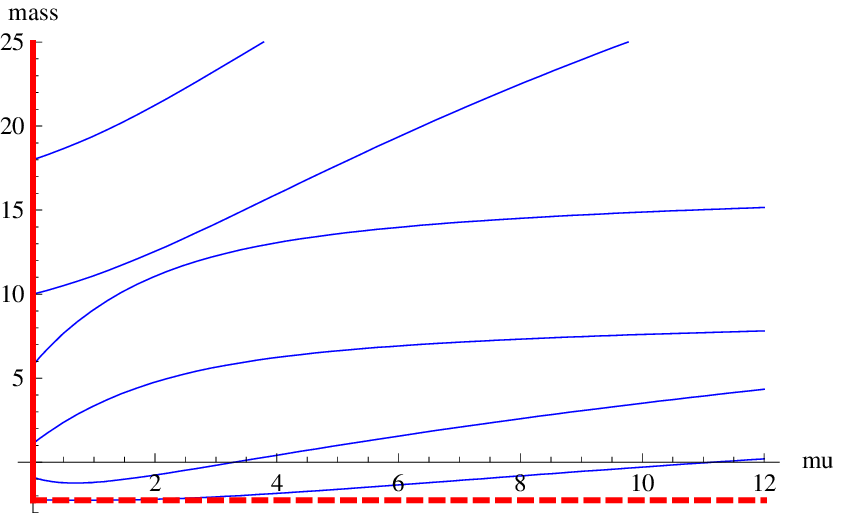}
}\vspace{0.5cm}
\subfigure[$\sigma=2$: Einstein]{
\psfrag{mu}{\footnotesize$\tilde{\mu}$}
\psfrag{mass}{\footnotesize$\tilde{M}^2/|W|^2$}
\includegraphics[width=7cm]{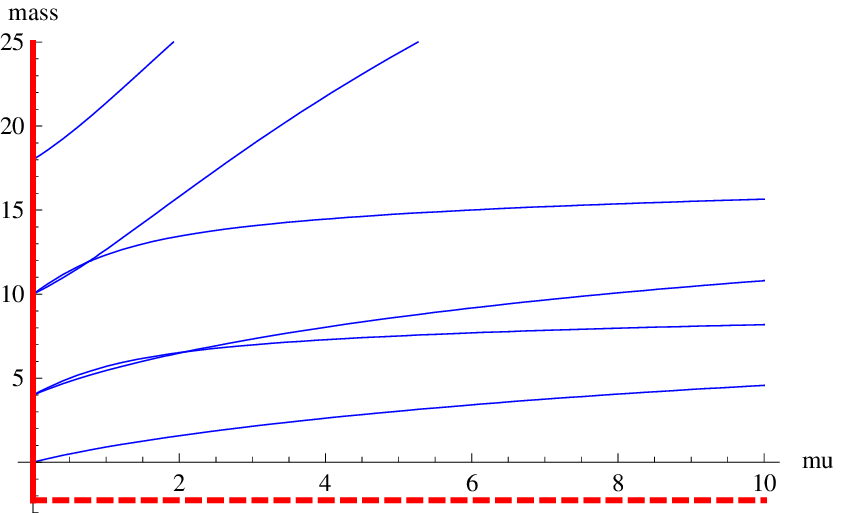}
}\hspace{0.1cm}
\subfigure[$\sigma=13$]{
\psfrag{mu}{\footnotesize$\tilde{\mu}$}
\psfrag{mass}{\footnotesize$\tilde{M}^2/|W|^2$}
\includegraphics[width=7cm]{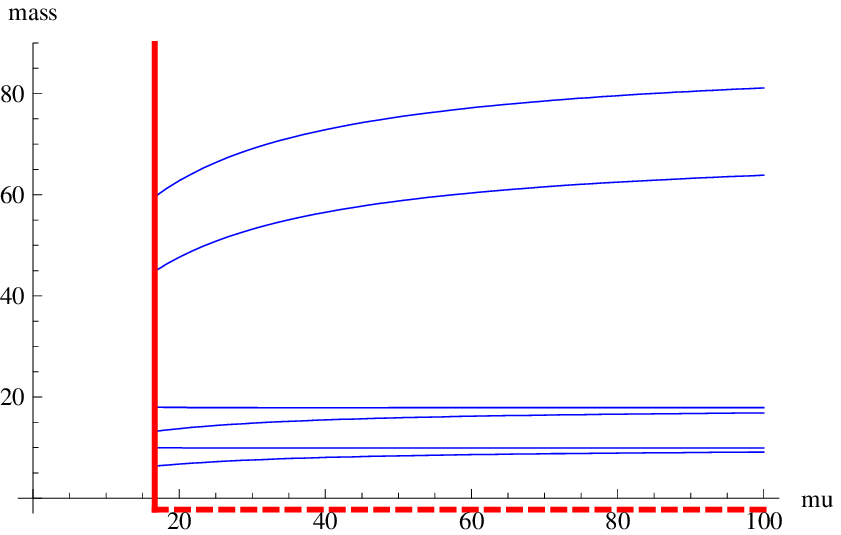}
}
\caption{Mass spectrum of  $\frac{\text{Sp(2)}}{\text{S}(\text{U(2)}\times \text{U(1)})}$.}
\label{plotSp2qSU2U1}
\end{figure}

We choose the expansion forms in \eqref{expansionSU3} as follows:
\eq{\spl{\label{basisSO5qSU2U1}
Y^{(2-)} & : \qquad a (e^{12}+e^{34}), -a e^{56} \, ; \\
Y^{(3+)} & : \qquad a^{3/2} (e^{235}+e^{246}+e^{145}-e^{136}) \, ,
}}
and the standard volume $\vols=-\int a^3 \, e^{123456}$. We find the following superpotential

\eq{\spl{
\mathcal{W}_{\E} &= \frac{i e^{-i\theta}e^{-\bg{\Phi}}}{4 \kappa_{10}^2} \vols a^{-1/2} \left( - \tilde{f} \sigma +\frac{8 \tilde{m}i}{5} \sigma^{1/2} z^0-\frac{3\tilde{m}i}{5}(2 \sigma t^1 + t^2)- 2(2 t^1+t^2)z^0+i\tilde{m} (t^1)^2 t^2 \right. \\
 & + \left. \sigma^{1/2}\left( \frac{3}{2} -\frac{5}{4} \sigma \right) (t^1)^2  - \left(\sigma^{-1/2} -\frac{3}{2}\sigma^{1/2}\right) t^1 t^2 \right) \, ,
}}
and K\"ahler potential
\eq{
\mathcal{K} = -\ln \left((t^1+\bar{t}^1)^2 (t^2+\bar{t}^2)\right) - \ln \left(4 (z^0+\bar{z}^0)^4 \right)+3 \ln(8\kappa_{10}^2 M_P^2 \vols^{-1}e^{4\bg{\Phi}/3}) \, .
}
This time the solution has next to the overall scale $a$ two free parameters:
the ``shape'' $\sigma=c/a$ and the orientifold tension $\tilde{\mu}$. In Figure \ref{plotSp2qSU2U1} we display
plots for several values of $\sigma$: $\sigma=1$ is the nearly-K\"ahler point while for $\sigma=2/5$ and $\sigma=2$
the lower bound for $\tilde{\mu}$ from \eqref{boundtra} is exactly zero. These were extreme points in \cite{tomtwistor}
since outside the interval $[2/5,2]$ the lower bound is above zero and solutions without orientifolds are no longer possible.
Moreover, for $\tilde{\mu}=0$ also $m=0$ and these solutions can be lifted to M-theory. We also display a plot for large $\sigma$, here
$\sigma=13$. We see that the lower bound for $\tilde{\mu}$ is indeed positive so that there must be net orientifold charge. The behaviour is however
already like the universal behaviour for $\tilde{\mu} \rightarrow \infty$. Again we see that in all cases all masses are above the Breitenlohner-Freedman bound
and by choosing $\tilde{\mu}$ large enough they are all positive.

\begin{figure}[tp]
\centering
\subfigure[$\sigma=-2$]{
\psfrag{mu}{\footnotesize$\tilde{\mu}$}
\psfrag{mass}{\footnotesize$\tilde{M}^2/|W|^2$}
\includegraphics[width=7cm]{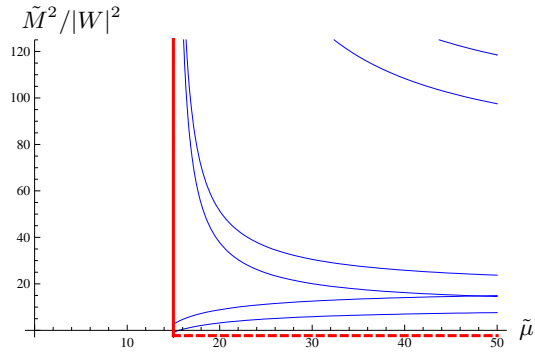}
}\hspace{0.2cm}
\caption{Mass spectrum of the continuation of $\frac{\text{Sp(2)}}{\text{S}(\text{U(2)}\times \text{U(1)})}$ to negative $\sigma$.}
\label{plotSp2qSU2U1negative}
\end{figure}

Again we would like to get $|\mathcal{W}_1^-| L_{int} \ll 1$ to decouple the Kaluza-Klein modes. From eq.~\eqref{SO5qSU2U1sol} we see that this can be formally obtained by putting $\sigma \rightarrow -2$, i.e.\ we need to analytically continue to negative values for $\sigma$. From \cite{feng} we learn that $\sigma<0$ is indeed possible, but the model cannot be described as a left-invariant SU(3)-structure on the coset $\frac{\text{Sp(2)}}{\text{S}(\text{U(2)}\times \text{U(1)})}$ anymore.
Rather it is a twistor bundle on a four-dimensional hyperbolic space. The precise agreement between the results of
\cite{tomtwistor} (which is based on \cite{feng}) and \cite{klt} (wherever they overlap) suggests that the analytic
continuation is possible. Strictly speaking, however, one should check that also the {\em mass spectrum} can be analytically continued
to negative values for $\sigma$. Although this seems plausible to us, verifying it directly would require using entirely different technology,
and lies beyond the scope of the present paper. In deriving the plot of Figure \ref{plotSp2qSU2U1negative} for $\sigma=-2$, we have assumed
that such analytic continuation of the mass spectrum is possible. We see that two mass eigenvalues stay light, while the others blow up if $\mathcal{W}_1^- \rightarrow 0$ and join the Kaluza-Klein masses.
In this limit the light modes have $\tilde{M}^2/|W|^2=(-38/49, 130/49)$.

\subsection{IIA on $\frac{\text{SU(3)}}{\text{U(1)}\times \text{U(1)}}$}
\label{lowenSU3U1U1}

In this case we choose the expansion forms in \eqref{expansionSU3} as follows:
\eq{\spl{\label{basisSU3qU1U1}
Y^{(2-)} & : \qquad -a e^{12}, a e^{34}, -a e^{56} \, ; \\
Y^{(3+)} & : \qquad a^{3/2} (e^{235}+e^{246}+e^{136}-e^{145}) \, ,
}}
and the standard volume $\vols=\int a^3 \, e^{123456}$.

Using the expression (\ref{WSU3}) for the superpotential in the $\text{SU}(3)$ case and the expansion given in (\ref{expansionSU3}), we derive the superpotential
\eq{\spl{
\mathcal{W}_{\E} = & -\frac{i e^{-i\theta} e^{-\bg{\Phi}}}{4\kappa_{10}^2} \vols a^{-1/2} \Bigg( \tilde{f} \rho \sigma -\frac{8\tilde{m}i}{5}(\rho \sigma)^{1/2} z^0+\frac{3\tilde{m}i}{5}(\rho \sigma t^1 + \sigma t^2 + \rho t^3) \\
& +\frac{1}{4}(\rho \sigma)^{-1/2}\Big( (3\sigma+3\rho\sigma-5\sigma^2)t^1 t^2
 + (3\rho-5\rho^2+3\rho \sigma) t^1 t^3 + (-5+3\rho+3\sigma)t^2 t^3 \Big) \\
& - 2z^0(t^1+t^2+t^3) -i\tilde{m} t^1 t^2 t^3 \Bigg) \, .
}}
The K\"ahler potential is evaluated as in section \ref{WKAnalysis} and reads
\eq{
\mathcal{K} = -\ln \left(\prod_{i=1}^3 (t^i+\bar{t}^i) \right) - \ln \left(4 (z^0+\bar{z}^0)^4 \right)+3 \ln(8\kappa_{10}^2 M_P^2 \vols^{-1}e^{4\bg{\Phi}/3}) \, .
}

\begin{figure}[tp]
\centering
\subfigure[$\rho=\sigma=1$: nearly-K\"ahler. Lines indicated with $2$ have multiplicity 2.]{
\psfrag{mu}{\footnotesize$\tilde{\mu}$}
\psfrag{mass}{\footnotesize$\tilde{M}^2/|W|^2$}
\psfrag{m2}{\footnotesize 2}
\includegraphics[width=7cm]{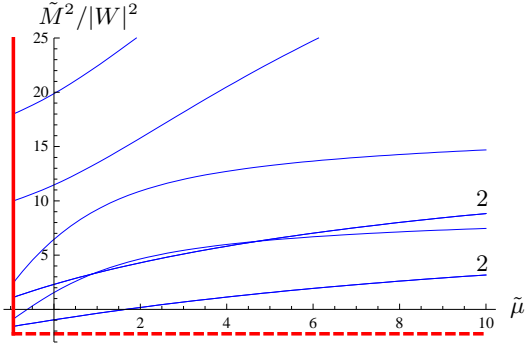}
}\hspace{0.2cm}
\subfigure[$\rho=1$ and $\sigma=\frac{2}{5}$.]{
\psfrag{mu}{\footnotesize$\tilde{\mu}$}
\psfrag{mass}{\footnotesize$\tilde{M}^2/|W|^2$}
\includegraphics[width=7cm]{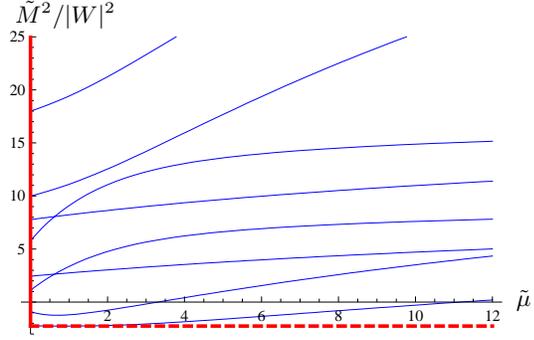}
}\hspace{0.2cm}
\subfigure[$\rho=1$ and $\sigma=2$.]{
\psfrag{mu}{\footnotesize$\tilde{\mu}$}
\psfrag{mass}{\footnotesize$\tilde{M}^2/|W|^2$}
\includegraphics[width=7cm]{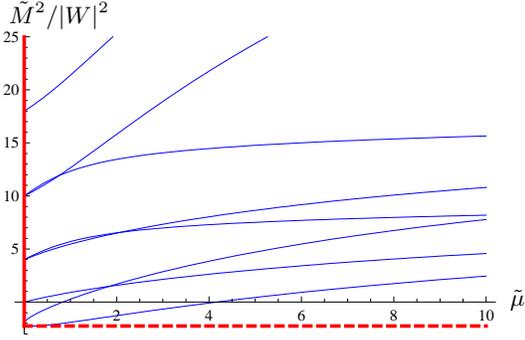}
}\hspace{0.2cm}
\centering
\subfigure[$\rho=\frac{5}{2}$ and $\sigma=\frac{1}{2}$.]{
\psfrag{mu}{\footnotesize$\tilde{\mu}$}
\psfrag{mass}{\footnotesize$\tilde{M}^2/|W|^2$}
\includegraphics[width=7cm]{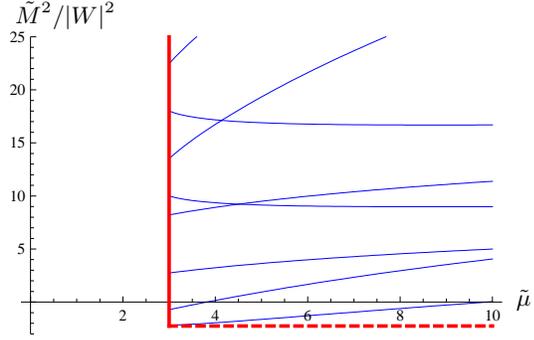}
}
\caption{Mass spectrum of $\frac{\text{SU(3)}}{\text{U(1)}\times \text{U(1)}}$.}
\label{plotSU3qU1U1}
\end{figure}

The model has this time two shape parameters: $\rho=b/a$ and $\sigma=c/a$. We display the mass spectrum for a number of selected values of
these parameters in Figure \ref{plotSU3qU1U1}. There is a symmetry under permuting $(a,b,c)$ which translates into a symmetry
under $\rho \leftrightarrow \sigma$ and $(\rho,\sigma,\tilde{\mu}) \leftrightarrow (\rho/\sigma,1/\sigma,\sigma \tilde{\mu})$. Applying these symmetries leads
to identical mass spectra. Moreover, the mass spectra for $\rho=1$ are apart from two more eigenvalues identical to the mass spectra
of $\frac{\text{Sp(2)}}{\text{S}(\text{U(2)}\times \text{U(1)})}$. We also display an example with $\sigma,\rho \neq 1$.

\begin{figure}[tp]
\centering
\subfigure[$\rho=\sigma=-\frac{1}{2}$.]{
\psfrag{mu}{\footnotesize$\tilde{\mu}$}
\psfrag{mass}{\footnotesize$\tilde{M}^2/|W|^2$}
\includegraphics[width=7cm]{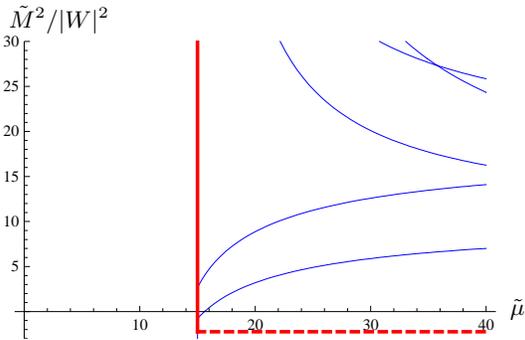}
}\hspace{0.2cm}
\subfigure[$\rho=-\frac{3}{4}$ and $\sigma=-\frac{1}{4}$.]{
\psfrag{mu}{\footnotesize$\tilde{\mu}$}
\psfrag{mass}{\footnotesize$\tilde{M}^2/|W|^2$}
\includegraphics[width=7cm]{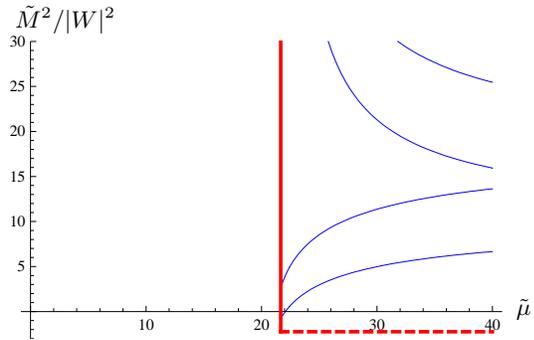}
}
\caption{Mass spectrum of $\frac{\text{SU(3)}}{\text{U(1)}\times \text{U(1)}}$ for negative $\sigma$ and $\rho$.}
\label{plotSU3qU1U1negative}
\end{figure}

In the plots of Figure \ref{plotSU3qU1U1negative} we have analytically continued to
$\rho<0,\sigma<0$ in order to approach the NCY limit, which we obtain for $\rho+\sigma=-1$.
Again, two eigenvalues stay light with $\tilde{M}^2/|W|^2=(-38/49, 130/49)$ in the limit
while the other eigenvalues blow up to the Kaluza-Klein scale.

\subsection{IIA on SU(2)$\times$SU(2)}

The expansion forms are given by
\eq{\label{expansionsu2su2}\spl{
Y_1^{2-} & =  a e^{14}  \, , \qquad Y_2^{2-} = b e^{25}  \, , \qquad Y_3^{2-} = c e^{36} \, , \\
Y_1^{3+} & = e^{x^1x^2y^3} =
\frac{-h}{4 c_1 (a+b+c)}(e^{123} + e^{456} + e^{126} + e^{345} + e^{315} + e^{264} + e^{156} + e^{234}) \, , \\
Y_2^{3+} & = e^{x^1y^2x^3} =
\frac{h}{4 c_1 (-a+b+c)}(e^{123} + e^{456} - e^{126} - e^{345} - e^{315} - e^{264} + e^{156} + e^{234}) \, , \\
Y_3^{3+} & = e^{y^1x^2x^3} =
\frac{-h}{4 c_1 (a-b+c)}(-e^{123} - e^{456} + e^{126} + e^{345} - e^{315} - e^{264} + e^{156} + e^{234}) \, , \\
Y_4^{3+} & = -e^{y^1y^2y^3} =
\frac{h}{4 c_1 (a+b-c)}(e^{123} + e^{456} + e^{126} + e^{345} - e^{315} - e^{264} - e^{156} - e^{234}) \, ,
}}
and the standard volume $\vols = -\int_M  abc \,e^{1\ldots 6}$.
One finds for the superpotential:
\eq{\spl{
\mathcal{W}  & = \frac{i e^{-i\theta}e^{-\bg{\Phi}}}{4\kappa_{10}^2 } \vols a^{-1/2}
\Bigg\{ \frac{3}{2} \tilde{c}_1 + i\tilde{m} \left( t^1t^2t^3 - \frac{3}{5} (t^1 + t^2 + t^3) - \frac{2}{5} (z^1 + z^2 + z^3 + z^4)     \right) \\
& + \frac{3}{2} \tilde{c}_1 (t^1t^2 + t^2t^3 + t^1t^3) \\
& + \frac{\tilde{c}_1}{\tilde{h}^2}\Big\{4\left[t^2t^3(1 - \rho^2 - \sigma^2) + t^1t^3\rho^2(-1 + \rho^2 - \sigma^2)
+ t^1t^2\sigma^2(-1 - \rho^2 + \sigma^2) \right] \\
& +\left[ t^1 (-1 + \rho^2 + \sigma^2) + t^2 \rho^2(1 - \rho^2 + \sigma^2) + t^3 \sigma^2 (1 + \rho^2 - \sigma^2)  \right](z^1 + z^2 + z^3 + z^4) \\
& +\rho\sigma\left[-2 t^1 + t^2(1+\rho^2-\sigma^2) + t^3(1-\rho^2+\sigma^2)\right](z^1 + z^2 - z^3 -z^4) \\
& + \sigma\left[t^1(1+\rho^2-\sigma^2) -2 \rho^2 t^2 + t^3(-1+\rho^2+\sigma^2)\right](z^1 - z^2 + z^3 -z^4) \\
& + \rho\left[t^1(1-\rho^2+\sigma^2) + t^2(-1+\rho^2+\sigma^2) -2 \sigma^2 t^3  \right](z^1 - z^2 - z^3 +z^4) \Big\}\Bigg\}  \, .
}}
The K\"{a}hler potential reads:
\eq{
\mathcal{K} = -\ln \left(\prod_{i=1}^3 (t^i+\bar{t}^i) \right) - \ln \left(4 \prod_{i=1}^4 \left( z^i+\bar{z}^i\right) \right) +3 \ln(8\kappa_{10}^2 M_P^2 \vols^{-1}e^{4\bg{\Phi}/3}) \, .
}

\begin{figure}[tp]
\centering
\subfigure[$\rho=\sigma=1$: nearly-K\"ahler. Lines indicated with $2$ have multiplicity $2$.]{
\psfrag{mu}{\footnotesize$\tilde{\mu}$}
\psfrag{mass}{\footnotesize$\tilde{M}^2/|W|^2$}
\psfrag{m2}{\footnotesize$\text{2}$}
\includegraphics[width=7cm]{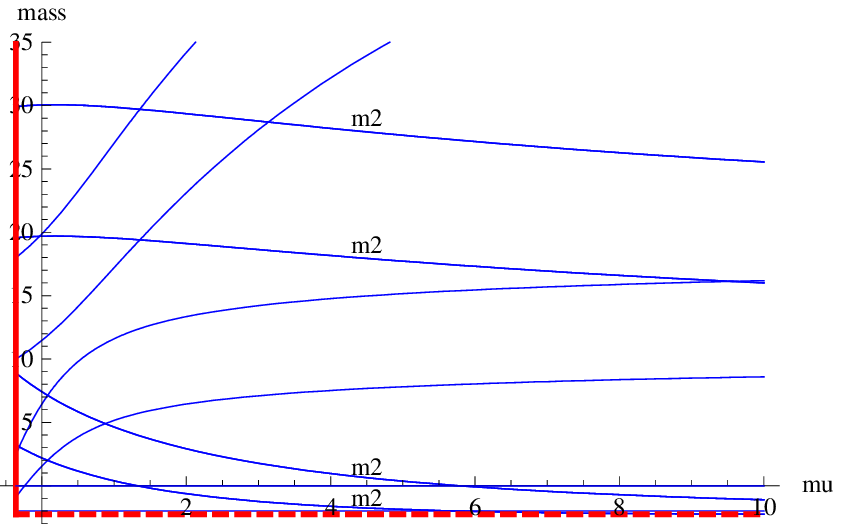}
}\hspace{0.2cm}
\subfigure[$\sigma=1$ and $\rho=\frac{2}{5}$.]{
\psfrag{mu}{\footnotesize$\tilde{\mu}$}
\psfrag{mass}{\footnotesize$\tilde{M}^2/|W|^2$}
\includegraphics[width=7cm]{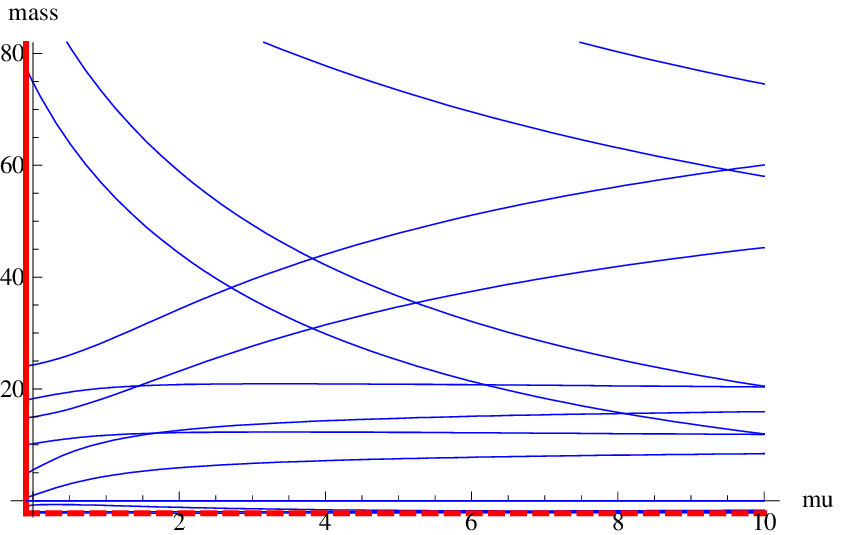}
}
\caption{Mass spectrum of SU(2)$\times$SU(2).}
\label{plotSU2SU2}
\end{figure}

There are again two shape parameters $\rho=b/a$ and $\sigma=c/a$ and the same symmetries $\rho \leftrightarrow \sigma$,
$(\rho,\sigma,\tilde{\mu}) \leftrightarrow (\rho/\sigma,1/\sigma,\sigma \tilde{\mu})$ as in the previous model. In Figure \ref{plotSU2SU2}
we display the mass spectrum for some values of the parameters. This time there will always be one unstabilized massless
axion ($\tilde{M}^2$=0) and a corresponding tachyonic complex structure modulus with $\tilde{M}^2/|W|^2=-2$.

In the limit $\mathcal{W}_1^- \rightarrow 0$, $\mathcal{W}_2^-$ blows up just as the lower bound for $\tilde{\mu}$.
So in principle we could decouple the Kaluza-Klein modes this way, however it is quite difficult to study
this singular limit.

\subsection{IIA on $\frac{\text{SU(3)}\times \text{U(1)}}{\text{SU(2)}}$}

We display the general results here and comment on the special case $5 c_1^2 - 4 e^{2\Phi} m^2 = 0$
in appendix \ref{SU3U1qSU2special}.
We choose the expansion forms in \eqref{expansionSU3} as follows:
\eq{
\label{SU3U1qSU2exp}
\spl{
Y^{(2-)} & : \qquad -a [(e^{13}-e^{24})-\rho(e^{14}+e^{23})], \, a e^{56} \, ; \\
Y^{(3+)} & : \qquad a^{3/2} [(e^{13}-e^{24})+\rho^{-1} (e^{14}+e^{23})]\wedge e^6, \,  a^{3/2} (e^{125}+e^{345}) \, ,
}}
and the standard volume $\vols=\int a^3 (1 + \rho^2) e^{123456}$.
The superpotential and K\"{a}hler potential read:

\eq{\spl{
\mathcal{W}_{\E} & = -\frac{i e^{-i\theta}e^{-\bg{\Phi}}}{4 \kappa_{10}^2}  \vols a^{-1/2} \left( \tilde{f} \sigma  +\frac{3 i \tilde{m}}{5} \sigma (2 t^1 +\frac{1}{\sigma} t^2) \right. \\ & + \sqrt{\frac{3}{2}} (1+\rho^2)^{-\frac{1}{4}} \left (-t^1 t^2 + \frac{\sigma}{2} (t^1)^2 \right)-i \tilde{m} (t^1)^2 t^2  \\
 &  \left.  -\frac{4 \sqrt{2} i \tilde{m}}{5 \rho} (1+\rho^2)^{\frac{1}{4}} z^1 + \frac{2 \sqrt{2} i \tilde{m}}{5} \sigma (1+\rho^2)^{-\frac{3}{4}} z^2 +\frac{2 \sqrt{3}}{\rho}  z^1 t^1 -\sqrt{3}(1+\rho^2)^{-1} t^2 z^2  \right)  \, ,
}}
and
\eq{\spl{
\mathcal{K} =& -\ln \left((t^1+\bar{t}^1)^2(t^2+\bar{t}^2) \right) - \ln \left(4 \frac{1}{\rho^2(1+\rho^2)} (z^1+\bar{z}^1)^2 (z^2+\bar{z}^2)^2\right) \\
 & +3\ln(8 \kappa_{10}^2 M_P^2 \vols^{-1}e^{4\bg{\Phi}/3}) \, .
}}

\begin{figure}[tp]
\centering
\subfigure[$\rho=\sigma=1$.]{
\psfrag{mu}{\footnotesize$\tilde{\mu}$}
\psfrag{mass}{\footnotesize$\tilde{M}^2/|W|^2$}
\psfrag{m2}{\footnotesize$\text{m2}$}
\includegraphics[width=7cm]{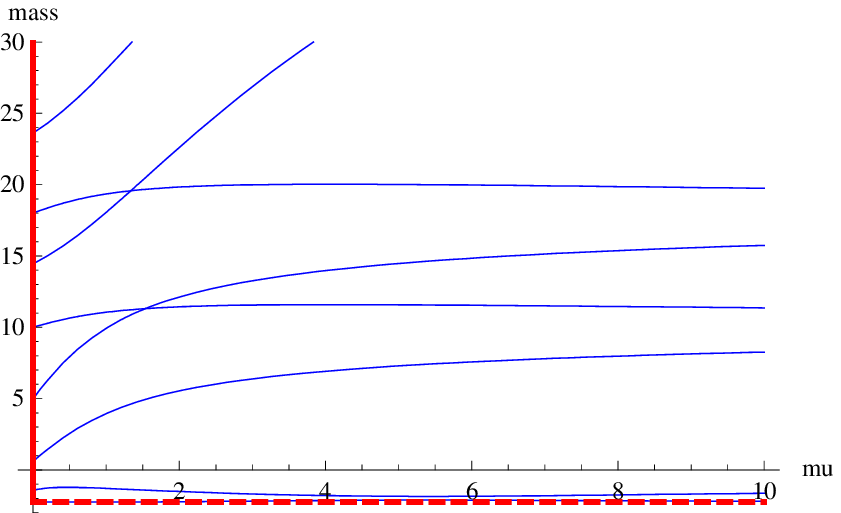}
}\hspace{0.2cm}
\subfigure[$\rho=\frac{1}{2}$ and $\sigma=2$.]{
\psfrag{mu}{\footnotesize$\tilde{\mu}$}
\psfrag{mass}{\footnotesize$\tilde{M}^2/|W|^2$}
\includegraphics[width=7cm]{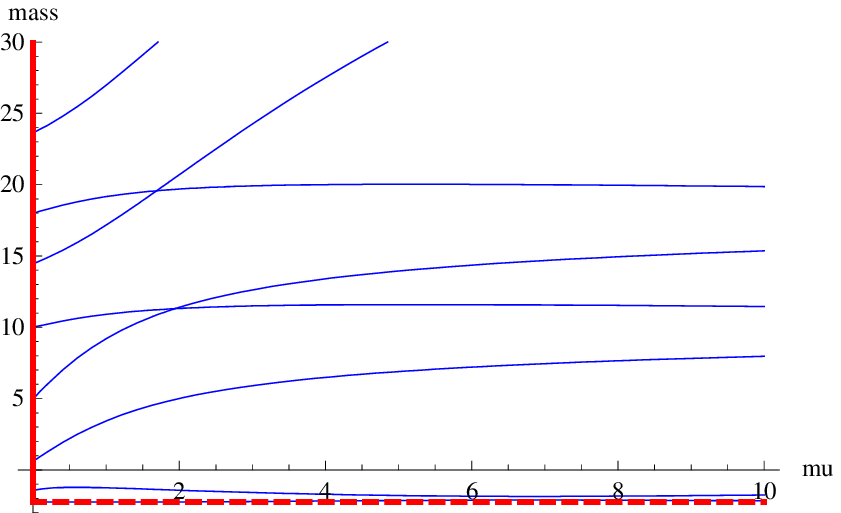}
}
\caption{Mass spectrum of $\frac{\text{SU(3)}\times \text{U(1)}}{\text{SU(2)}}$.}
\label{plotSU3U1qSU2}
\end{figure}

This model has two shape parameters $\rho=b/a$ and $\sigma=c/a$, and a symmetry under $(\rho,\sigma,\tilde{\mu}) \leftrightarrow (1/\rho,\sigma/\rho,\rho \tilde{\mu})$.
In Figure \ref{plotSU3U1qSU2}, we show the mass spectrum for some values of the parameters. The mass spectrum at $\mu=0$ turns out
to be independent of the parameters $\rho,\sigma$. There always seem to be two negative $\tilde{M}^2$ eigenvalues.

\section{Application to inflation in type IIA}
\label{inflation}

In the previous two sections, we have derived the low energy effective actions for the
AdS$_{4}$ compactifications studied in this paper. While an extensive analysis of the physical
properties of these low energy effective actions is beyond the scope of this paper, we would nevertheless
like to take a first look at our models in the context of the recent interesting
 work \cite{kachrunogo}. Extending the earlier work
\cite{kachruinflation}, the authors of \cite{kachrunogo} proved a no-go theorem against a period of slow-roll inflation
in type IIA compactifications on Calabi-Yau manifolds
with standard RR and NSNS-fluxes, D6-branes and O6-planes at large volume and with small string coupling. More precisely, they show that the slow-roll parameter $\epsilon$ is at least $\frac{27}{13}$ whenever the potential is positive, ruling out slow-roll inflation in a near-de Sitter regime, as well as meta-stable dS vacua.
As emphasized in \cite{kachrunogo}, however, the inclusion of other ingredients such as NS5-branes,
geometric fluxes and/or non-geometric fluxes evade the assumptions that underly this no-go theorem\footnote{In \cite{silversteindS} de Sitter vacua in type IIA were found using some of these additional ingredients. Furthermore a concrete string inflationary model on nilmanifolds with D4-branes was presented in \cite{Silverstein:2008sg}.}. Our coset models could thus be candidates for circumventing the no-go theorem as they all have geometric fluxes. So let us study this in some more detail.

The proof of this no-go theorem is remarkably simple and uses only the scaling properties of the scalar potential with respect to
the volume modulus
\eq{
\rho = \left(\frac{\text{Vol}}{\vols}\right)^{1/3} \, ,
}
where $\vols=|\int e^{123456}|$ is a standard volume,
and the dilaton modulus
\eq{
\tau = e^{-\Phi} \sqrt{\text{Vol}} \, ,
}
 as well as the signs of the various contributions to the potential. Concretely, if one denotes by $V_{3}$, $V_{p}$, $V_{D6}$ and $V_{O6}$ the potential contributions due to, respectively, $H_{3}$-flux, $F_{p}$-flux, D6-branes and O6-planes, the full potential has the schematic form
 \eq{\spl{
\label{scalingFHO}
 V = & V_3+ \sum_{p} V_{p} +V_{D6}+V_{O6} \\
 = &\frac{A_{3}(\phi_{i})}{\rho^{3}\tau^{2}} + \sum_{p}\frac{A_{p}(\phi_{i})}{\rho^{p-3}\tau^{4}} + \frac{A_{D6}(\phi_{i})}{\tau^{3}} -  \frac{A_{O6}(\phi_{i})}{\tau^{3}}
}}
with \emph{positive} coefficients, $A_{i}$, that depend on the other moduli, $\phi_{i}$. One can check that our potentials have indeed
this behaviour. This implies
\begin{equation}
 -\rho\frac{\partial V}{\partial \rho} -3\tau \frac{\partial V}{\partial \tau}=9V+\sum_{p}pV_{p}\geq 9V. \label{estimate}
\end{equation}
{} From this inequality and using
\begin{equation}
 \epsilon \equiv \frac{\mathcal{K}^{i\bar{\jmath}} V_{,i} V_{,\bar{\jmath}}}{V^{2}} \geq \frac{M_{P}^{2}}{2}\Big[\Big(\frac{\partial \ln V}{\partial{\hat{\rho}}} \Big)^{2} +\Big(\frac{\partial \ln V}{\partial{\hat{\tau}}} \Big)^{2}\Big],   \label{epsilondefinition}
\end{equation}
where the hatted fields are the canonically normalized volume modulus and dilaton, one derives the bound \cite{kachrunogo}
\begin{equation}
 \epsilon \geq \frac{27}{13} \qquad \textrm{whenever } V>0. \label{jjj}
\end{equation}
This forbids slow-roll inflation  everywhere in moduli space. Moreover, for a vacuum, the right-hand side of eq.~\eqref{estimate}
should vanish so that  de Sitter vacua are ruled out (as well as Minkowski vacua whenever $V_{p}> 0$ for at least one of $p=2,4,6$).

In our compactifications, this no-go theorem no longer needs to hold, because some of these compactifications have geometric fluxes with schematic potentials
\begin{equation}
\label{scalingf}
 V_{f}\propto \pm \rho^{-1}\tau^{-2}.
\end{equation}
Such a contribution to the potential would weaken (\ref{estimate}) to
\begin{equation}
  -\rho\frac{\partial V}{\partial \rho} -3\tau \frac{\partial V}{\partial \tau}=9V+\sum_{p}pV_{p} -2V_{f}. \label{estimate2}
\end{equation}
If $V_{f}$ turns out to be negative, the above expression would still be at least $9V$ just as before, and the no-go theorem expressed in the form (\ref{jjj}) would still hold. Thus, if geometric fluxes \emph{alone} are to circumvent this no-go theorem, they can do so at most if they are positive:
\begin{equation}
\label{avoid}
 V_{f}>0 \qquad \textrm{(Necessary condition for evading the no-go theorem)}.
\end{equation}

In fact, we can immediately find the geometric part of the potential from the Einstein-Hilbert term in the ten-dimensional action:
\eq{
V_f = - \frac{1}{2} M_P^4 \kappa_{10}^2 e^{2 \Phi} \text{Vol}^{-1} R = - \frac{1}{2} M_P^4 \kappa_{10}^2 \tau^{-2} R \, ,
}
where $R$ is the scalar curvature of the internal manifold. For cosets/group manifolds $R$ can be calculated from \eqref{riccicoset}.
This expression has indeed the expected scaling behaviour since $R \propto g^{-1} \propto \rho^{-1}$.
It follows that the condition \eqref{avoid} for avoiding the no-go theorem can be rephrased as
\eq{
R < 0 \, .
}
Let us display the scalar curvature for some of our coset models:
\subeq{\al{
\frac{\text{G}_2}{\text{SU(3)}}: & \qquad R  = \frac{10}{k_1} \, , \\
\frac{\text{Sp(2)}}{\text{S}(\text{U(2)}\times \text{U(1)})}: & \qquad R = \frac{6}{k_1} + \frac{2}{k_2} - \frac{k_2}{2(k_1)^2}  \, , \\
\frac{\text{SU(3)}}{\text{U(1)}\times \text{U(1)}}: & \qquad R = 3 \left(\frac{1}{k_1}+\frac{1}{k_2}+\frac{1}{k_3} \right) - \frac{1}{2} \left(\frac{k_1}{k_2k_3}+\frac{k_2}{k_1k_3}+\frac{k_3}{k_1k_2} \right)\, , \\
\frac{\text{SU(3)}\times \text{U(1)}}{\text{SU(2)}}: & \qquad
R = \frac{1}{\sqrt{1+\rho^2}} \left(\frac{6}{k_1} - \frac{3 \rho k_2}{4(1+\rho^2)k_1^2}\left|\frac{u_2}{u_1}\right| \right)\, ,
}}
where $k_i>0$ are the K\"ahler moduli and $u_i$ the complex structure moduli that enter the expansion of $J$ and $\Im \Omega$ in the basis \eqref{basisG2qSU3}, \eqref{basisSO5qSU2U1}, \eqref{basisSU3qU1U1} and \eqref{SU3U1qSU2exp}, respectively (where we put $a=1$).
We see that for $\frac{\text{G}_2}{\text{SU(3)}}$ the curvature is always positive, so inflation is still excluded, however for
the other models there are values of the moduli such that $R<0$. For SU(2)$\times$SU(2) we did not display the curvature, because taking generic values
of the complex structure and K\"ahler moduli, its expression is quite complicated and not very enlightening. However, also in that case it is possible
to choose the moduli such that $R<0$.

Note that this does not yet guarantee that the $\epsilon$ parameter is  indeed small, it just says that the theorem that requires it to be at least $27/13$ no longer applies. Hence, a logical next step would be to calculate $\epsilon$ in this region, ideally by taking also all other moduli into account
(see the general expression (\ref{epsilondefinition})) and try to make $\epsilon$ small or zero. These would
be necessary conditions for, respectively, inflation or de Sitter vacua. They are not sufficient however,
because for inflation, we would also need the $\eta$ parameter to be small and further obtain a satisfactory inflationary model
which could end in a meta-stable de Sitter vacuum etc. For a meta-stable de Sitter vacuum, on the other hand, one would also have to check that
the matrix of second derivatives only has negative eigenvalues.\footnote{The eta-parameter, or, more generally, the matrix of second derivatives of the potential, played the key role in the recent works \cite{etasmall} (see also the older paper \cite{Rami}), in which difficulties for inflation or meta-stable de Sitter vacua in various string inspired supergravity potentials are discussed. Note that the no-go theorem of \cite{kachrunogo} has no direct connection to these works, as it makes the much stronger statement that a small epsilon parameter or a de Sitter critical point of the potential do not exist, whereas this existence was assumed in
\cite{etasmall}.}

We remark that (\ref{estimate}) also played a r\^{o}le in the (failed) F-term uplifting
attempts by \cite{kalloshuplift}, so one might also reconsider that in the present context.

For our coset models, we have completely explicit expressions for the low energy effective theory so we have the necessary
tools to address all these questions and we hope to come back to them in future work.

\section{Conclusions}
\label{conclusions}
In this paper, we studied a number of type IIA SU(3)-structure compactifications on nilmanifolds and cosets, which are tractable enough to allow for an explicit derivation of the low energy effective theory. In particular, we calculated the mass spectrum of the light scalar modes, using $\mathcal{N}=1$ supergravity techniques. For the torus and the Iwasawa solution, we have also performed an explicit Kaluza-Klein reduction, which led to the same result,
supporting the validity of the effective supergravity approach, with superpotential \eqref{suppoteinstein} and K\"ahler potential \eqref{kahlereinstein}, also in the presence of geometric fluxes. Furthermore we have demonstrated that this superpotential and K\"ahler potential
 lead to sensible results in type IIB string theory with static SU(2)-structure as well.
For the
nilmanifold examples we have found that  there are always three unstabilized moduli corresponding to axions in the RR sector. On the other hand, in the
coset models, except for SU(2)$\times$SU(2), all moduli are stabilized.

It would be interesting to study the uplifting of these models to de Sitter space-times. This might be accomplished by
incorporating a suitable additional uplifting term in the potential along the lines of, e.g,  \cite{kklt}.
Although a negative mass squared for a light field
in AdS does not necessarily signal an instability, after the uplift all fields should have positive mass squared.
Unless the uplifting potential can change the sign of the squared masses,   it is thus desirable that they
are all positive even before the uplifting.  We find that this can be arranged in the coset models $\frac{\text{G}_2}{\text{SU(3)}}$, $\frac{\text{Sp(2)}}{\text{S}(\text{U(2)}\times \text{U(1)})}$ and $\frac{\text{SU(3)}}{\text{U(1)}\times \text{U(1)}}$ for suitable values of the orientifold charge.

An alternative approach towards obtaining  meta-stable de Sitter vacua could also be to search for non-trivial de Sitter minima in the original flux potential away from the AdS vacuum. In such a case, one would have to re-investigate the spectrum of the light fields and the issue of the
Kaluza-Klein decoupling.

We discussed this Kaluza-Klein decoupling in section \ref{consistency} for the original AdS vacua and found that it requires going to the nearly-Calabi Yau limit.
For our nilmanifolds, this can be easily arranged by tuning the parameters, while for our coset models it is somewhat harder. Indeed, we
found that for $\frac{\text{Sp(2)}}{\text{S}(\text{U(2)}\times \text{U(1)})}$ and $\frac{\text{SU(3)}}{\text{U(1)}\times \text{U(1)}}$ one
has  to make a continuation to negative values of the ``shape'' parameters. Strictly speaking, this can no longer be described as a left-invariant SU(3) structure on a coset anymore, but it can still be described in terms of a twistor bundle over a four-dimensional hyperbolic space.
It would be interesting to study these models in more detail, as there are more examples of this type. Another class
of vacua may be obtained by quotienting out the internal manifold by a discrete group $\Gamma$, where $\Gamma$ is a subgroup of SU(3).  This possibility may be of interest for model-building.

Another promising avenue would be to include space-time filling D-branes supporting the matter and gauge structure of the Standard Model. A lot is already known on model building
with intersecting D6-branes, see, e.g.,~\cite{uranga,blumenhagen} for reviews and many references. In our models it is indeed possible to insert D6-branes that do not break the supersymmetry by having them wrap special Lagrangian cycles: for a discussion in the context of AdS$_4$ compactifications see \cite{adsbranes}. We further remark that a superpotential for D-brane moduli, which fits nicely together with the superpotential \eqref{suppoteinstein}, was given in \cite{branesuppot}, but it is not complete in that it does not describe the charged fields coming from open strings ending on different D-branes, which are of course exactly the important ones in reproducing the Standard Model.

We have also discussed how geometric fluxes leading to negative scalar curvature  circumvent the assumptions that underlie the no-go theorem \cite{kachrunogo} against modular
inflation in type IIA string theory. We found that we can arrange for negative curvature in all the coset models except for $\frac{\text{G}_2}{\text{SU(3)}}$.
Circumventing this no-go theorem is just a first step towards a successful inflationary model. It  would certainly be worthwhile to study this possibility in more detail, perhaps including
extra ingredients such as NS5-branes and other types of branes \cite{silversteindS,Silverstein:2008sg}.

\begin{acknowledgments}
We would like to thank Ram Brustein, Jan Louis and Luca Martucci for useful discussions.
This work is supported in part by the European Community's Human Potential Programme under contract MRTN-CT-2004-005104 ``Constituents,
fundamental forces and symmetries of the universe'' and the Excellence Cluster ``The Origin and the Structure of the Universe'' in Munich.
C.~C., P.~K. and M.~Z.~are supported by the German Research Foundation (DFG) within the Emmy-Noether-Program (Grant number ZA 279/1-2).
\end{acknowledgments}

%%%%%%%%%%%%%%%%%%%%%%%%%%%%%%%%%%%%%%%%%%%%%%%%%%%%%%%%%%%

\appendix

\section{Type II supergravity}\label{sec:appb}

The bosonic content of type II supergravity consists of a metric $g$, a dilaton $\Phi$, an NSNS 3-form $H$ and RR-fields
$F_{n}$. In the democratic formalism of \cite{democratic}, where the number
 of  RR-fields is doubled,  $n$ runs over $0,2,4,6,8,10$ in IIA and over $1,3,5,7,9$ in type IIB.
We write $n$ to denote
the dimension of the RR-fields; for example $(-1)^n$ stands for $+1$ in type IIA and $-1$ in type IIB.
After deriving the equations of motion from the action, the redundant RR-fields are to be removed by hand
by means of the duality condition:
\eq{
\label{Fduality}
F_{n} = (-1)^{\frac{(n-1)(n-2)}{2}} e^{\frac{n-5}{2}\Phi}\star_{10} F_{(10-n)} ~,
}
given here in the Einstein frame.
We will often collectively denote the RR-fields, and the corresponding potentials, with polyforms
$F=\sum_n F_{n}$ and $C=\sum_n C_{(n-1)}$, so that: $F=\d_H C$.

In the Einstein frame, the bosonic part of the bulk action  reads:
\eq{
S_{\text{bulk}} = \frac{1}{2 \kappa_{10}^2} \int \d^{10} x \sqrt{-g} \left[ R -\frac{1}{2} (\partial \Phi)^2 - \frac{1}{2}
e^{-\Phi}H^2  -\frac{1}{4} \sum_n e^{\frac{5-n}{2}\Phi} F_{n}^2 \right] \, ,
}
where for an $l$-form $A$ we define
\eq{
A^2 = A \cdot A = \frac{1}{l!} \, A_{M_1 \ldots M_l} A_{N_1 \ldots N_l} g^{M_1N_1} \cdots g^{M_lN_l} \, .
}
Since \eqref{Fduality} needs to be imposed by hand this is strictly-speaking only a pseudoaction.
Note that the doubling of the RR-fields leads to factors of $1/4$ in their kinetic terms.

The contribution from the calibrated (supersymmetric) sources can be written as:
\al{\label{dbia}
S_{\text{source}} = \int \langle C, j \rangle
-\sum_n e^{\frac{n}{4}\Phi} \int\langle  \Psi_n,j\rangle \, ,
}
with
\eq{\label{dbib}
\Psi_n = e^A \d t \wedge \frac{e^{-\Phi}}{(n-1)! \hat{\epsilon}_1{}^T \epsilon_1} \hat\epsilon_1{}^T \gamma_{M_1 \ldots M_{n-1}} \hat\epsilon_2 \, \d X^{M_1} \wedge \ldots \wedge \d X^{M_{n-1}} \, ,
}
with $\hat{\epsilon}_{1,2}$ nine-dimensional internal supersymmetry generators.
For space-filling sources in compactifications to AdS$_4$ this becomes \cite{adsbranes}
\eq{
\Psi_n = \text{vol}_4 \wedge \left. e^{4A-\Phi} \Im \Psi_{1 \E} \right|_{n-4} \, ,
}
with $\Psi_{1\E}$ the pure spinor $\Psi_1$ in the Einstein frame.

The dilaton equation of motion and the Einstein equation read
\begin{subequations}
\begin{align}
\label{dilaton}
0&= \nabla^2 \Phi +\frac{1}{2}e^{-\Phi} H^2 - \frac{1}{8} \sum_n (5-n) e^{\frac{5-n}{2}\Phi}F_{n}^2
+\frac{\kappa_{10}^2}{2} \sum_n (n-4)e^{\frac{n}{4}\Phi}  \star \! \langle \Psi_n,j\rangle
\, , \\
\label{einstein}
0&= R_{MN} + g_{MN} \left( \frac{1}{8}e^{-\Phi} H^2 + \frac{1}{32} \sum_n (n-1)e^{\frac{5-n}{2}\Phi}F_{n}^2\right)\\
&~~~~~~~~~~~-\frac{1}{2}\partial_M \Phi \partial_N \Phi - \frac{1}{2}e^{-\Phi} H_{M} \cdot H_{N}
-\frac{1}{4} \sum_n e^{\frac{5-n}{2}\Phi}F_{n\,M} \cdot F_{n\,N} \nn\\
&~~~~~~~~~~~- 2 \kappa_{10}^2 \sum_n e^{\frac{n}{4}\Phi}\star \! \langle  \left(-\frac{1}{16} n g_{MN} + \frac{1}{2} g_{P(M} dx^P \otimes \iota_{N)}\right)\Psi_n,j\rangle\nonumber \, ,
 \, ,
\end{align}
\end{subequations}
where we defined for an $l$-form $A$
\eq{
A_{M} \cdot A_{N} = \frac{1}{(l-1)!} A_{MM_2 \ldots M_l} A_{NN_2 \ldots N_l} g^{M_2N_2} \cdots g^{M_lN_l} \, .
}
The Bianchi identities and the equations of motion
for the RR-fields, including  the contribution from the `Chern-Simons' terms of the sources,
take the form
\begin{subequations}
\label{eomBian}
\begin{align}
\label{Bianchis}
0&= \d F+H\wedge F  + 2 \kappa_{10}^2 \,  j\, , \\
\label{eomF}
0&= \d\left( e^{\frac{5-n}{2}\Phi}\star F_{n} \right) -e^{\frac{3-n}{2}\Phi}H\wedge  \star F_{(n+2)}
-2 \kappa_{10}^2 \, \alpha(j)
\, .
\end{align}
\end{subequations}
Finally, for the equation of motion for $H$ we have:
\eq{
\label{eomH}
0= \d (e^{-\Phi} \star \! H)
- \frac{1}{2} \sum_n e^{\frac{5-n}{2}\Phi}\star F_{n} \wedge F_{(n-2)}
+ \left. 2 \kappa_{10}^2 \sum_n e^{\frac{n}{4}\Phi}\Psi_n \wedge \alpha(j)\right|_8\, .
}
In the above equations we can redefine $j$ in order to absorb the factor of $2 \kappa_{10}^2$,
\eq{
(2 \kappa_{10}^2) j \rightarrow j \, ,
}
which we do in this paper.

%
%The Einstein equation and the dilaton equation of motion
%read
%\begin{subequations}
%\begin{align}
%\label{dilaton}
%0&= \nabla^2 \Phi +\frac{1}{2}e^{-\Phi} H^2 - \frac{1}{8} \sum_n (5-n) e^{\frac{5-n}{2}\Phi}F_{n}^2
%-\frac{\kappa_{10}^2}{2} \sum_n (n-4)e^{\frac{n}{4}\Phi}  \star \! \langle \Psi_n,j\rangle
%\, , \\
%\label{einstein}
%0&= R_{MN} + g_{MN} \left( \frac{1}{8}e^{-\Phi} H^2 + \frac{1}{32} \sum_n (n-1)e^{\frac{5-n}{2}\Phi}F_{n}^2\right)\\
%&~~~~~~~~~~~-\frac{1}{2}\partial_M \Phi \partial_N \Phi - \frac{1}{2}e^{-\Phi} H_{M} \cdot H_{N}
%-\frac{1}{4} \sum_n e^{\frac{5-n}{2}\Phi}F_{n M} \cdot F_{n N} \nn\\
%&~~~~~~~~~~~- 2 \kappa_{10}^2 \sum_n e^{\frac{n}{4}\Phi}\star \! \langle  \left(-\frac{1}{16} n g_{MN} + \frac{1}{2} %g_{P(M} dx^P \otimes \iota_{N)}\right)\Psi_n,j\rangle\nonumber
% \, .
%\end{align}
%\end{subequations}
%
%The Bianchi identities and equations of motion
%for the RR-fields, including  the contribution from the `Chern-Simons' terms of the sources,
%take the form
%\begin{subequations}
%\label{eomBian}
%\begin{align}
%0&= \d F+H\wedge F  + 4 \kappa_{10}^2 \,  j\, , \\
%0&= \d\left( e^{\frac{5-n}{2}\Phi}\star F_{n} \right) -e^{\frac{3-n}{2}\Phi}H\wedge  \star F_{(n+2)}
%-4 \kappa_{10}^2 \, \alpha(j)
%\, .
%\end{align}
%\end{subequations}
%Finally, for the equation of motion for $H$ we have:
%\eq{
%\label{eomH}
%0= \d (e^{-\Phi} \star \! H)
%- \frac{1}{2} \sum_n e^{\frac{5-n}{2}\Phi}\star F_{n} \wedge F_{(n-2)}
%+ \left. 2 \kappa_{10}^2 \sum_n e^{\frac{n}{4}\Phi}\Psi_n \wedge \alpha(j)\right|_8\, .
%}
%
The equations of motion resulting from $S_{\text{bulk}} +S_{\text{source}} $ were given in this form (in the string frame) in \cite{kt}, where it was shown that, under certain mild assumptions, imposing the supersymmetry equations together with the Bianchi identities for the forms, is enough to guarantee that the dilaton and Einstein equations are also satisfied.

\section{Structure groups}
\label{structure}

In this paper we have assumed the following $\mathcal{N}=1$ compactification ansatz for the ten-dimensional supersymmetry generators \cite{granaN1}
\eq{\label{spinansatz}\spl{
\epsilon_1 & =\zeta_+\otimes \eta^{(1)}_+ \, + \, \zeta_-\otimes \eta^{(1)}_- \ , \\
\epsilon_2 & =\zeta_+\otimes \eta^{(2)}_\mp \, + \, \zeta_-\otimes \eta^{(2)}_\pm \ ,
}}
for IIA/IIB, where $\zeta_\pm$ are four-dimensional and $\eta^{(1,2)}_\pm$ six-dimensional Weyl spinors. The Majorana conditions for $\epsilon_{1,2}$ imply the four- and six-dimensional reality conditions $(\zeta_+)^*=\zeta_-$ and $(\eta^{(1,2)}_+)^*=\eta^{(1,2)}_-$. This reduces the structure of the {\em generalized} tangent bundle to SU(3)$\times$SU(3) \cite{gualtieri}. The structure of the tangent bundle itself on the other hand is a subgroup of SU(3) since there is at least one invariant internal spinor. What subgroup
exactly depends on the relation between $\eta^{(1)}$ and $\eta^{(2)}$. Combining the terminology of \cite{granaN1} and \cite{andriot} the following classification can be made:
\begin{itemize}
\item strict SU(3)-structure: $\eta^{(1)}$ and $\eta^{(2)}$ are parallel everywhere;
\item static SU(2)-structure: $\eta^{(1)}$ and $\eta^{(2)}$ are orthogonal everywhere;
\item intermediate SU(2)-structure: $\eta^{(1)}$ and $\eta^{(2)}$ at a fixed angle, but neither a zero angle nor a right angle;
\item dynamic SU(3)$\times$SU(3)-structure: the angle between $\eta^{(1)}$ and $\eta^{(2)}$ varies, possibly becoming a zero angle or a right angle at a special locus.
\end{itemize}
Since for static and intermediate SU(2)-structure there are two independent internal spinors the structure of the tangent bundle reduces to SU(2), while
for dynamic SU(3)$\times$SU(3)-structure no extra constraints beyond SU(3) are imposed on the topology of the tangent bundle since the two internal spinors $\eta^{(1)}$ and $\eta^{(2)}$ might not be everywhere independent.

In \cite{effective} it was realized that in type IIB strict SU(3) compactifications to AdS$_4$ are
impossible\footnote{That is at a pure classical level. Taking non-perturbative corrections into account the authors of \cite{kklt} indeed constructed an AdS$_4$ vacuum with SU(3)-structure. See also \cite{effective} for a discussion.}. We will review the argument in section \ref{structureSU2}, while we will also show that,
conversely, in type IIA static SU(2) compactifications are impossible (which was previously noted in \cite{blt}). But in fact we will show more:  intermediate SU(2)-structure AdS$_4$ vacua with {\em left-invariant} pure spinors are impossible in {\em both} type IIA and type IIB. The way out of this no-go theorem is that in type IIA we must allow $e^{2A-\Phi} \eta^{(2)\dagger}_+ \eta^{(1)}_+$ to vary along the internal manifold, while in type IIB we need a genuine dynamic SU(3)$\times$SU(3)-structure that changes type to static SU(2) on a non-zero locus. So the most interesting but also the most complicated case, the dynamic SU(3)$\times$SU(3)-structure is still possible, but we will leave this to further work. Note that in \cite{kt,andriot} examples of constant intermediate SU(2)-structure on {\em Minkowski}
compactifications were provided.

In this paper we will mostly focus on the effective theory around strict SU(3) vacua in type IIA
and also give one example of a T-dual static SU(2) vacuum in type IIB.
So let us discuss the conventions for these cases in some more detail in the next subsections.

\subsection{SU(3)-structure}\label{structureSU3}

A real non-degenerate two-form $J$ and a complex decomposable
three-form $\Omega$ completely specify an SU(3)-structure on the six-dimensional
 manifold $\mathcal{M}$ iff:
\begin{subequations}\label{OmegaJ}
\begin{align}
\Omega\wedge J&=0 \, , \\
\Omega\wedge\Omega^*&=\frac{4i}{3}J^3\neq 0 ~,
\label{a1}
\end{align}
\end{subequations}
and the associated metric \eqref{su3metric} is positive definite.
Up to a choice of orientation, the volume normalization can be taken such that
\begin{align}
\frac{1}{6}J^3=-\frac{i}{8}\Omega\wedge\Omega^*=\mathrm{vol}_6~.
\label{voln}
\end{align}

When the internal supersymmetry generators of \eqref{spinansatz} are proportional,
\eq{\label{su3ansatz}
\eta^{(2)}_+ = (b/a) \eta^{(1)}_+ \, ,
}
with $|\eta^{(1)}|^2 = |a|^2, |\eta^{(2)}|^2=|b|^2$,
they define an SU(3)-structure as follows. First let us define a normalized
spinor $\eta_+$ such that $\eta^{(1)}_+= a \eta_+$ and $\eta^{(2)}_+ = b \eta_+$
and moreover we choose the phase of $\eta$ such that $a=b^*$. Note that in compactifications
to AdS$_4$ the supersymmetry imposes $|a|^2=|b|^2$ such that $b/a=e^{i\theta}$ is just a phase.
Now we can construct
$J$ and $\Omega$ as follows
\eq{
\label{defJOm}
J_{mn}= i \eta^{\dagger}_+ \gamma_{mn} \eta_+ \, , \qquad \Omega_{mnp}= \eta^{\dagger}_- \gamma_{mnp} \eta_+ \, .
}

The intrinsic torsion of  $\mathcal{M}$ decomposes into five modules (torsion classes)
${\cal W}_1,\dots,{\cal W}_5$. These also appear in the
SU(3) decomposition of the exterior derivative of $J$, $\Omega$. Intuitively,
 this is because the intrinsic torsion parameterizes the
failure of the manifold to be of special holonomy, which can also
be thought of as the deviation from closure of $J$, $\Omega$.
More specifically we have:
\eq{\spl{
\d J&=\frac{3}{2}\Im(\mathcal{W}_1\Omega^*)+\mathcal{W}_4\wedge J+\mathcal{W}_3 \, , \\
\d \Omega&= \mathcal{W}_1 J\wedge J+\mathcal{W}_2 \wedge J+\mathcal{W}_5^*\wedge \Omega ~,
\label{torsionclassesv}
}}
where $\mathcal{W}_1$ is a scalar, $\mathcal{W}_2$ is a primitive (1,1)-form, $\mathcal{W}_3$ is a real
primitive $(1,2)+(2,1)$-form, $\mathcal{W}_4$ is a real one-form and $\mathcal{W}_5$ a complex (1,0)-form.
For the vacua of interest to us only the classes $\mathcal{W}_1$, $\mathcal{W}_2$ are non-vanishing and they are
purely imaginary, which we will indicate with a minus superscript. Indeed, we can readily see that eq.~(\ref{torsionclasses}) follows from eq.~(\ref{torsionclassesv})
above, upon setting $\mathcal{W}_{3,4,5}$ to zero and imposing $\mathcal{W}_{1,2}=\mathcal{W}^-_{1,2}=i\Im\mathcal{W}^-_{1,2}$.

\begin{sloppypar}
Note that by definition $\mathcal{W}_2$ is primitive, which
means
\eq{
\label{primitive}
\mathcal{W}_2 \wedge J \wedge J = 0 \, .
}
One interesting property of a primitive (1,1)-form is
\eq{
\star \left(\mathcal{W}_2 \wedge J\right) = - \mathcal{W}_2 \, ,
\label{primprop}
}
which can be shown using $J^{mn}\mathcal{W}_{2mn}=0$ (which follows from
the primitivity) and $J_m{}^nJ_p{}^q\mathcal{W}_{nq}=\mathcal{W}_{mp}$
(which follows from the fact that $\mathcal{W}_{2}$ is of type (1,1)).
\end{sloppypar}

Let us now calculate the part of $\d\mathcal{W}_2^-$ proportional
to $\Re\Omega$:
\begin{align}
\d \mathcal{W}_2^-=\alpha\,\Re\Omega + (2,1)+(1,2)~,
\label{l1}
\end{align}
for some $\alpha$. Taking the exterior
derivative of $\Omega\wedge \mathcal{W}_2^-=0$ and using (\ref{l1}) as well as
the eqs.~(\ref{a1}), (\ref{torsionclassesv}),
we arrive at:
\begin{align}
\mathcal{W}_2^-\wedge \mathcal{W}_2^-\wedge J=\frac{2i}{3}\alpha J^3~.
\end{align}
We can now use \eqref{primprop} to show
\eq{
\mathcal{W}_2^-\wedge \mathcal{W}_2^-\wedge J = 2 |\mathcal{W}_2^-|^2 \text{vol}_6 \, ,
}
from which we obtain $\alpha=-i|\mathcal{W}_2|^2/8$, in accordance with (\ref{c2expr}).

{} From the SU(3)-structure \eqref{a1}, we can read off the metric as follows \cite{hitchinold}.
{} From $\Re \Omega$ alone we can construct an almost complex structure. First we define
\eq{
\tilde{\mathcal{I}}^l{}_k = -\varepsilon^{lm_1\dots m_5} (\Re \Omega)_{km_1m_2} (\Re \Omega)_{m_3m_4m_5} \, ,
}
where $\varepsilon^{m_1\dots m_6}=\pm1$ is the totally antisymmetric symbol in six dimensions, and
then properly normalize it
\eq{
\mathcal{I} = \frac{\tilde{\mathcal{I}}}{\sqrt{-\text{tr}\, \frac{1}{6}\,\tilde{\mathcal{I}}^2}} \, ,
}
so that $\mathcal{I}^2 = -\bbone$. Note that
\eq{
\label{hitchinfunc}
H(\Re \Omega)=\text{tr} \, \frac{1}{6} \, \tilde{\mathcal{I}}^2
}
is called the Hitchin functional. The metric can then be constructed from $\mathcal{I}$ and $J$ via:
\eq{
\label{su3metric}
g_{mn}=\mathcal{I}_m{}^lJ_{ln}~.
}

Finally, let us mention some useful formul\ae{} for $J$ and $\Omega$ as defined in \eqref{defJOm}
\eq{\spl{
\gamma_m\eta_-&=-iJ_m{}^n\gamma_n\eta_- \, , \\
\gamma_{mn}\eta_+&=-iJ_{mn}\eta_++\frac{1}{2}\Omega_{mnp}\gamma^p\eta_- \, , \\
\gamma_{mnp}\eta_-&=3iJ_{[mn} \gamma_{p]}\eta_--\Omega^*_{mnp}\eta_+~.
}}

\subsection{SU(3)$\times$SU(3)-structure and static SU(2)-structure }\label{structureSU2}

It turns out that in order to study static SU(2)-structure, it is most convenient to use
the generalized geometry formalism. The supersymmetry generators $\eta^{(1)}$ and $\eta^{(2)}$ from
\eqref{spinansatz} are then collected into two spinor bilinears, which using the
Clifford map, can be associated with two polyforms of definite degree
\eq{\label{polys}
\slashchar{\Psi}_+ = \frac{8}{|a||b|}\eta^{(1)}_+ \otimes \eta^{(2)\dagger}_+ \, , \qquad
\slashchar{\Psi}_- = \frac{8}{|a||b|}\eta^{(1)}_+ \otimes \eta^{(2)\dagger}_- \, .
}
It can be shown that these are associated to pure spinors of $SO(6,6)$ and that they satisfy the normalization
\eq{
\label{norm}
\langle \Psi_+, \Psi_+^* \rangle = \langle \Psi_-, \Psi_-^* \rangle \neq 0 \, ,
}
with the Mukai pairing $\langle \cdot, \cdot \rangle$ given by
\eq{
\label{mukai}
\langle \phi_1, \phi_2 \rangle = \phi_1 \wedge \alpha(\phi_2)|_{\text{top}} \, .
}
The operator $\alpha$ acts by inverting the order of indices on forms.
The Mukai pairing has the following useful property:
\eq{
\label{mukaiprop}
\langle e^{b} \phi_1, e^{b} \phi_2 \rangle = \langle \phi_1, \phi_2 \rangle \, ,
}
for an arbitrary two-form $b$.
Since there are two compatible invariant pure spinors
the structure of the generalized tangent bundle is reduced to $SU(3) \times SU(3)$.
In order to obtain similar equations in IIA and IIB one redefines
\eq{\label{A/B}
\Psi_1 = \Psi_{\mp} \, , \qquad \Psi_2 = \Psi_{\pm} \, ,
}
with upper/lower sign for IIA/IIB. We collect all the RR-fields of the democratic formalism into one polyform and make the following compactification ansatz
\eq{
\label{Fansatz}
F = \hat{F} + \text{vol}_4 \wedge \tilde{F} \, ,
}
with $\text{vol}_4$ the four-dimensional (AdS$_4$) volume form. In fact, in this paper we will drop the
hat and hope that it is clear from the context whether we mean the full $F$ or only the internal part.

With these definitions the supersymmetry conditions (in string frame) take the following concise form in both
IIB and IIA \cite{granaN1}
\begin{subequations}
\label{gensusy}
\begin{align}
\label{gensusy1}
& \d_H \left( e^{4A-\Phi} \Im \Psi_1 \right) = 3 e^{3A-\Phi} \Im (W^* \Psi_2) + e^{4A} \tilde{F} \, , \\
\label{gensusy2}
& \d_H \left[ e^{3A-\Phi} \Re (W^* \Psi_2) \right] = 2 |W|^2 e^{2A-\Phi} \Re \Psi_1 \, , \\
& \d_H \left[ e^{3A-\Phi} \Im (W^* \Psi_2) \right] = 0 \, ,
\end{align}
\end{subequations}
where we used $|a|^2=|b|^2 \propto e^A$.
{} From the above, the equations of motion for $F$ follow as integrability conditions,
as well as the following equation:
\eq{
\d_H \left( e^{2A-\Phi} \Re \Psi_1 \right) = 0 \, .
}
Here $W$ is defined in terms of the AdS Killing spinors
\eq{
\nabla_\mu \zeta_- = \pm \frac{1}{2} W \gamma_\mu \zeta_+ \, ,
}
for IIA/IIB.

These equations should be supplemented with the Bianchi identities for the
RR-fluxes \eqref{Bianchis} where the (localized or smeared) sources $j$ have to be calibrated
\begin{subequations}
\begin{align}
\label{calcond1}
& \langle \Re \Psi_1, j \rangle = 0 \, , \\
\label{calcond2}
& \langle \Psi_2, \mathbb{X} \cdot j \rangle = 0 \, , \quad \forall \mathbb{X} \in \Gamma(T_M \oplus T^\star_M) \, .
\end{align}
\end{subequations}
Analogously to the SU(3)-case, an easy way to solve these calibration conditions is to choose
\eq{\label{osource}
j = -k \, \Re \Psi_1 \, ,
}
for some function $k$, which is positive for net D-brane charge and negative for net orientifold charge.
{} Applying an exterior derivative on \eqref{gensusy1}, taking (\ref{gensusy2}), (\ref{Bianchis}), (\ref{Fansatz}) into account,
it can be shown that
\eq{
\pm \d_H \left\{ \alpha\left[ \star \d_H\left(e^{3A-\Phi} \Im \Psi_1\right)\right]\right\}= - e^{4A} j - 6|W|^2 e^{A-\Phi} \Re \Psi_1 \, ,
}
for IIA/IIB.

With the SU(3)-structure ansatz (\ref{su3ansatz}) we get from (\ref{polys})
\eq{
\label{SU3pure}
\Psi_- = - \Omega \, , \qquad \Psi_+ = e^{-i\theta} e^{iJ}  \, ,
}
where $J$ and $\Omega$ are defined in \eqref{defJOm}. For IIA we arrive at (\ref{ltsol}) and (\ref{ltsolc}) after plugging (\ref{A/B}) into (\ref{gensusy}).
For IIB on the other hand, where
the above definitions of $\Psi_1$ and $\Psi_2$ are switched, it is immediately obvious from \eqref{gensusy2}
that there is no SU(3)-structure solution possible. Indeed, the left-hand side is a four-form, which would put
the zero- and two-form part of the right-hand side to zero, making \eqref{norm} impossible to be satisfied, unless $W=0$ --
implying the vanishing of the AdS$_4$ curvature. We can go even further and show that in fact no intermediate structure
is possible for type IIB unless it is really a dynamic SU(3)$\times$SU(3)-structure that changes type
to static SU(2) somewhere in the manifold. Indeed, for intermediate structure $\Psi_1$ is a pure spinor
of type 0, which means that the lowest form in $\Psi_1$ is a zero-form.
According to Gualtieri \cite{gualtieri} the generic form of such a pure spinor is
\eq{
e^{2A-\Phi} \Psi_1 = c e^{i \omega + b} \, ,
}
where $\omega$ and $b$ are arbitrary real two-forms. From \eqref{gensusy2} we find, unless $W=0$,
\eq{
\Re c = 0 \, , \qquad c\,  \omega  \quad \text{exact} \, .
}
It follows that
\eq{
\Im c \, \langle e^{2A-\Phi} \Psi_1 , e^{2A-\Phi} \Psi_1^* \rangle = \frac{8}{3!} (c \, \omega)^3
}
is exact. At the same time it is proportional to the volume form, which is non-exact.
The only way to satisfy \eqref{norm} is to have $\Im c=0$ at least somewhere, which means that the type
changes on that locus to static SU(2).

For IIB we are interested in the static SU(2)-structure case for which
\eq{
\eta^{(2)}_+ = V^i \gamma_i \eta^{(1)}_- \, .
}
It is convenient to define the following SU(2)-structure quantities
\eq{
\omega_2 = J + 2i V \wedge V^* \, , \qquad \Omega_2 = \iota_{V^*} \Omega \, .
}
where $J$ and $\Omega$ form the SU(3)-structure associated to $\eta_+=|a|^{-1} \eta^{(1)}_+$ as in \eqref{defJOm}.
We then find for the pure spinors
\begin{subequations}
\label{SU2pure}
\begin{align}
\Psi_+ & = - e^{2 V \wedge V^*} \Omega_2 \, , \\
\Psi_- & = - 2 \, V \wedge e^{i \omega_2} \, .
\end{align}
\end{subequations}
Plugging this ansatz in \eqref{gensusy}, one finds equations for the SU(2)-structure
quantities $V$, $\omega_2$ and $\Omega_2$, but it should be less complicated to try to solve these equations
directly in terms of the pure spinors.

In the same vain as above it follows from \eqref{gensusy2} that IIA compactifications to AdS$_4$ are
incompatible with static SU(2)-structure, as already noted in \cite{blt}. Indeed, this
equation would, unless $W=0$, force $\Re \Psi_1|_1=0$, making it impossible to satisfy \eqref{norm}.
We can extend the argument to intermediate structure with left-invariant pure spinors. In this
case $\Psi_2=\Psi_+$ starts with a zero-form instead of a two-form. However, because of the assumption of left-invariance
the zero-form on the left-hand side of \eqref{gensusy2} is constant and thus zero upon acting with the exterior derivative.
Again we find then $\Re \Psi_1|_1=0$ and \eqref{norm} is not satisfied. The conclusion is that we can only have intermediate SU(3)$\times$SU(3)-structure if $\d \left(e^{3A-\Phi} \Psi_+|_0\right) \neq 0$.

\subsubsection*{Nilmanifold 5.1}

Let us now apply the above formalism to the solution of
section \ref{nilIIBsol}. We can compute the pure spinors using \eqref{SU2pure}.
One can also check that
\eq{
- \d_H \left\{ \alpha\left[ \star \d_H\left(\Im \Psi_+\right)\right]\right\}=-4 \, \beta^2 \Re \Psi_+ \, ,
}
which leads to an orientifold source as in \eqref{osource} with
\eq{
k= - \frac{5}{2} \beta^2  \, .
}

\section{How to dress smeared sources with orientifold involutions}\label{smearedori}

Suppose we are given a form $j$ representing the Poincar\'e dual of smeared orientifolds. How do we decide what the
orientifold involutions should be? Let us first give an example for a {\em localized} orientifold in flat space. If we
have an orientifold along the directions $\Sigma=(x^1,x^2,x^3)$ then the corresponding source is
\eq{
j = T_{\text{O}p}\, j_{\Sigma} = -T_{\text{O}p} \, \delta(x^1,x^2,x^3) \, \d x^4 \wedge \d x^5 \wedge \d x^6 \, ,
}
where $T_{\text{O}p}\,<0$ for an orientifold and $j$ is the Poincar\'e dual of $\Sigma$ satisfying
\eq{
\label{jdual}
\int_\Sigma \phi = \int_{\mathcal{M}} \langle \phi,j_{\Sigma} \rangle = - \int_{\mathcal{M}} \phi \wedge j_{\Sigma} \, ,
}
for an arbitrary form $\phi$ \footnote{The definition with the Mukai pairing is the one appropriate for generalizing
to D-branes with world-volume gauge flux as explained in \cite{deforms}. Here it will just give an extra minus sign}.
In this case the orientifold involution is of course
\eq{
O6: \qquad x^4 \rightarrow -x^4 \, , \quad x^5 \rightarrow -x^5 \, , \quad x^6 \rightarrow -x^6 \, .
}
Suppose we now introduce many orientifolds and completely smear them in the directions $(x^1,x^2,x^3)$ obtaining
\eq{
j = -T_{\text{Op}}\, c \, \d x^4 \wedge \d x^5 \wedge \d x^6 \, ,
}
where $c$ is a constant representing the orientifold density. We have now lost information about the exact location
but we would still like to associate the orientifold involution
\eq{
O6: \qquad \d x^4 \rightarrow -\d x^4 \, , \quad \d x^5 \rightarrow -\d x^5 \, , \quad \d x^6 \rightarrow -\d x^6 \, .
}

An important observation is that $\d x^4 \wedge \d x^5 \wedge \d x^6$ is not just any form, it is a {\em decomposable} form, i.e.\
it can be written as a wedge product of three one-forms. These one-forms span the annihilator space of $T_\Sigma$, the tangent
space of $\Sigma$. So if we are given a smeared orientifold current $j$ we should write it as a sum of decomposable
forms and then associate to each term an orientifold involution as above.

Let us now study more formally how we could write $j$ as a sum of decomposable forms and whether the decomposition is
unique. First, let us introduce a basis of forms $e^i \in \mathbb{V}^\star$ that span (locally) $T_{\mathcal{M}}$. Indeed, for
the case of group manifolds we have such a basis, which is even defined globally. For the cosets left-invariant forms in
this basis are also globally defined.

Now, let $\mathbb{V}$ be a $d$-dimensional vector space and
 $\mathbb{V}^{\star}$  its dual. A (real/complex) $p$-form $j\in \Lambda^p\mathbb{V}^{\star}$ is called {\it simple}
or {\it decomposable} if it can be written as a wedge product of $p$ one-forms.\footnote{Note that
a (real/complex) form of fixed dimension is a pure spinor if and only if it is simple. In fact, we could
regard the notion of pure spinor as a generalization of the notion of decomposable forms to polyforms.} What we are
interested in is that there is a one-to-one correspondence between $(d-p)$-planes (our orientifold planes) and
decomposable $p$-forms (up to a proportionality factor). This isomorphism is called the {\em Pl\"ucker map}. A discussion of the criteria
for having a simple form can be found in e.g.~\cite{gh} pp. 209-211. We will use here the criterion based on
\eq{
j^\perp=\{X \in \mathbb{V} : \iota_X j = 0 \} \subset \mathbb{V} \, ,
}
and
\eq{
W = \text{Ann}(j^\perp) \subset \mathbb{V}^\star \, .
}
In \cite{gh} it is shown that $j$ is simple if and only if $\text{dim} \, W = p$. Using this the following alternative criterion
is shown:

{\it Theorem}:  A $p$-form
$j\in \Lambda^p\mathbb{V}^{\star}$ is simple if and only if for every $(p-1)$-polyvector $\xi\in \Lambda^{p-1}\mathbb{V}$,
\begin{equation}\label{pluecker}
\iota_{\xi}j\wedge j=0~,
\end{equation}
where $\iota_{\xi}j$ is the one-form contraction of $j$ with $\xi$.

Now for the special case of three-forms in six dimensions there is
another useful theorem due to Hitchin \cite{hitchinold}.

{\it Theorem}:  Consider a real three-form $j \in \Lambda^3 \mathbb{V}^{\star}$ and calculate its
Hitchin functional $H(j)$ defined in \eqref{hitchinfunc}. Then
\begin{itemize}
\item $H(j)>0$ if and only if $j=j_1 + j_2$ where $j_1,j_2$ are unique (up to ordering) real decomposable three-forms and $j_1 \wedge j_2 \neq 0$;
\item $H(j)<0$ if and only if $j=\alpha + \bar{\alpha}$ where $\alpha$ is a unique (up to complex conjugation) complex decomposable three-form and $\alpha \wedge \bar{\alpha} \neq 0$.
\end{itemize}

Now we have two base-independent characterizations of $j$: the Hitchin functional $H(j)$ and $\text{dim} \, W$.
To get a feeling of the relation between both we prove the following

{\it Theorem}: If $H(j) \neq 0$ then $\text{dim} \, W=6$ (but not the other way round!).

Indeed, let us first consider $H(j)>0$. We use
the above decomposition and try to find $X$ such that $\iota_X (j_1 + j_2)=0$. From this relation follows that
$\iota_X (j_1 \wedge j_2)=\iota_X j_1 \wedge j_2 - j_1 \wedge \iota_X j_2=-\iota_X j_2 \wedge j_2+j_1 \wedge \iota_X j_1$,
which is zero because $j_1$ and $j_2$ are simple and \eqref{pluecker}. On the other hand it must be non-zero since $j_1 \wedge j_2 \propto \text{vol}_6 \neq 0$.
It follows that there is no such solution for $X$ and thus $j^\perp$ is empty. Then $\text{dim} \, W = 6$. Analogously for $H(j)<0$, but we use now the
above theorem for {\em complex} simple forms.

Using these two characterizations we can classify the possible $j$ and decompose it in simple terms:

\begin{itemize}
\item if $H(j)>0$ it follows immediately that $j$ is a sum of exactly two real simple terms;
\item if $H(j)<0$ then $j$ is a sum of exactly two (conjugate) complex simple terms and thus of exactly four real simple terms. This will in fact be almost always the case for the orientifold sources in this paper.
\item if $H(j)=0$ we have three cases. Either \eqref{pluecker} is satisfied (equivalently $\text{dim} \, W=3$) and $j$ is simple,
either  $\text{dim} \, W = 5$ and then $j$ will be a sum of two simple terms $j_1$ and $j_2$ such that $j_1 \wedge j_2=0$, or
$\text{dim} \, W=6$ and $j$ will be
a sum of three simple terms. All this is easy to prove by looking at possible types of sums of two and three simple terms.
\end{itemize}

An important remark is in order: while the Hitchin theorem states that for $H(j) \neq 0$ the two real/complex forms in the
decomposition of $j$ are unique (up to ordering/complex conjugation), the choice of one-forms out of which these
forms are made is {\em not} unique. In the case of $H(j)<0$ it is the freedom of choosing a basis of complex one-forms
belonging to a complex structure, which is SL(3,$\mathbb{C}$). As a consequence the choice of the four real forms in which $j$ is decomposed is
{\em not} unique. Indeed, suppose we choose one basis of complex one-forms and associated $x$ and $y$ coordinates:
$e^{z^i}= e^{x^i} - i e^{y^i}$. Then $j$ can be written as the sum of the following four terms:
\eq{
j = \Re (e^{z^1z^2z^3})=e^{x^1x^2x^3} - e^{x^1y^2y^3} - e^{y^1x^2y^3} - e^{y^1y^2x^3} \, ,
}
which leads to the following orientifold involutions:
\eq{\label{standinv}\spl{
O6: & \qquad e^{x^1} \rightarrow -e^{x^1} \, , \quad e^{x^2} \rightarrow -e^{x^2} \, , \quad e^{x^3} \rightarrow -e^{x^3} \, , \\
O6: & \qquad e^{x^1} \rightarrow -e^{x^1} \, , \quad e^{y^2} \rightarrow -e^{y^2} \, , \quad e^{y^3} \rightarrow -e^{y^3} \, , \\
O6: & \qquad e^{y^1} \rightarrow -e^{y^1} \, , \quad e^{x^2} \rightarrow -e^{x^2} \, , \quad e^{y^3} \rightarrow -e^{y^3} \, , \\
O6: & \qquad e^{y^1} \rightarrow -e^{y^1} \, , \quad e^{y^2} \rightarrow -e^{y^2} \, , \quad e^{x^3} \rightarrow -e^{x^3} \, .
}}
If we perform a SL(3,$\mathbb{C}$) transformation, $j$ takes exactly the same form, but now in the {\em new} basis.
So alternatively we could have chosen four orientifold involutions taking the same form as the old ones,
but now in the {\em new} basis, which is rotated. This means that our choice of orientifold involutions is not
unique. We must then further choose them such that the structure constant tensor of the group or coset is even, and $\Re \Omega$ and $J$ are odd.

In the case of $H(j)>0$ the argument does not apply because the remaining freedom GL(3,$\mathbb{R}$)$\times$GL(3,$\mathbb{R}$)
leaves the two terms of the decomposition {\em separately} invariant and the choice of orientifold involutions is unique.

\subsection*{Application to SU(2)$\times$SU(2)}

Let us now apply the above procedure to the model of section \eqref{SU2SU2}. Calculating the Hitchin
functional $H(j^6)$ of \eqref{SU2SU2source} we find that it is negative so that it contains four orientifold involutions.
We must now fix the freedom of choosing them such that $\Re \Omega$ and $J$ are odd, and the structure constant tensor
$f$ is even. Some reflection should make clear that if $\Re \Omega$ is to be odd it should be a sum of the
same four terms as $j^6$, but with different coefficients. In fact, we could reverse the procedure and choose a complex
basis $e^{z^i}$ in which $\Omega$ and $J$ take their standard form:
\eq{
\Omega = e^{z^1z^2z^3} \, , \qquad J = - \frac{i}{2} \sum_i e^{z^i\bar{z}^{i}} \, .
}
Then $\Re \Omega$ and $J$ are automatically odd under the associated orientifold involutions \eqref{standinv}.
However, this should of course also be the orientifold involutions that follow from $j^6$. This will
be the case if and only if $j^6$ has the same terms as $\Re \Omega$ (but with different coefficients) or
equivalently $j^6$ should take the form
\eq{
\label{wantedj}
j^6 = \Re \left( c^0 e^{z^1z^2z^3} + c^{11} e^{\bar{z}^1z^2z^3} + c^{22} e^{z^1\bar{z}^2z^3} + c^{33} e^{z^1z^2\bar{z}^3}\right) \, ,
}
with all coefficients $c$ real. To accomplish this we still have the freedom to make a base transformation such that
$\Omega$ and $J$ invariant, i.e.~an SU(3)-transformation. A priori, $j^6$ is an arbitrary three-form which transforms
under SU(3) as
\eq{
20 = 1 + \bar{1} + 3 + \bar{3} + 6 + \bar{6} \, .
}
However, we know that $j^6$ has to satisfy the calibration conditions \eqref{calcond},
which remove the $3+\bar{3}$ representation and only leave the
form proportional to $\Re \Omega$ out of $1+\bar{1}$. Here the $6$ is the $(3\times 3)_S$ i.e.~the symmetric product
of two fundamental representations of SU(3). It follows that the most general $j^6$ satisfying the calibration conditions
looks like
\eq{\spl{
j^6 & = c_0 \Re \Omega + \Re \left[c^{ki} g_{(k|\bar{\jmath}}\d\bar{z}^{\bar{\jmath}} \wedge \iota_{z^{i)}} \Omega\right] \\
& = c_0 \Re \Omega + \Re \Big[ c^{11} e^{\bar{z}^1 z^2 z^3}+ c^{22} e^{z^1 \bar{z}^2 z^3} + c^{33} e^{z^1 z^2 \bar{z}^3} \\
& + c^{12} \left(e^{\bar{z}^2 z^2 z^3}+e^{z^1 \bar{z}^1 z^3}\right)
+ c^{13} \left(e^{\bar{z}^3 z^2 z^3}+e^{z^1 z^2 \bar{z}^1}\right)
+ c^{23} \left(e^{z^1 \bar{z}^3 z^3}+e^{z^1 z^2 \bar{z}^2}\right)\Big] \, ,
}}
with $c_0$ real and the entries of the coefficient matrix
\eq{
C = \left( \begin{array}{ccc} c^{11} & c^{12} & c^{13} \\ c^{21} & c^{22} & c^{23} \\ c^{31} & c^{32} & c^{33} \end{array} \right) \, ,
}
complex. Now we have to find an SU(3)-transformation to put $j^6$ in the form \eqref{wantedj}. $c_0$ does not transform
but is luckily already of the right form, while the coefficient matrix transforms as
\eq{
C \rightarrow U C U^T \, .
}
{}From \eqref{wantedj} we see that we want to transform $C$ to a diagonal real matrix. In fact, since the above transformation
cannot change the determinant this is only possible if
\eq{
\text{det} \, C \in \mathbb{R} \, .
}
This is a condition we have to add to the calibration conditions. For the $j^6$ of \eqref{SU2SU2source} one can
check that it is indeed satisfied and it is possible to find the
complex coordinates with the required properties. Also, under the associated orientifold involution the structure constant
tensor $f$ is even as required. Note that alternatively, as we actually did in \eqref{complbasis},
we can also construct a complex basis associated to $\Omega$ such that $f$ is even.
This then automatically implies that $j$ is odd and that it is a sum of the same four terms as $\Re \Omega$.

\section{Kaluza-Klein reduction: calculational details}\label{appkk}

We assemble here the details on the calculation of the mass spectrum through Kaluza-Klein reduction.

\subsection{Solving the Bianchi identities}

Here we will obtain an expression of the fluctuations of the gauge flux in terms of the fluctuations of potentials
ensuring that the Bianchi identities are automatically satisfied. The analysis is complicated by the presence of a source.

We assume that the source does not fluctuate since it is associated to smeared orientifolds.
For the Bianchi identities of the background and the fluctuation we find then respectively
\subeq{\al{
\label{biback}
(\d + \bg{H}) \bg{F} & = - j \, , \\
\label{bifluc}
(\d + \bg{H}+\delta H) (\bg{F}+ \delta F) & = - j \, .
}}
The integrability equations read
\subeq{\al{
(\d + \bg{H}) j & = 0 \, , \\
(\d + \bg{H}+\delta H) j & = 0 \, ,
}}
from which follows
\eq{
\delta H \wedge j = 0 \,  .
}
This implies also
\eq{
(\d + \bg{H}) (e^{\delta B} \wedge j) = 0 \, ,
}
so that, subtracting \eqref{biback}, we can define (locally)
\eq{
- (e^{\delta B}-1) \wedge j = (\d+\bg{H})\delta \omega \, .\label{deltaOmega}
}
Now, for orientifold sources the left hand side of this equation always vanishes. This follows because
the pull-back of $\delta B$ to the orientifold, $\delta B|_\Sigma$, must be zero, which implies using
\eqref{jdual}:
\eq{
\delta B \wedge j = 0 \, ,
}
and the same for all powers of $\delta B$. Then, we can also choose $\delta \omega=0$.

The difference between \eqref{biback} and \eqref{bifluc} gives the Bianchi identity for the fluctuations
\eq{
\left(\d + \bg{H}+ \delta H\right) \delta F + \delta H \wedge \bg{F} = 0 \, ,
}
which can be rewritten as
\eq{
\left(\d + \bg{H}\right) \left(e^{\delta B}\delta F \right)+ \delta H \wedge e^{\delta B} \bg{F} = 0 \, .
}
One can easily show that (with $\delta F_0=0$) this Bianchi identity can be satisfied by introducing potentials $\delta C$
and putting
\eq{
e^{\delta B} \delta F = (\d + \bg{H}) \delta C - (e^{\delta B}-1) \bg{F} + \delta \omega \, ,
}
where we can set $\delta \omega=0$ so that we obtain eq.~(\ref{Fsolvedpot}).

Expanding this expression we find for the IIA-fluctuations
\eq{\spl{\label{IIAfluc}
\delta F_0 & = 0 \, , \\
\delta F_2 & = \d \delta C_1 - m \delta B \, , \\
\delta F_4 & = \d \delta C_3 + \bg{H} \wedge \delta C_1 - \delta B \wedge (\bg{F}_2 + \delta F_2) - \frac{1}{2} m (\delta B)^2 \, , \\
\delta F_6 & = \d \delta C_5 + \bg{H} \wedge \delta C_3 - \delta B \wedge (\bg{F}_4 + \delta F_4) - \frac{1}{2} (\delta B)^2 \wedge (\bg{F}_2 + \delta F_2) - \frac{1}{3!} m (\delta B)^3 \, ,
}}
and for the IIB-fluctuations
\eq{\spl{\label{IIBfluc}
\delta F_1 & = \d \delta C_0 \, , \\
\delta F_3 & = \d \delta C_2 + \bg{H} \wedge \delta C_0- \delta B \wedge (\bg{F}_1 + \delta F_1) \, , \\
\delta F_5 & = \d \delta C_4 + \bg{H} \wedge \delta C_2 - \delta B \wedge (\bg{F}_3 + \delta F_3) - \frac{1}{2} (\delta B)^2\wedge(\bg{F}_1 + \delta F_1)  \, .
}}
For the Kaluza-Klein reduction we will only need the terms linear in the fluctuations while for an analysis
of finite fluctuations using the K\"ahler potential and superpotential we need higher orders too.

Also, in the Kaluza-Klein reduction we will only need fluctuations of the physical fields
$\delta F_2, \delta F_4$ since the higher-form fluxes are removed from the equations of motion using
\eqref{Fduality}, while in the superpotential approach, which is formulated in the democratic formalism, we should work with
the internal part of $\delta F_6$ instead of the external part of $\delta F_4$. As we explain in section \ref{noteC3} we could
actually have done this in the Kaluza-Klein approach too.

\subsection{Expansion/Truncation}

For the Kaluza-Klein reduction on T$^6$ we will expand the fluctuations of the various fields in the following basis:
\subeq{\label{expall}\al{
\label{ExpansionB}
\delta B(x,y) = & b^{i,\vec{n}} (x) \mathcal{Y}^{(2)}_{i,\vec{n}} (y)+b_1^{i,\vec{n}} (x) \mathcal{Y}^{(1)}_{i,\vec{n}} (y) + b_2^{\vec{n}}(x) \mathcal{Y}^{(0)}_{\vec{n}}(y) \, , \\
\label{Expansionphi}
\delta \phi (x,y) = & \delta \phi^{\vec{n}}(x) \mathcal{Y}^{(0)}_{\vec{n}}(y) \, , \\
\label{ExpansionC1}
\delta C^{(1)}(x,y) = & c^{(1)i,\vec{n}} (x) \mathcal{Y}^{(1)}_{i,\vec{n}} (y) + c^{(1)\vec{n}}_1 (x) \mathcal{Y}^{(0)}_{\vec{n}}(y) \, , \\
\label{ExpansionC3}
\delta C^{(3)} (x,y) = & c^{(3)i,\vec{n}}(x) \mathcal{Y}^{(3)}_{i,\vec{n}} (y) + c^{(3)i,\vec{n}}_1(x) \mathcal{Y}^{(2)}_{i,\vec{n}} (y) + c^{(3)i,\vec{n}}_2 (x) \mathcal{Y}^{(1)}_{i,\vec{n}} (y) \nonumber \\ & + c^{(3)\vec{n}}_3 (x) \mathcal{Y}^{(0)}_{\vec{n}}(y) \, , \\
\label{Expansiong}
\delta g(x,y) = & h^{i,\vec{n}}(x) \mathcal{X}^{(2)}_{i,\vec{n}} (y) + h_1^{i,\vec{n}} (x) \mathcal{Y}^{(1)}_{i,\vec{n}} (y) + h_2^{\vec{n}} (x) \mathcal{Y}^{(0)}_{\vec{n}}(y) \, .
}}
The functions $\mathcal{Y}^{(l)}_{i,\vec{n}}(y)$ are the $l$-eigenforms of the Laplacian operator and are given by
\eq{
\mathcal{Y}^{(l)}_{i,\vec{n}}(y) = Y^{(l)}_i e^{i\vec{p}\cdot\vec{y}} \, , \qquad \vec{p}=\frac{\vec{n}}{R} \, , \quad \vec{n} \in\mathbb{Z}^6 \, ,
}
where the $Y^{(l)}_i$ form a basis of harmonic $l$-forms on T$^6$.
$\mathcal{X}^{(2)}$ are symmetric two-tensors
\eq{
\mathcal{X}^{(2)}_{i,\vec{n}}(y) = X^{(2)}_i e^{i\vec{p}\cdot\vec{y}} \, , \qquad \vec{p}=\frac{\vec{n}}{R} \, , \quad \vec{n} \in\mathbb{Z}^6 \, ,
}
Since we will restrict our analysis to the zero modes ($\vec{p}=0$), we only keep $\mathcal{Y}^{(l)}_{i,\vec{n}=0}(y)= Y^{(l)}_i$
and $\mathcal{X}^{(2)}_{i,\vec{n}=0}(y) = X^{(2)}_i$ in the expansions above and derivatives only act on the external fields.
For the Iwasawa manifold we will use for the expansion forms $Y_i^{(l)}$ left-invariant forms, which will not necessarily be all
harmonic. When exterior derivatives act on these forms terms will be generated of the order of the geometric fluxes.

\subsection{IIA on AdS$_4\times$T$^6$}
\label{torusKK}

We would now like to perform a Kaluza-Klein reduction on the AdS$_4\times$T$^6$ solution described
in section \ref{sixtorusex}.

A basis for the harmonic $l$-forms $Y_i^{(l)}$ is simply given
by all exterior products of the form $\d y^{m_1}\wedge\dots\wedge \d y^{m_l}=e^{m_1 \ldots m_l}$, $1\leq l\leq 6$.
Hence:
\eq{
b_l=\left( \begin{array}{c} 6\\l \end{array}\right)~,
}
where $b_l$ denotes the real dimension of the $l$th cohomology group of T$^6$.  However, we must then impose the orientifold projection, which means
that suitable expansion forms must be even or odd under {\em all} the orientifold involutions. For the torus we find from \eqref{torusinv}:
\begin{center}
  \begin{tabular}{|c|c|c|}

    \hline
    type & basis & name \\
    \hline
    \hline
    odd 2-form  &  $e^{12},~e^{34},~e^{56}$ & $Y^{(2-)}_i$\\
    \hline
    even 3-form  &  $e^{135},~e^{146},~e^{236},~e^{245}$ & $Y^{(3+)}_i$\\
    \hline
    odd 3-form  &  $e^{136},~e^{145},~e^{235},~e^{246}$  & $Y^{(3-)}_i$ \\
    \hline
    even 4-form &  $e^{1234},~e^{1256},~e^{3456}$ & $Y^{(4+)}_i$ \\
    \hline
    even symmetric 2-tensor & $e^{1} \otimes e^{1}, e^{2} \otimes e^{2},\ldots, e^{6} \otimes e^{6}$ & $X^{(2)}_i$\\
    \hline
   \end{tabular}
\end{center}

Under the orientifold projection we find from \eqref{oriall} that $\Phi,g,F_0,C_3$ are even, while $B, C_1$ are odd.
This simplifies the expansion \eqref{expall} considerably
\subeq{\label{expTrunc}
\al{
\label{ExpB}
\delta B(x,y) &= b^i (x) Y^{(2-)}_i \, , \\
\label{Expphi}
\delta \Phi (x,y) &= \Phi(x) \, , \\
\label{ExpC3}
\delta C^{(3)} (x,y) &= c^{(3)i}(x) Y^{(3+)}_i + c^{(3)}_3 (x) \, , \\
\label{Expg}
\delta g(x,y) &= h^i(x) X^{(2)}_i + h_2 (x) \, .
}}
Note in particular that the orientifold projection removes all four-dimensional gauge fields, which in fact
holds for all type IIA models for which the orientifolds project out all one-forms and even two-forms.

We find then from (\ref{IIAfluc}) the linear fluctuations of the field strengths
\subeq{\al{
\delta F_2 &=  -m \delta B \, , \\
\delta F_4 &= \d \delta C_3 \, .
}}

To derive the mass matrix for the four-dimensional fields we proceed as follows. We first compute the variation of all the equations of motion (\ref{dilaton}),(\ref{einstein}),(\ref{eomF}) and (\ref{eomH}) to first order. Remember that we should use \eqref{Fduality} to remove the redundant
RR-fields so that the only RR-fluctuations are the ones above. Next we plug in the background values and the truncated expansion of the fields (\ref{expTrunc}). Since we are only considering the zero internal modes we use that for the torus derivatives only act on the external fields.

Let us consider first the equations for the RR-fields and $H$. It will turn out that these do not mix with the dilaton and the metric. Applying the steps described above we get from the equation of motion for $H$ the following equation, which
has (external, internal) index structure $(0,2)$:
\begin{align}
\label{eqb}
0 &= \Delta (b^i Y^{(2-)}_i) - \star (\bg{F}_4 \wedge \d c^{(3)}_3) - m \star(\star\bg{F}_4\wedge b^i Y^{(2-)}_i) + m^2 b^i Y^{(2-)}_i \, .
\end{align}
{} From the variation of the equation of motion of $F_4$ we get a $(0,3)$-equation and a $(1,6)$-equation
\subeq{
\label{eqc3}
\al{
\label{eqc3_0}
0 &= \Delta (c^{(3)i} Y^{(3+)}_i) - \star(\bg{H} \wedge \d  c^{(3)}_3 ) \, , \\
\label{eqc3_3}
0 &= \d\star\d c^{(3)}_3 + \d b^i  \wedge Y^{(2-)}_i \wedge \bg{F}_4 + \bg{H} \wedge \d c^{(3)i} \wedge Y^{(3+)}_i \, ,
}}
and from $F_2$ a $(4,5)$- and $(3,6)$-equation
\subeq{\label{eqc1}
\al{
\label{eqc1a}
0 &= \bg{H} \wedge \star\left[h^i X_i^{(2)}\cdot \bg{F}_4\right] \, , \\
\label{eqc1b}
0 &= \bg{H} \wedge \star(\d c^{(3)i} \wedge Y^{(3+)}_i) \, ,
}}
where we used in the upper equation the variation of the $\star$
\begin{align}
(\delta \star) F_l &= \left(\frac{1}{2} g^{MN} \delta g_{MN}\right) \star F_l - \star [\delta g \cdot F_l]  \, ,
\end{align}
where
\begin{align}
[\delta g \cdot F_l]_{M_1 \ldots M_l} &= l \cdot \delta g_{[M_1 | A} g^{AB} F_{B | M_2 \ldots M_l ]} \, .
\end{align}
The equations (\ref{eqc1}) are automatically satisfied using the orientifold projection. Indeed, the right-hand sides should have contained an even internal five-form respectively six-form under all orientifold involutions, which do not exists, so they must vanish.

Next, we integrate (\ref{eqc3_3}) and put the integration constant to zero because it would
correspond to changing the background value of $f$. This procedure corresponds to dualizing $c_3^{(3)}$ as explained in \cite{louismicu,gl}. We
come back to it in section \ref{noteC3}.

To proceed we make a choice of expansion basis for the even three-forms
\subeq{\label{expformsthree}\al{
& Y_0^{(3+)} = \Im \Omega \, , \\
& Y_i^{(3+)} \, , \quad i=1,2,3 : \quad \text{3 real (2,1)+(1,2) forms} \, ,
}}
and the odd two-forms
\subeq{\al{\label{expformstwo}
& Y_0^{(2-)} = J \, , \\
& Y_i^{(2-)} \, , \quad i=1,2 : \quad \text{2 primitive real 2-forms} \, ,
}}
where a primitive two-form is defined in \eqref{primitive}.

In the end we obtain the following result for the eigenvalues
$\tilde{M}^2 = M^2 + 2/3 \Lambda$:
\begin{center}
  \begin{tabular}{|c|c|}

    \hline
    mass eigenmode & mass (in units $m^2/25$) \\
    \hline
    \hline
    $b^i, \quad i=1,2$  &  $10$ \\
    \hline
    $c^{i}, \quad i=1,2,3$ & $0$ \\
    \hline
    $b^0 - 4 c^{(3)0}$  &  $10$ \\
    \hline
    $3 b^0 + c^{(3)0}$  &  $88$ \\
    \hline
   \end{tabular}
\end{center}

Next we consider the dilaton and Einstein equation.
With the same procedure as above, we arrive at the following equations
\begin{align}
0 &= (\Delta + \frac{67 m^2}{25}) \delta \Phi + \frac{7 m^2}{25}  \sum_{i=1}^6 h^i  \, ,
\end{align}
and
\begin{align}
\label{intEinstein}
0 &= \Delta h^i +\frac{8 m^2}{25} h^i + \frac{7 m^2}{50} g_{ii} \delta \Phi +\frac{m^2}{50} g_{ii} \sum_{j=1}^6 h^j + \frac{2 m^2}{5} g_{ii} h^{i-(-1)^i}  \, .
\end{align}
The result of diagonalizing the mass matrix is
\begin{center}
  \begin{tabular}{|c|c|}

    \hline
    mass eigenmode & mass (in units $m^2/25$) \\
    \hline
    \hline
    $-h_{z^1\bar{z}^1}+h_{z^2\bar{z}^2}=-h^1-h^2+h^3+h^4$  &  $18$ \\
    \hline
    $-h_{z^1\bar{z}^1}+h_{z^3\bar{z}^3}=-h^1-h^2+h^5+h^6$  &  $18$ \\
    \hline
    $-3 \, \delta \Phi + 7\sum h_i$  &  $18$ \\
    \hline
    $7 \, \delta \Phi + \sum h_i$ &  $70$ \\
    \hline
    $\Re h_{z^1z^1}=-h^1+h^2$ & $-2$ \\
    \hline
    $\Re h_{z^2z^2}=-h^3+h^4$ & $-2$ \\
    \hline
    $\Re h_{z^3z^3}=-h^5+h^6$ & $-2$ \\
    \hline
   \end{tabular}
\end{center}

The external part of the Einstein equation on the other hand becomes
\eq{
\label{extEinstein}
\frac{1}{2} \Delta_L h_{\mu\nu} + \nabla_{(\mu}\nabla^{\rho}h_{\nu)\rho}
-\frac{1}{2}\nabla_{(\mu}\nabla_{\nu)}h^{P}{}_{P} + \frac{3}{25} m^2 h_{\mu\nu} - \frac{3}{20} m^2 g_{\mu\nu} \sum h_i - \frac{21}{100} m^2 g_{\mu\nu} \delta \Phi=0 \, .
}
At this point we have to take into account that so far we worked in the {\em ten}-dimensional Einstein frame. From \eqref{convg} we find
that the conversion to the {\em four}-dimensional Einstein frame is as follows
\eq{
g_{\E\mu\nu} = c \, \sqrt{g_6} \, g_{\mu\nu} \, ,
}
where the constant factor $c=M_P^{-2} \kappa_{10}^{-2} \vols$ does not matter here, so that
\eq{
c^{-1} h_{\E\mu\nu} = \sqrt{g_6} \, h_{\mu\nu} + \frac{1}{2} \sqrt{g_6} \, g_{\mu\nu} \sum_i h_i  \, .
}
Plugging this into \eqref{extEinstein} and using \eqref{intEinstein} we find for $h_{\E\mu\nu}$ exactly
equation \eqref{metric} with $M^2=0$ so that $h_{\E\mu\nu}$ indeed describes a {\em massless} graviton.

\subsection{IIA on the Iwasawa manifold}
\label{iwasawaKK}

For the solution on the Iwasawa manifold of section \ref{iwasawaex} with $m=0$ we get the
same even and odd forms under the orientifold involution as on the torus (but now in the left-invariant
basis appropriate for the Iwasawa). Again $\Phi,g,F_0,C_3$ are even, while $B, C_1$ are odd,
resulting in the same expansion \eqref{expTrunc} as for the torus.
This time we get from (\ref{IIAfluc}) for the linear fluctuations of the field strengths
\subeq{\al{
\delta F_2 &= 0 \, , \\
\delta F_4 &= \d \delta C_3 - \delta B \wedge \bg{F}_2 \, .
}}

Expanding the equation of motion for $H$ around the Iwasawa solution, we obtain
\eq{\spl{
0 = & \Delta b^i \, Y^{(2-)}_i + b^i \left( \star_6 \d \star_6 \d Y^{(2-)}_i \right)  - c^{(3)i} \star_6 (\star_6 \d Y_i^{3+} \wedge \bg{F}_2) \\
& + b^i \star_6 \left[ \star_6 \left( Y_i^{(2-)} \wedge \bg{F}_2 \right) \wedge \bg{F}_2 \right]
+ f c^{(3)i} \star_6 \d Y_i^{3+} - b^i f \star_6 \left( Y_i^{(2-)}\wedge \bg{F}_2\right)  \, ,
}}
while the equation of motion for $F_4$ splits in $(1,6)$ and $(4,3)$ index structures
\subeq{\al{
\label{iwc3}
0 & = \d \star_4 \d c_3^{(3)}  + \frac{1}{2} f \d \left( \delta g^{\mu}{}_{\mu} - \delta g^{m}{}_m - \delta \Phi\right) \, , \\
0 & = \Delta c^{(3)i} \, Y^{(3+)}_i + c^{(3)i} \left( \star_6 \d \star_6 \d Y^{(3+)}_i \right) + f b^i \star_6 \d Y_i^{(2-)}
- b^i \star_6 \d \star_6 \left( Y_i^{(2-)} \wedge \bg{F}_2\right) \, .
}}
In a similar way as in the torus case, we integrate \eqref{iwc3}, put the integration constant to zero and plug the
result for $\d c_3^{(3)}$ in the other equations.

As expansion forms we take the same three-forms as in eq.~\eqref{expformsthree}, while for the
two-forms we take this time %(see the Iwasawa vacuum in \eqref{iwasawasol})
\subeq{\al{
& Y_0^{(2-)} = \beta^2 e^{56} \, , \\
& Y_1^{(2-)} = e^{12}+e^{34} \, , \\
& Y_2^{(2-)} = e^{12}-e^{34} \, .
}}
Note that this time $Y_0^{(3+)}$ and $Y_0^{(2-)}$ are not closed.
Introducing $m_T$ such that $\beta=\frac{2}{5} e^{\Phi} m_T $ (this is of course the Romans mass
of the T-dual torus solution), we get the following masses:
\begin{center}
  \begin{tabular}{|c|c|}
    \hline
    mass eigenmode & mass (in units $m_T^2/25$) \\
    \hline
    \hline
    $c^{i}, \quad i=1,2,3$ & $0$ \\
    \hline
    $b^0+b^1$  &  $10$ \\
    \hline
    $b^2$  &  $10$ \\
    \hline
    $8c^{(3)0} + 5 b^0+3b^1$  &  $10$ \\
    \hline
    $c^{(3)0}-b^0+2b^1$  &  $88$ \\
    \hline
   \end{tabular}
\end{center}
Due to T-duality the mass eigenvalues are the same as for the torus solution.

The equation for the variation of the dilaton around the background reads
\eq{
0 = (\Delta + \frac{27 m_T^2}{25})\delta \phi -\frac{9 m_T^2}{25} \sum_{i=5}^6 h^i + \frac{3 m_T^2}{25} \sum_{i=1}^4 h^i \, .
}
For the Einstein equation we find:
\subeq{\al{
0 &= \Delta h^i +\frac{49 m_T^2}{50} h^i +\frac{53 m_T^2}{50} h^{i-(-1)^i} -\frac{11 m_T^2}{50} \sum_{j=1}^4 h^j - \frac{33 m_T^2}{50} \delta \phi \, \quad \text{for} \quad i=5,6 \, , \\
0 &= \Delta h^i +\frac{8 m_T^2}{25} h^i +\frac{2 m_T^2}{5} h^{i-(-1)^i} - \frac{3 m_T^2}{10} \sum_{j=5}^6 h^j +\frac{m_T^2}{10}\sum_{j=1}^4 h^j + \frac{3 m_T^2}{10} \delta \phi \, \quad \text{for} \quad i=1,2,3,4 \, .
}}
Here we used that
\eq{\label{variationR}
\delta R_{mn} = \frac{1}{2} \Delta_L \delta g_{mn} + \nabla_{(m}\nabla^{P} \delta g_{n)P}-\frac{1}{2}\nabla_m \nabla_n \delta g^Q{}_Q \, ,
}
where $\Delta_L$ is the Lichnerowicz operator defined in (\ref{Lichnerowicz}) and all covariant derivatives and contractions are with respect to the background metric. In (\ref{variationR}) the last two terms are vanishing.

Diagonalizing the mass matrix we find the following eigenmodes:
\begin{center}
  \begin{tabular}{|c|c|}
    \hline
    mass eigenmode & mass (in units $m_T^2/25$) \\
    \hline
    \hline
    $-h_{z^1\bar{z}^1}+h_{z^2\bar{z}^2}= -h^1-h^2+h^3+h^4 $  &  $18$ \\
    \hline
    $11 h_{z^1\bar{z}^1}+ 5 h_{z^3\bar{z}^3}= 11 (h^1+h^2)+ 5 (h^5+h^6)$  &  $18$ \\
    \hline
    $5 \delta \Phi  - 3 (h^1+h^2)$  &  $18$ \\
    \hline
    $3 \delta \Phi  -3 (h^5+h^6) + (h^1 + h^2 + h^3 + h^4)$ &  $70$ \\
    \hline
    $\Re h_{z^1z^1}=-h^1+h^2$ & $-2$ \\
    \hline
    $\Re h_{z^2z^2}=-h^3+h^4$ & $-2$ \\
    \hline
    $\Re h_{z^3z^3}=-h^5+h^6$ & $-2$ \\
    \hline
   \end{tabular}
\end{center}
Once again, we find the same masses as in the torus example.

\subsection{A note on integrating out $\d c_3^{(3)}$}
\label{noteC3}

Both in the torus and in the Iwasawa analysis we integrated out $d c_3^{(3)}$. In general one gets from the
part of the equation of motion of $F_4$ with $(1,6)$ index structure
\eq{\spl{
\label{F4extvar}
e^{\frac{1}{2} \Phi} \star_4 \d c_3^{(3)} \wedge \text{vol}_6 = &
+ \frac{1}{2} e^{\frac{1}{2}\Phi} f \left( \delta g^{\mu}{}_{\mu} - \delta g^{m}{}_m - \delta \Phi\right) \wedge \text{vol}_6 \\
& + c^{(3)i} \bg{H} \wedge Y^{(3+)}_i - b^i  \wedge Y^{(2-)}_i \wedge \bg{F}_4 + \delta f   \, ,
}}
where the integration constant $\delta f$ corresponds to a variation of the background flux $f$, which we put to zero.

This describes the external part of $F_4$, which equivalently can be described by the internal part of $F_6$.
{} Indeed, from varying
\eq{
F_6 = e^{\frac{1}{2} \Phi} \star F_4 \, ,
}
which we got from \eqref{Fduality}, follows
\eq{
\delta F_{6,\text{int}} = \frac{1}{2} e^{\frac{1}{2}\Phi} f \left( \delta g^{\mu}{}_{\mu} - \delta g^{m}{}_m - \delta \Phi\right) \wedge \text{vol}_6
+ e^{\frac{1}{2} \Phi} \star \d c_3^{(3)} \, ,
}
so that plugging in \eqref{F4extvar} we find
\eq{
\delta F_{6,\text{int}} = c^{(3)i} \bg{H} \wedge Y^{(3+)}_i - b^i  \wedge Y^{(2-)}_i \wedge \bg{F}_4 \, .
}
This corresponds to the part of $\delta F_6$ in \eqref{IIAfluc} that is first order in the fluctuations.
We conclude that instead of introducing $d c_3^{(3)}$, the external part of $F_4$, we might as well
have worked with the internal part of $F_6$. That is exactly what we will do in the superpotential analysis.

\section{Effective supergravity}\label{appeff4d}

The superpotential for SU(3)$\times$ SU(3)-structure was derived in various ways in \cite{granasup,grimmsup,effective} (based on \cite{GVW,gl}).
Here we will follow the approach of \cite{effective}, which calculated the superpotential and the (conformal)
K\"ahler potential in the superconformal formalism of \cite{supconf}.

The bosonic part of the effective four-dimensional superconformal action takes the following form
\eq{
\label{confaction}
S = \int \d^4 x \sqrt{-g_4} \left( \frac{1}{2} \mathcal{N} R + 3 \, \mathcal{N}_{I\bar{J}} \, g^{\mu\nu} D_{\mu} X^I D_{\nu} X^{*\bar{J}}
+ \frac{1}{3} \, \mathcal{W}_I \left(\mathcal{N}^{-1}\right)^{I\bar{J}} \mathcal{W}^*_{\bar{J}} + \cdots \right) \, ,
}
where the vector multiplet sector, including D-terms, has been omitted. Here the $X^I$ are the $n+1$ scalars and
$D_\mu X^I = \partial_\mu X^I - \frac{1}{3} i A_{\mu} X^I$, where $A_{\mu}$ is the gauge field associated to
the U(1)-transformations, generated by $\alpha$ (see \eqref{Weylsym}), in the complex Weyl transformation. From dimensional reduction of the ten-dimensional
supergravity action the conformal K\"ahler potential $\mathcal{N}$ and the superpotential $\mathcal{W}$ were
found and read (here we reinstate dimensionful coupling constants)
\subeq{\al{
\mathcal{N} & = \frac{1}{\kappa_{10}^2} \int_M\d^6y\sqrt{\det h}\, e^{2A-2\Phi} = \frac{1}{8 \kappa_{10}^2}\Big(i\int_{M}e^{-4A}\langle\mathcal{Z},\bar{\mathcal{Z}}\rangle\Big)^{1/3}\Big(i\int_{M}e^{2A}\langle t,\bar{t}\rangle\Big)^{2/3}\ , \\
\label{superpotential} \mathcal{W} & = \frac{1}{4 \kappa_{10}^2} \int_M\langle \mathcal{Z},F+i\,\d_H(\Re \mathcal{T})\rangle\ .
}}
Here $\mathcal{Z}$, $\Re \mathcal{T}$ and $t$ are defined through
\subeq{\al{
\mathcal{Z} & = -i e^{3A - \Phi} \Psi_2  \, , \\
t & = e^{-\Phi} \Psi_1 \, , \\
\Re \mathcal{T} & = \Im t = e^{-\Phi} \Im \Psi_1  \, .
}}

The dimensionally reduced action is naturally invariant under the following complex Weyl symmetry
\eq{
\label{Weylsym}
A \rightarrow A + \sigma \, , \quad g \rightarrow e^{-2\sigma} g \, , \quad \mathcal{Z} \rightarrow e^{3 \sigma +i \alpha} \mathcal{Z} \, ,
\quad \mathcal{N} \rightarrow e^{2 \sigma} \mathcal{N} \, .
}
Since the scalars $X^I$ transform as
\eq{
X^I \rightarrow e^{\sigma + \frac{i}{3} \alpha} X^I \, ,
}
we find that $\mathcal{Z}$ must be homogeneous of degree 3 in the $X^I$. To go to the usual Einstein frame, we must gauge-fix
the Weyl symmetry. We first explicitly isolate the unphysical degree of freedom, which is called the conformon, as follows
\eq{\label{Yextract}
X^I = Y x^I(\phi^i) \, , \qquad \mathcal{Z}=Y^3 \mathcal{Z}(\phi^i) \, \qquad \mathcal{N} = |Y|^2 e^{-\mathcal{K}/3}, \qquad \mathcal{W} = Y^3 M_P^{-3} \mathcal{W}_{\E}(\phi^i) \, ,
}
where $Y$ is the conformon, $\phi^i$ are the $n$ scalar degrees of freedom in the Einstein frame and $M_P$ the
four-dimensional Planck mass. $\mathcal{K}$ and $\mathcal{W}_{\E}$ will turn out to be the K\"ahler potential and the Einstein-frame superpotential after
gauge-fixing. Indeed, in the new coordinates the action \eqref{confaction} becomes
\eq{\spl{
S = & \int \d^4 x \sqrt{-g_4} \left[ \frac{1}{2} |Y|^2 e^{-\mathcal{K}/3} R -  |Y|^2 e^{-\mathcal{K}/3} \, \mathcal{K}_{i\bar{\jmath}} \, g^{\mu\nu}\, \partial_\mu \phi^i \partial_\nu \bar{\phi}^{\bar{\jmath}} + \cdots \right. \\
& \left. - M_P^{-6} |Y|^4 e^{\mathcal{K}/3} \left( \mathcal{K}^{i\bar{\jmath}} D_i \mathcal{W}_{\E} D_{\bar{\jmath}} \mathcal{W}^*_{\E} - 3 |\mathcal{W}_{\E}|^2 \right) + \cdots \right] \, ,
}}
where for the kinetic term of the scalars we omitted pieces that will vanish after the gauge-fixing.

We then impose the following gauge
\eq{\label{Eframe}
\mathcal{N} = |Y|^2 e^{-\mathcal{K}/3} = M_P^2 \, ,
}
which obviously gives us the usual Einstein-frame action
\eq{\label{4Daction}
S =  \int \d^4 x \sqrt{-g_4}  \left( \frac{M_P^2}{2} R  - M_P^2\mathcal{K}_{i \bar{\jmath}} \partial_{\mu} \phi^i \partial^{\mu}  \bar{\phi}^{\bar{\jmath}} - V (\phi,\bar{\phi}) \right)\, ,
}
and also leads to the standard expression for the potential
\eq{
V(\phi,\bar{\phi}) = M_P^{-2} e^{\mathcal{K}} \left( \mathcal{K}^{i\bar{\jmath}} D_i \mathcal{W}_{\E} D_{\bar{\jmath}} \mathcal{W}^*_{\E} - 3 |\mathcal{W}_{\E}|^2 \right) \, .
}
The U(1)-symmetry must also be gauged, but for more details on this we
refer to \cite{supconf}.

The K\"ahler potential reads
\eq{\spl{
\mathcal{K} & =  - \ln i \int_M e^{-4A} \langle \mathcal{Z}, \bar{\mathcal{Z}} \rangle - 2 \ln i \int_M e^{2A} \langle t , \bar{t} \rangle +3 \ln(8 \kappa_{10}^2|Y|^2)\, .
}}
Note that in \cite{hitchin} it is shown that $\Im t$ is a function of $\Re t$ so that $t$ can be seen as (non-holomorphically)
dependent on $\mathcal{T}$. To take this relation properly into account we use the fact that the K\"ahler potential for the $t$-sector
may be written as
\eq{
\mathcal{K}_t = - 2 \ln 4 \int_M e^{2A} H(\Im t) \, ,
}
where $H(\Im t)$ is the Hitchin functional \cite{hitchinold,hitchin,granasup}. For stable  pure spinors of $SO(6,6)$ it is defined as follows
\eq{\label{HitchinFunction}
H(\Im t)= \sqrt{-\frac{1}{12}\mathcal{J}^{\Pi}{}_{\Sigma} \mathcal{J}^{\Sigma}{}_{\Pi}} \, \, .
}
where $\mathcal{J}_{\Pi\Sigma} = \langle \Im t, \Gamma_{\Pi\Sigma} \Im t \rangle$ is a generalized complex structure and $\Pi,\Sigma=1,\ldots,12$. The generalized $SO(6,6)$ gamma matrices $\Gamma^{\Sigma}$ act on forms as
\eq{
\Gamma_\Sigma = \iota_m \quad \text{for} \quad m=\Sigma=1,\ldots,6 \, \quad \text{and} \quad  \Gamma_\Sigma = e^m \wedge \quad \text{for} \quad m+6=\Sigma = 7,\ldots,12 \, .
}
In the case of SU(3)-structure $\Im t=-\Im \Omega$ and the Hitchin functional reduces to \eqref{hitchinfunc}.

Note that if we make an expansion of the warp factor $A$ in harmonic modes
\eq{
A = A^0 + \sum_{\vec{n} \neq 0} A^{\vec{n}} \mathcal{Y}^{(0)}_{\vec{n}}(y) \, = A^0 + \tilde{A},
}
the Weyl transformation \eqref{Weylsym} only acts on $A^0$ since $\sigma$ is constant in the
internal coordinates (while of course it can depend on the four-dimensional coordinates).
Suppose $A$ and $\Phi$ are constant over the internal space (so $\tilde{A}$=0).
A good choice of $Y$ in \eqref{Yextract} would be
\eq{\label{choiceY}
Y = e^{A - \Phi/3} M_P \, ,
}
where the $M_P$ is introduced for convenience as it allows $\mathcal{K}$ to be dimensionless upon
imposing the Einstein gauge \eqref{Eframe}.
With this choice we find for the superpotential and the K\"ahler potential
\subeq{\label{canKW}\al{
\mathcal{K} & =  - \ln i \int_M \langle \Psi_2, \bar{\Psi}_2 \rangle - 2 \ln i \int_M \langle t , \bar{t} \rangle +3 \ln(8 \kappa_{10}^2 M_P^2)\, , \\
\mathcal{W}_{\E} & = \frac{-i}{4 \kappa_{10}^2} \int_M\langle \Psi_2,F+i\,\d_H(\Re \mathcal{T})\rangle\ .
}}
Note that another choice $Y'=f Y$ would amount
to a K\"ahler transformation
\eq{
\label{ktrans}
\mathcal{W}'_{\E} = f^{-3} \mathcal{W}_{\E} \, , \qquad
\mathcal{K}' = \mathcal{K} + 3 \ln f + 3 \ln f^* \, .
}

{}From the four-dimensional Einstein-frame action (\ref{4Daction}) we compute the equation of motion for the scalar fields
\eq{
\Delta \phi^k + M_P^{-2} (\mathcal{\hat{K}}^{-1} \hat{M})^k{}_i \phi^{i} = 0 \, ,
}
where $\bg{M}_{ij} = \frac{1}{2} \frac{\partial^2 V}{\partial \phi^i \partial \phi^{j}}|_{\text{background}}$ is the mass matrix and $\bg{\mathcal{K}}_{ij}$ is the K\"ahler metric in real coordinates in the background. Therefore, to compare the results for the masses in the analysis with the superpotential and the K\"ahler potential with the results from the Kaluza-Klein reduction we need to diagonalize the matrix $M_P^{-2} \mathcal{\bg{K}}^{-1} \hat{M}$.
%Using \eqref{canKW} we find the following dimensionless prefactor
%\eq{
%\label{extraFactor}
%32 \, M_P^2 \kappa_{10}^2 \vols^{-1} \, .
%}
We also have to take into account that the results from the Kaluza-Klein reduction were in the {\em ten-dimensional}
Einstein frame, while here we get the result in the {\em four-dimensional} Einstein frame:
\eq{\spl{
& g_s = e^{\frac{\Phi}{2}} g_{\E_{10}} \, , \\
& g_s = M_P^2 \mathcal{N}^{-1} g_{\E_4} \, ,
}}
and thus
\eq{
\label{convg}
g_{\E_{10}} = M_P^2 e^{-\Phi/2} \mathcal{N}^{-1} g_{\E_4} = M_P^2 \kappa_{10}^2 e^{-2A} \text{Vol}^{-1}_{\E}  \, g_{\E_4} \, ,
}
where in the last expression we assumed $A$ and $\Phi$ constant over the internal space.
The conversion for the mass is
\eq{
\label{convM}
m^2_{\E} = \kappa_{10}^{2} M_P^{2} e^{-2A} \text{Vol}^{-1}_{\E} \, m^2_{\E_{10}} \, .
}
%Comparing \eqref{extraFactor} and \eqref{convM}, putting the background values of $A$ and $\Phi$
%to zero we find the sought after factor of 32!

\section{$\mathcal{N}=2$ for IIA on $\frac{\text{SU(3)}\times \text{U(1)}}{\text{SU(2)}}$}
\label{SU3U1qSU2special}

{}From \eqref{aaa} we see that $5 \tilde{c}_1^2 = 4 \tilde{m}^2$
is a special point in that we have only two orientifolds, one along $345$ and one along $125$.
As a result there are more odd/even forms:
\eq{\spl{
Y^{(1+)}&=e^5, \qquad Y^{(1-)}=e^6 \, , \\
Y^{(2+)}&=e^{12}+e^{34}, \quad Y_1^{(2-)}=e^{13}-e^{24}, \quad Y_2^{(2-)}= e^{14}+e^{23}, \quad Y_3^{(2-)}= e^{56} \, , \\
Y_1^{(3-)}&=e^{145}+e^{235}, \quad Y_2^{(3-)}=e^{135}-e^{245}, \quad Y_3^{(3-)}=e^{126} + e^{346} \, , \\
Y_1^{(3+)}&=e^{146}+e^{236}, \quad Y_2^{(3+)}=e^{136}-e^{246}, \quad Y_3^{(3+)}=e^{125}+e^{345} \, , \\
Y^{(5+)}&=e^{12345}, \quad Y^{(5-)}=e^{12346} \, ,
}}
where we did not display the four-forms, which are dual to
the two-forms, because we do not need them for the expansion below.
However we find that $Y_1^{(3+)} \wedge Y_2^{(2-)} \neq 0$ and $Y_2^{(3+)} \wedge Y_1^{(2-)} \neq 0$
so that not all fluctuations expanded in these forms turn out to be consistent. Indeed, suppose we make the following expansion
\eq{\spl{
& J_c = J - i \delta B = t^i Y_i^{(2-)} \, , \\
& e^{-\Phi} e^{\delta B} \Im \Psi_1 + i \delta C = z^1 Y^{(1-)} + z^{1+i} Y_i^{(3+)} + z^5 Y^{(5-)} \, ,
}}
we see first of all that if $z^1 \neq 0$ or $z^5 \neq 0$ we do not have strict SU(3)-structure anymore but
rather intermediate SU(2) and secondly that the compatibility between the two pure spinors is not automatic anymore.
Thirdly there is a $\delta B$ fluctuation that affects both pure spinors. It is best to absorb $\delta B$ into the pure spinors
$\Psi_{1,2}'=\Psi_{1,2} e^{\delta B}$. The generalization of the strict SU(3) compatibility condition $J \wedge \Omega=0$
is \cite{granasup}
\eq{
\langle \Im \Psi'_1, \mathbb{X} \cdot \Psi'_2 \rangle = 0 \, ,
}
for arbitrary $\mathbb{X}$. Working this out we find the following constraint
\eq{
-2i \left[ (\Re z^2) t^2 + (\Re z^3) t^1 \right]
- \Re z^1 \left[ (t^1)^2 + (t^2)^2 \right] + \Re z^5 = 0 \, .
}
One way to solve this constraint is to put
\eq{
t^2= - \rho t^1 \, , \qquad \Re z^3 = \rho \Re z^2 \, , \qquad z^1=z^5=0 \, ,
}
which brings us to the restricted set of fluctuations in \eqref{SU3U1qSU2exp}.

However, in this model there can be three extra fluctuations in the NSNS-sector. One physical one, involving
only $\Im t_1, \Im t_2$, satisfying to linear order around the background
\eq{
\Im t^1 = \rho \, \Im t^2
}
and inducing $\delta B_{(2,0)}+ \delta B_{(0,2)}$. There are also two spurious ones
\eq{
\delta \Omega = -4 (g\bar{\chi}) \wedge J \, ,  \qquad \delta J = \iota \chi \Omega + \iota \bar{\chi} \bar{\Omega} \, ,
}
with $\chi = E^5 - \frac{i \sigma}{2\sqrt{1+\rho^2}} E^6$, and the other
involving $\Re z^1$ and $\Re z^5$ such that
\eq{
- \Re z^1 \left[ (t^1)^2 + (t^2)^2 \right] + \Re z^5 = 0 \, .
}
Indeed, one can check that these do not affect the metric nor the $B$-field, only the pure spinors
defining the structure \cite{granasup}. As such, one would not expect them to appear in the low energy effective action.

The presence of a spurious fluctuation that leads to leaving the strict SU(3)-structure case and that is still allowed under the orientifold projection, indicates
that the theory is in fact $\mathcal{N}=2$ and therefore outside the scope of this paper. Indeed, suppose $\eta$ was the original
internal spinor generating the supersymmetry and satisfying $\sigma^*(\eta_\pm)= \eta_\mp$ under all orientifolds (for the required transformation behaviour of
the internal spinors under supersymmetric orientifolds see \cite{kt}), there is now a second one
$\eta' = \gamma^{\underline{5}} \eta$ also satisfying $\sigma^*(\eta'_\pm)=\eta'_\mp$ under all orientifolds. It follows that we can make the $\mathcal{N}=2$
spinor ansatz
\eq{\spl{
& \epsilon_1 = \xi^{(1)}_+ \otimes \eta_+ + \xi^{(2)}_+ \otimes \eta'_+ + \, (\text{cc}) \, , \\
& \epsilon_2 = \xi^{(1)}_- \otimes \eta_+ + \xi^{(2)}_- \otimes \eta'_+ + \, (\text{cc}) \, .
}}
Note that this spinor ansatz is different from the usual $\mathcal{N}=2$ ansatz as found in e.g.\ \cite{granasup}.
The spurious deformations are not physical, so
they are not really in the spectrum. However, their partners in the RR-fields are. In total, the extra sector contains four scalars:
one from $\delta B_{(2,0)}+\delta B_{(0,2)}$, from $\delta C_{1}$, from $\delta C_{3}$ and from $\delta C_{5}$.
We also have four vectors: one from the metric (along $e^5$), from $C_{3}$, from $C_{5}$ and from $B$. This makes up the bosonic content
of a massive gravitino multiplet in $\mathcal{N}=2$. Note that since there is an extra internal spinor, we also have an extra gravitino.
%

%%%%%%%%%%%%%%%%%%%%%%%%%%%%%%%%%%%%%%%%%%%%%%%%%%%%%%%%%%%

\end{document}